\documentclass[rmp,aps,twocolumn]{revtex4}

\usepackage{graphics}
\usepackage{epsfig}

\newcommand{\be}{\begin{equation}}
\newcommand{\ee}{\end{equation}}
\newcommand{\ba}{\begin{eqnarray}}
\newcommand{\ea}{\end{eqnarray}}
\newcommand{\ban}{\begin{eqnarray*}}
\newcommand{\ean}{\end{eqnarray*}}
\newcommand{\braket}[2]{\mbox{$ \langle #1 | #2 \rangle $}}

\newcommand{\ket}[1]{\mbox{$ | #1 \rangle $}}
\newcommand{\bra}[1]{\mbox{$ \langle #1 | $}}

\newcommand{\si}{\sigma}

\newcommand{\demi}{\frac{1}{2}}

\newcommand{\one}{\leavevmode\hbox{\small1\normalsize\kern-.33em1}}

\newcommand{\Tr}{\mbox{Tr}}
\newcommand{\prob}{\mbox{Prob}}

\newcommand{\moy}[1]{\langle #1 \rangle}

\begin{document}

\title{The Security of Practical Quantum Key Distribution}
\author{Valerio Scarani$^{\,1,2}$, Helle Bechmann-Pasquinucci$^{\,3,4}$, Nicolas J. Cerf$^{\,5}$,
Miloslav Du\v{s}ek$^{\,6}$, Norbert L\"utkenhaus$^{\,7,8}$, Momtchil
Peev$^{\,9}$}
\affiliation{$^1$ Centre for Quantum Technologies and Department of Physics,
National University of Singapore, Singapore
\\ $^2$ Group of Applied Physics, University of Geneva, Geneva, Switzerland
\\ $^3$ University of Pavia, Dipartimento di Fisica ``A.\ Volta'', Pavia, Italy 
\\ $^4$ UCCI.IT, Rovagnate (LC), Italy 
\\ $^5$ Quantum Information and Communication, Ecole Polytechnique, Universit\'e Libre de Bruxelles,
Brussels, Belgium
\\ $^6$ Department of Optics, Faculty of Science, Palack\'y University, Olomouc, Czech Republic
\\ $^7$ Institute for Quantum Computing \& Department for Physics and Astronomy, University of Waterloo, Waterloo, Canada
\\ $^8$ Max Planck Research Group, Institute for Optics, Information and Photonics, University of Erlangen-Nuremberg, Erlangen, Germany
\\ $^9$ Quantum Technologies, Smart Systems Division, Austrian Research Centers GmbH – ARC, Vienna, Austria}
\date{\today}

\begin{abstract}

Quantum key distribution (QKD) is the first quantum information task to
reach the level of mature technology, already fit for
commercialization. It aims at the creation of a secret key between
authorized partners connected by a quantum channel and a classical
authenticated channel. The security of the key can in principle be
guaranteed without putting any restriction on the eavesdropper's power.

The first two sections provide a concise up-to-date
review of QKD, biased toward the practical side. The rest of the paper presents the essential theoretical tools that have been developed to assess the security of the main experimental platforms (discrete variables, continuous variables and distributed-phase-reference protocols).

\end{abstract}

\maketitle

\tableofcontents

\section{Introduction}
\label{secintro}

\subsection{Cryptography}
\label{secintrocrypto}

Cryptography is a field of applications that provide privacy, authentication and confidentiality to users. An important subfield is that of secure communication, aiming at allowing confidential communication between different parties such that no unauthorized party has access to the content of the messages. This field has a long history of successes and failures, as many methods to encode messages emerged along the centuries, always to be broken some time later.

History needs not repeat forever, though. In 1917, Vernam invented the so-called \textit{One-Time Pad} encryption, which uses a symmetric, random secret key shared between sender and receiver \cite{ver26}. This scheme cannot be broken in principle, provided the parties do not reuse their key. Three decades later, Shannon proved that the Vernam scheme is optimal: there is no encryption method that requires less key \cite{sha49}. This means that the key is being used up in the process. To employ this scheme, therefore, the communicating parties must have a secure method to share a key as long as the text to be encrypted. Because of this limitation, which becomes severe in case huge amounts of information have to be securely transmitted, most cryptographic applications nowadays are based on other schemes, whose security cannot be proved in principle, but is rather based on our experience that some problems are hard to solve. In other words, these schemes can be broken, but with a substantial amount of computational power. One can therefore set a security parameter to a value, such that the amount of required computational power lies beyond the amount deemed to be available to an adversary; the value can be adjusted in time, along with technological advances.

The picture has changed in the last two decades, thanks to unexpected inputs from \textit{quantum physics}. In the early 1980s, Bennett and Brassard proposed a solution to the key distribution problem based on quantum physics \cite{bb84}; this idea, independently re-discovered by Ekert a few years later \cite{eke91}, was the beginning of \textit{quantum key distribution (QKD)} which was to become the most promising task of \textit{quantum cryptography}\footnote{Quantum cryptography is often identified with QKD, but actually comprises all possible tasks related to secrecy that are implemented with the help of quantum physics. The first appearance of a link between secrecy and quantum physics was Wiesner's idea of \textit{quantum money}, which dates back to the early 1970s although was published a decade later \cite{wiesner83a}. To our knowledge, there is nothing else before Bennett's and Brassard's first QKD protocol. In 1999, two new tasks were invented and both were given the same name, \textit{quantum secret sharing}. In one case, the protocol is a multi-partite generalization of key distribution \cite{hil99,kar99}; in the other case it refers to the sharing of secret quantum information, i.e. the goal is for the authorized partners to share quantum information (instead of a list of classical random variables) known only to them \cite{cle99,cre05}. Other examples of cryptographic tasks are bit commitment or oblivious transfer; for these tasks, contrary to the case of QKD and secret sharing, quantum physics cannot guarantee unconditional security \cite{lo97,may97,lo97b} and therefore their interest seems limited --- though new paradigms like ``bounded-storage models'' may change this perception in the future \cite{dam05,dam07,weh08}.}. Since then, QKD devices have constantly increased their key generation rate and have started approaching maturity, needed for implementation in realistic settings.

In an intriguing independent development, ten years after the advent of QKD, Peter Shor discovered that large numbers can in principle be factorized efficiently if one can perform coherent manipulations on many quantum systems \cite{sho94,sho97}. Factorizing large numbers is an example of a mathematical task considered classically hard to solve and for this reason related to a class of cryptographic schemes which are currently widely used. Though quantum computers are not realized yet, the mere fact that they could be built brought into awareness that the security of some cryptographic schemes may be threatened\footnote{This issue will be discussed in more detail in Sec. \ref{sscompare}.}.

This review focuses therefore on the cryptographic task of key distribution, and in particular on its realization using quantum physics. Note that a secret key serves many useful purposes in cryptography other than message encryption: it can be used, for example, to authenticate messages, that is, to prove that a message has been indeed sent by the claimed sender.

\subsection{Basics of Quantum Key Distribution (QKD)}
\label{ssbasic}

In this paragraph, we introduce the basic elements of quantum key distribution (QKD), for the sake of those readers who would not be familiar with the field. Alternative presentations of this material are available in many sources, ranging from books with rather general scope \cite{lo98,eke01,leb06,sca06b} to other review articles specific to the topic \cite{gis02,dus06,lo08}. 

\subsubsection{Generic setting}

\begin{figure}[ht]
\includegraphics[scale=0.5]{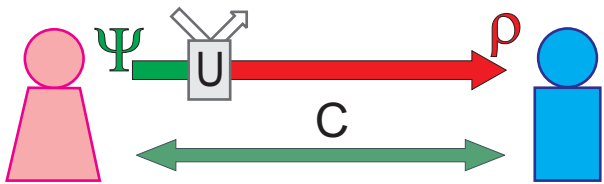}
\caption{(Color online) The setting of QKD: Alice and Bob are connected by a quantum channel, on which Eve can tap without any restriction other than the laws of physics; and by an authenticated classical channel, which Eve can only listen to.}\label{fig:basic}
\end{figure}

The generic settings of QKD are schematically represented in Fig.~\ref{fig:basic}. The two \textit{authorized partners}, those that want to establish a secret key at a distance, are traditionally called Alice and Bob. They need to be connected by two channels: a {\em quantum channel}, allowing them to share quantum signals; and a {\em classical channel}, on which they can send classical messages forth and back.

The classical channel needs to be \textit{authenticated}: this means that Alice and Bob identify themselves; a third person can listen to the conversation but cannot participate in it. The quantum channel, however, is open to any possible manipulation from a third person. Specifically, the task of Alice and Bob is to guarantee security against an adversarial \textit{eavesdropper}, usually called Eve\footnote{The name, obtained from assonance with the English term ``eavesdropping'', is remarkably suited for someone whose task is to mess things up!}, tapping on the quantum channel and listening to the exchanges on the classical channel.

By ``security'' we mean that``a non-secret key is never used'': either the authorized partners can indeed create a secret key (a common list of secret bits known only to themselves), or they abort the protocol\footnote{No physical principle can prevent an adversary to cut the channels, thus blocking all transfer of information between Alice and Bob. Stepping back then, one can imagine the following eavesdropping strategy (suggested to one of us by A. Beveratos): Eve systematically cuts all QKD channels, until Alice and Bob, who after all want to communicate, opt for less secure methods --- and then Eve gets the information. There is obviously a point of humor in this idea but, given that Eve has no hope if QKD is used correctly, this strategy may be the most effective indeed.}. Therefore, after the transmission of a sequence of symbols, Alice and Bob must estimate how much information about their lists of bits has leaked out to Eve. Such an estimate is obviously impossible in classical communication: if someone is tapping on a telephone line, or when Eve listens to the exchanges on the classical channel for that matters, the communication goes on unmodified. This is where quantum physics comes into the game: \textit{in a quantum channel, leakage of information is quantitatively related to a degradation of the communication}. The next paragraph delves a bit deeper into the physical reasons for this statement.

\subsubsection{The origin of security}
\label{sssorigin}

The origin of security of QKD can be traced back to some fundamental principles of quantum physics. One can argue for instance that any action, by which Eve extracts some information out of quantum states, is a generalized form of measurement; and a well-known tenet of quantum physics says that \textit{measurement in general modifies the state of the measured system}. Alternatively, one may think that Eve's goal is to have a perfect copy of the state that Alice sends to Bob; this is however forbidden by the \textit{no-cloning theorem} \cite{woo82}, which states that one cannot duplicate an unknown quantum state while keeping the original intact. Both these arguments appear already in the seminal paper \cite{bb84}; they lead to the same formalization. A third physical argument can be invoked, which is usually considered rather as a fact than as a principle, but a very deep one: quantum correlations obtained by separated measurements on members of entangled pairs \textit{violate Bell's inequalities} and cannot therefore have been created by pre-established agreement. In other words, the outcomes of the measurements did not exist before the measurements; but then, in particular, Eve could not know them \cite{eke91}. This argument supposes that QKD is implemented with entangled states.

The fact that security can be based on general principles of physics suggests the possibility of \textit{unconditional security}, i.e. the possibility of guaranteeing security without imposing any restriction on the power of the eavesdropper (more on this notion in Sec. \ref{sssuncond}). Indeed, at the moment of writing, unconditional security has been proved for several QKD protocols.

\subsubsection{The choice of light}
\label{sslight}

In general, quantum information processing can be implemented with any system, and one indeed finds proposal to implement quantum computing with ions, atoms, light, spins... Abstractly, this is the case also for QKD: one could imagine to perform a QKD experiment with electrons, ions, molecules; however, \textit{light} is the only practical choice. Indeed, the task of key distribution makes sense only if Alice and Bob are separated by a macroscopic distance: if they are in the same room, they have much easier ways of generating a common secret key.

Now, as well known, light does not interact easily with matter; therefore quantum states of light can be transmitted to distant locations basically \textit{without decoherence}, in the sense that little perturbations are expected in the definition of the optical mode. The problem with light is scattering, i.e. \textit{losses}: quite often, the photons just don't arrive. The way losses affect QKD varies with the protocol and the implementation; we shall deal with these issues in detail later, but it's useful to give here a rapid overview. First and quite obviously, losses impose bounds on the secret key rate (that cannot scale with the distance better than the transmittivity of the line) and on the achievable distance (when losses are so large that the signal is lost in spurious events, the ``dark counts''). Second: losses may leak information to the eavesdropper, according to the nature of the quantum signal: for coherent pulses it is certainly the case, for single photons it is not, the case for entangled beams is more subtle. A third basic difference is determined by the detection scheme. Indeed, implementations that use photon counters rely on post-selection: if a photon does not arrive, the detector does not click and the event is simply discarded\footnote{Note that this is possible because the task is to distribute a random key. In the days of booming of quantum information, some authors considered the possibility of sending directly the message on the quantum channel \cite{bei02,bos02}. This task has been called ``Quantum Secure Direct Communication'' and has generated some interest. However, it was soon recognized (even by some of the original authors) that the idea suffers of two major defaults with respect to standard QKD: (i) It is obviously not robust against losses: you cannot afford losing a significant amount of the message. (ii) It allows no analog of privacy amplification: if an eavesdropper obtains information, it is information \textit{on the message itself} and cannot of course be erased.}. On the contrary, implementations that use homodyne detection always give a signal, therefore losses translate as additional noise.

In summary, \textit{QKD is always implemented with light} and there is no reason to believe that things will change in the future. As a consequence, the quantum channel is any medium that propagates light with reasonable losses: typically, either an optical fiber, or just free space provided Alice and Bob have a line of sight.

\subsubsection{The BB84 protocol}

All the points and concepts introduced above will be dealt in more depth and detail in the main sections of this review. Let us first practice the generic ideas on a very concrete example: the first QKD protocol, published by Bennett and Brassard in 1984 and called therefore BB84 \cite{bb84}.

Suppose Alice holds a source of single photons. The spectral properties of the photons are sharply defined, so the only degree of freedom left is polarization. Alice and Bob align their polarizers and agree to use either the Horizontal/Vertical ($+$) basis, or the complementary basis of linear polarizations i.e. +45/-45 ($\times$). Specifically, the coding of bits is
\ba
\begin{array}{lcl}
\ket{H}&\mbox{ codes for }&0_+\\ \ket{V}&\mbox{ codes for }&1_+\\ \ket{+45}&\mbox{ codes for }&0_\times\\ \ket{-45}&\mbox{ codes for }&1_\times \end{array}\,.
\ea We see that both bit values 0 and 1 are coded in two possible ways, more precisely in non-orthogonal states, because
\ba
\ket{\pm 45}&=&\frac{1}{\sqrt{2}}\big(\ket{H}\pm\ket{V}\big)\,.
\ea
Given this coding, the BB84 protocol goes as follows:
\begin{enumerate}
\item Alice prepares a photon in one of the four states above and sends it to Bob on the quantum channel. Bob measures it in either the $+$ or the $\times$ basis. This step is repeated $N$ times. Both Alice and Bob have now a list of $N$ pairs $(\mbox{bit,basis})$.

\item Alice and Bob communicate over the classical channel and compare the ``basis'' value of each item and discard those instances in which they have used different bases. This step is called \textit{sifting}. At its end, Alice and Bob have a list of approximately $N/2$ bits, with the promise that for each of them Alice's coding matched Bob's measurement. This list is called \textit{raw key}.

\item Alice and Bob now reveal a random sample of the bits of their raw keys and estimate the error rate in the quantum channel, thus in turn Eve's information. In the absence of errors, the raw key is identical for Alice and Bob \textit{and} Eve has no information: in this case, the raw key is already the secret key. If there are errors however, Alice and Bob have to correct them and to erase the information that Eve could have obtained\footnote{Historical note: the procedure that erases the information of the eavesdropper was not discussed in \cite{bb84} and appears for the first time a few years later \cite{ben88}.}. Both tasks can be performed by communication on the classical channel, so this part of the protocol is called \textit{classical post-processing}. At the end of this processing, Alice and Bob share either a truly secret key or nothing at all (if Eve's information was too large).
\end{enumerate}

\subsubsection{An example of eavesdropping}

A particularly simple eavesdropping strategy is the one called \textit{intercept-resend}. To obtain information, Eve does the same as Bob: she intercepts the photon coming from Alice and measures it either in the $+$ or in the $\times$ basis. But Bob is waiting for some signal to arrive. Let's then suppose that Eve resends the same photon to Bob (Eve is limited only by the laws of physics, therefore in particular she can perform a quantum non-demolition measurement). If Eve has measured in the basis of Alice's preparation, the photon is intact: on such instances, Eve has got full information on Alice's bit without introducing any errors. However, when Eve has chosen the wrong basis, her result is uncorrelated with Alice's bit; moreover, she has modified the state so that, even if Bob uses the same basis as Alice, half of the times he will get the wrong result.

In average over long keys then, this particular attack gives Eve full information on half of the bits of the raw key ($I_E=0.5$) at the price of introducing an error rate $Q=0.25$. Can a secure key be extracted under such conditions? One has to know how to quantify the length of the final key that can be extracted. For this particular case, under some assumptions on the classical post-processing it holds \cite{csi78}
\ba
r&=&\max\{I(A:B)-I_E,0\}\,.
\ea where $I(A:B)=H(A)+H(B)-H(AB)$ is the mutual information between Alice's and Bob's raw keys ($H$ is Shannon entropy). Assuming that both bit values are equally probable, i.e. $H(A)=H(B)=1$, one has $I(A:B)=1-h(Q)$ where $h$ is binary entropy. Having these elements, one can plug in the values obtained for the intercept-resend attack and find that $I(A:B)<I_E$: Eve has more information on Alice's string than Bob, therefore no secret key can be extracted\footnote{This conclusion is valid for all protocols: no secret key can be extracted if the observed statistics are compatible with Eve performing the intercept-resend attack \cite{cur04b}. The reason is that this attack ``breaks'' the quantum channel into two pieces, in which case the correlations between Alice and Bob can always be obtained with classical signals; and no secrecy can be distributed with classical communication.}.

Another simple exercise consists in supposing that Eve perform the intercept-resend attack only on a fraction $p$ of the photons sent by Alice, and leaves the others untouched. Then obviously $Q=p/4$ and $I_E=p/2=2Q$; this leads to conclude that, if $Q\gtrsim 17\%$, a secure key cannot be extracted from the BB84 protocol --- at least, if the classical post-processing is done according to the assumptions of \cite{csi78}.

\subsubsection{Beyond the example: the field of QKD}

The basic example that we have just presented calls for a number of important questions: 
\begin{itemize}
\item The adversary is clearly not restricted to perform the intercept-resend attack. What is the maximal amount of information Eve can possibly obtain, if she is allowed to do anything that is compatible with the laws of physics? This is the question about the possibility of proving unconditional security.
\item The BB84 protocol is just a particular protocol. What about other forms of coding and/or of processing the data?
\item The protocol supposed that the quantum signal is a \textit{qubit} --- explicitly, a bimodal single photon, i.e. an elementary excitation of the light field in only two modes (polarization in the explicit example). How close can an implementation come to this? And after all, should any implementation of QKD actually aim at coming close to this?
\item In a real device, information may leak out in channels that are neglected in a theoretical description. What are the potential threats in an implementation?
\end{itemize}
The whole field of QKD has developed along the answer to these and similar questions.

\subsection{Scope of this review}

\subsubsection{Focus}

The label ``quantum cryptography'' applies nowadays to a very wide range of interests, going from abstract mathematical considerations to strictly technological issues.

This review focuses somewhere in the middle of this range, in the realm where theoretical and experimental physics meet, that we call \textit{practical QKD}. There, theorists cannot pursue pure formal elegance and are compelled to complicate their models in order to take real effects into account; and experimentalists must have a serious grasp on theoretical issues in order to choose the right formulas and make the correct claims about the security of their devices. Specifically, we want to address the following two concerns:
\begin{enumerate}
\item On the one hand, the theoretical tools have reached a rather satisfactory level of development; but from outside the restricted group of experts, it has become almost impossible to follow this development, due also to the fact that quite a few strong security claims made in the past had to be revisited in the light of better understanding. As theorists involved in the development of security proofs, we want to provide an updated review of the status of such proofs.
\item On the other hand, several competing experimental platforms exist nowadays. It is desirable to have a synthetic view of those, highlighting the interest and possible shortcomings of each choice. Also, we want to raise the awareness of the complexity of any comparison: ``physical'' figures of merit like the secret key rate or the maximal achievable distance are in competition with ``practical'' figures of merit like stability and cost.
\end{enumerate}
Along the review, we shall make reference also to some strictly mathematical or strictly technological progresses, but without any claim of exhaustiveness.

\subsubsection{Outline}

The review is structured as follows. Section \ref{secelem}
introduces all the basic elements of practical QKD. Section
\ref{secssk} is devoted to the rate at which a secret key is
produced: this is the fundamental parameter of QKD, and depends
both on the speed and efficiency of the devices, and on the
intrinsic security of the protocol against eavesdropping. The next
three sections provide a detailed analysis, with a consistent set
of explicit formulas, for the three main families of protocols:
those based on discrete-variable coding (Section \ref{secdiscr}),
those based on continuous-variable coding (Section \ref{seccv})
and the more recent distributed-phase-reference coding (Section
\ref{secmixed}). In Section \ref{seccompare}, we put everything
together and sketch some directions for comparison of different
experimental platforms. Finally, in Section \ref{secpersp}, we
discuss future perspectives for QKD, both as a field in itself and
in the broader context of key distribution.

\section{The Elements of Practical QKD}
\label{secelem}

\subsection{Milestones}
\label{ssmilestones}

\subsubsection{Foundations: 1984-1995}

QKD unfolded with the presentation of the first complete protocol \cite{bb84}, which was based on earlier ideas by Wiesner \cite{wiesner83a}. In the BB84 protocol, bits are coded in two complementary bases of a two level system (qubit); this qubit is sent by Alice to Bob, who measures it. The no-cloning theorem is explicitly mentioned as the reason for security. This work was published in conference proceedings and was largely unknown to the community of physicists. It was not until 1991, when Artur Ekert, independently from the earlier developments, published a paper  on quantum key distributions, that the field gained a rapid popularity \cite{eke91}. Ekert's argument for security had a different flavor: an eavesdropper introduces ``elements of reality'' into the correlations shared by Alice and Bob; so, if they observe correlations that violate a Bell inequality, the communication cannot have been completely broken by Eve. Shortly later, Bennett, Brassard and Mermin argued\footnote{The argument is correct under some assumptions; only around the year 2006 it was fully realized that Ekert's view is qualitatively different and allows to reduce the set of assumptions about Alice's and Bob's devices; see \ref{sssdevindep}. This is also why the Ekert protocol was not implemented as such in an experiment until very recently \cite{lin08}.} that entanglement-based protocols, such as E91, are equivalent to prepare\&measure protocols, such as the BB84 protocol \cite{ben92a}. The same year 1992 witnessed two additional milestones: the invention of the B92 protocol \cite{ben92} and the very first in-principle experimental demonstration \cite{bennett92a}. One can reasonably conclude the foundational period of QKD with the definition of \textit{privacy amplification}, the classical post-processing needed to erase Eve's information from the raw key \cite{ben95}.

\subsubsection{The theory-experiment gap opens: 1993-2000}

After these foundational works, the interest and feasibility of QKD became apparent to many. Improved \textit{experimental demonstrations} took place, first in the lab with a growing distance of optical fiber next to the optical table \cite{tow93,bre94,fra94}, then in installed optical fibers \cite{mul95}, thereby demonstrating that QKD can be made sufficiently robust for a real-world implementation. In this development, an obvious milestone is the invention of the so-called \textit{Plug\&Play setups} by the Geneva group \cite{mul97,rib98}. By the year 2000, QKD over large distances was demonstrated also with entangled photons \cite{jen00,nai00,tit00}.

Theorists became very active too. \textit{New protocols} were proposed. For instance, the elegant six-state protocol, first mentioned back in 1984 as a possible extension of BB84 \cite{ben84b}, was rediscovered and studied in greater detail \cite{bru98,bec99}. But by far a more complex task was at stake: the derivation of \textit{rigorous security proofs} that would replace the intuitive arguments and the first, obviously sub-optimal estimates. The first such proof has been given by Mayers, who included even advanced features such as the analysis of finite key effects \cite{may96,may01}. However, this proof is not very intuitive, and other proofs emerged, starting with the basic principle of entanglement distillation ideas \cite{deutsch96a} which were put into a rigorous framework by Lo and Chau \cite{lo99b}. These entanglement based proofs would require the ability to perform quantum logic operations on signals. At present, we do not have the experimental capability to do so. Therefore the result by Shor and Preskill \cite{sho00} provided a step forward, as it combined the property of Mayers result of using only classical error correction and privacy amplification with a very intuitive way of proving the security of the BB84 protocol. That result uses the ideas of quantum error correction methods, and reduces the corresponding quantum protocol to an actual classically-assisted prepare-and-measure protocol.

As of the year 2000 therefore, both experimental and theoretical QKD had made very significant advances. However, almost inevitably, a gap had opened between the two: security proofs had been derived only for very idealized schemes; setups had been made practical without paying attention to all the security issues.

\subsubsection{Closing the gap: 2000 to present}

The awareness of the gap was triggered by the discovery of \textit{photon-number-splitting (PNS) attacks} \cite{bra00}, which had actually been anticipated years before \cite{ben92,hut95,dus99} but had passed rather unnoticed. The focus is on the source: the theoretical protocols supposed single-photon sources, but experiments were rather using attenuated laser pulses, with average photon numbers below one. In these pulses, photons are distributed according to the Poissonian statistics: in particular, there are sometimes two or more photons, and this opens an important loophole. Security proofs could be adapted to deal with the case \cite{lut00,ina07,got04}: the extractable secret key rate was found to scale much worse with the distance than for single-photon sources ($t^2$ compared to $t$, where $t$ is the transmittivity of the quantum channel).

It took a few years to realize that methods can be devised to reduce the power of PNS attacks while keeping the very convenient laser sources. One improvement can be made by a mere change of software by modifying the announcements of the BB84 protocol \cite{sca04}: in this SARG04 protocol, the key rate scales as $t^{3/2}$ \cite{koa05b,kra05}. Another significant improvement can be made by an easy change of hardware: by varying the quantum state along the protocol (\textit{decoy states}), one can perform a more complete test of the quantum channel \cite{hwa03}. When the decoy state idea is applied to laser sources, the key rate scales as $t$ \cite{lo05,wan05}.

Parallel to this development, the field of practical QKD\footnote{The whole field of QKD witnessed many other remarkable developments, especially in theoretical studies, which are not included in this paragraph but are mentioned in due place in the paper.} has grown in breadth and maturity. New families of protocols have been proposed, notably \textit{continuous-variable protocols} \cite{ralph,hillery,GP01,Cerf01,GG02,silberhorn} and the more recent \textit{distributed-phase-reference protocols} \cite{ino02,stu05}. Critical thinking on existing setups has lead to the awareness that the security against Eve tapping on the quantum channel is not all: one should also protect the devices against more commonplace \textit{hacking attacks} and verify that information does not leak out in \textit{side-channels} \cite{mak05}. Since a short time, QKD has also reached the \textit{commercial market}: at least three companies\footnote{idQuantique, Geneva (Switzerland), www.idquantique.com; MagiQ Technologies, Inc., New York., www.magiqtech.com; and Smartquantum, Lannion (France), www.smartquantum.com.} are offering working QKD devices. New questions can now be addressed: in which applications QKD can help \cite{all07}, how to implement a network of QKD systems\footnote{This is the aim of the European Network SECOQC, www.secoqc.net.}, how to certify QKD devices for commercial markets (including the verification that these devices indeed fulfill the specifications of the corresponding security proofs) etc.

\subsection{Generic QKD Protocol}

\subsubsection{Classical and quantum channels}
\label{introchannels}

As introduced in Sec. \ref{ssbasic}, Alice and Bob need to be connected by two channels. On the {\em quantum channel}, Alice can send quantum signals to Bob. Eve can interact with these signal, but if she does, the signals are changed because of the laws of quantum physics -- the essence of QKD lies precisely here.

On the {\em classical channel}, Alice and Bob can send classical messages forth and back. Eve can listen without penalty to all communication that takes place on this channel. However, in contrast to the quantum channel, the classical channel is required to be authenticated, so that Eve cannot change the messages that are being sent on this channel. Failure to authenticate the classical channel can lead to the situation where Eve impersonates one of the parties to the other, thus entirely compromising the security. Unconditionally secure authentication\footnote{Authentication schemes that do not rely on pre-shared secrecy exist, but are not unconditionally secure. Since we aim at unconditional security for QKD, the same level of security must in principle be guaranteed in all the auxiliary protocols. However, breaking the authentication code after one round of QKD does not threaten security of the key that has been produced; one may therefore consider authentication schemes that guarantee security only for a limited time, e.g.based on complexity assumptions.} of the classical channel requires Alice and Bob to pre-share an initial secret key or at least partially secret but identical random strings \cite{ren03}. QKD therefore does not create a secret key out of nothing: rather, it will expand a short secret key into a long one, so strictly speaking it is a way of key-growing. This remark calls for two comments. First, key growing cannot be achieved by use of classical means alone, whence QKD offers a real advantage. Second, it is important to show that the secret key emerging from QKD is {\em composable}, that is, it can be used like a perfect random secret key in any task (more in Sec. \ref{ssssecu}), because one has to use a part of it as authentication key for the next round. 

\subsubsection{Quantum information processing}
\label{sssprotoq}

The first step of a QKD protocol is the exchange and measurement of signals on the quantum channel. Alice's role is \textit{encoding}: the protocol must specify which quantum state $\ket{\Psi({\cal S}_n)}$ codes for the sequence of $n$ symbols ${\cal S}_n=\{s_1,...,s_n\}$. In most protocols, but not in all, the state $\ket{\Psi({\cal S}_n)}$ has the tensor product form $\ket{\psi(s_1)}\otimes...\otimes \ket{\psi(s_n)}$. In all cases, it is crucial that the protocol uses a set of non-orthogonal states\footnote{There is only one 
exception \cite{gol95} when Alice uses just two orthogonal states. Alice 
prepares a qubit in one of the two orthogonal superposition of two
spatially separated states, then -- at a random time instant --
she sends one component of this superposition to Bob. Only later
she sends the second component. Precise time synchronization
between Alice and Bob is crucial. See also Peres' criticism \cite{per96}, the authors' reply \cite{gol96} and a related discussion \cite{koa97}. Unconditional security has not been proved for this protocol.}, otherwise Eve could decode the sequence without introducing errors by measuring in the appropriate basis (in other words, a set of orthogonal states \textit{can} be perfectly cloned). Bob's role is twofold: his measurements allow of course to \textit{decode} the signal, but also to \textit{estimate the loss of quantum coherence} and therefore Eve's information. For this to be possible, non-compatible measurements must be used.

We have described the quantum coding of QKD protocols with the language of \textit{Prepare-and-Measure} (P\&M) schemes: Alice chooses actively the sequence ${\cal S}_n$ she wants to send, prepares the state
$\ket{\Psi({\cal S}_n)}$ and sends it to Bob, who performs some measurement. Any such scheme can be immediately translated into an \textit{entanglement-based} (EB) scheme: Alice prepares the entangled state \ba
\ket{{\mathbf{\Phi}}^n}_{AB}&=&\frac{1}{\sqrt{d_n}}\sum_{{\cal
S}_n}\,\ket{{\cal S}_n}_A\otimes\ket{\Psi({\cal S}_n)}_B  \label{eqentgb}\ea
where $d_n$ is the number of possible ${\cal S}_n$ sequences and the $\ket{{\cal S}_n}_A$ form an orthogonal basis. By measuring in this basis, Alice learns one ${\cal S}_n$ and prepares the corresponding $\ket{\Psi({\cal S}_n)}$ on the sub-system that is sent to Bob: from Bob's point of view, nothing changes. This formal translation obviously does not mean that both realizations are equally practical or even feasible with present-day technology. However, it implies that the security proof for the EB protocol translates immediately to the corresponding P\&M protocol and viceversa.

A frequently quoted statement concerning the \textit{role of entanglement} in QKD says that ``entanglement is a necessary condition to extract a secret key'' \cite{cur04b,aci05}. Two important comments have to be made to understand it correctly. First of all, this is not a statement about implementations, but about the quantum channel: it says that no key can be extracted from an entanglement-breaking channel\footnote{As the name indicates, a channel $\rho\rightarrow \rho'=C(\rho)$ is called \textit{entanglement-breaking} if $(\one\otimes C)\ket{\Psi}_{AB}$ is separable for any input $\ket{\Psi}_{AB}$. A typical example of such a channel is the one obtained by performing a measurement on half of the entangled pair.}. In particular, the statement does \textit{not} say that entanglement-based implementations are the only secure ones.

Second: as formulated above, the statement has been derived under the assumption that Eve holds a purification of $\rho_{AB}$, where $A$ and $B$ are the degrees of freedom that Alice and Bob are going to measure. One may ask a more general question, namely, how to characterize all the \textit{private states}, i.e. the states out of which secrecy can be extracted \cite{hor05,hor06,hor08}. It was realized that, in the most general situation, Alice and Bob may control some additional degrees of freedom $A'$ and $B'$; thus, Eve is not given a purification of $\rho_{AB}$, but of $\rho_{AA'BB'}$. In such situation, it turns out that $\rho_{AB}$ can even be separable; as for $\rho_{AA'BB'}$, it must be entangled, but may even be bound entangled. The reason is quite clear: $A'$ and $B'$ shield the meaningful degrees of freedom from Eve's knowledge. We do not consider this most general approach in what follows\footnote{In \cite{smi06}, the formalism of private states is used to study pre-processing, see \ref{12way}.}, because at the moment of writing no practical QKD scheme with shielding systems has been proposed.

\subsubsection{Classical information processing}
\label{sssprotoc}

Once a large number $N$ of signals have been exchanged and measured on the quantum channel, Alice and Bob start processing their data by exchanging communication on the classical channel. In all protocols, Alice and Bob estimate the statistics of their data; in particular, they can extract the meaningful parameters of the quantum channel: error rate in decoding, loss of quantum coherence, transmission rate, detection rates... This step, called \textit{parameter estimation}, may be preceded in some protocols by a \textit{sifting} phase, in which Alice and Bob agree to discard some symbols (typically, because Bob learns that he has not applied the suitable decoding on those items). After parameter estimation and possibly sifting, both Alice and Bob hold a list of $n\leq N$ symbols, called \textit{raw keys}. These raw keys are only partially correlated and only partially secret. Using some \textit{classical information post-processing} (see \ref{12way}), they can be transformed into a fully secure key $K$ of length $\ell \leq n$. The length $\ell$ of the final secret key depends of course on Eve's information on the raw keys.

\subsubsection{Secret fraction and secret key rate}
\label{ssssecret}

In the asymptotic case $N\rightarrow\infty$ of infinitely long keys, the meaningful quantity is the {\em secret fraction}\footnote{Often, especially in theoretical studies, this quantity is called ``secret key rate''. In this paper, we reserve this term to (\ref{eqK}), which is more meaningful for practical QKD.} \ba
r&=&\lim_{N\rightarrow\infty} \ell/n\,.\ea The secret fraction is clearly the heart of QKD: this is the quantity for which the security proofs (\ref{sssproofs}) must provide an explicit expression. However, a more prosaic parameter must also be taken into account as well in practical QKD: namely, the {\em raw-key rate} $R$, i.e. the length of the raw key that can be produced per unit time. This rate depends partly on the protocol: for instance, it contains the \textit{sifting factor}, i.e. the fraction of exchanged symbols that is discarded in a possible sifting phase. But, surely enough, its largest dependence is on the details of the setup: repetition rate of the source, losses in the channel, efficiency and dead time of the detectors, possible duty cycle, etc. In conclusion, in order to assess the performances of practical QKD systems, it is natural to define the \textit{secret key rate} as the product
\ba K&=&R\, r\,. \label{eqK}\ea The whole Section \ref{secssk} will be devoted to a detailed discussion of this quantity.

As mentioned, these definitions hold in the asymptotic regime of infinitely long keys. When \textit{finite-key corrections} are taken into account, a reduction of the secret fraction is expected, mainly for two reasons. On the one hand, parameter estimation is made on a finite number of samples, and consequently one has to consider the worst possible values compatible with statistical fluctuations. On the other hand, the yield of the classical post-processing contains terms that vanish only in the asymptotic limit; intuitively, these correction take care of the fact that security is never absolute: the probability that Eve knows a $n$-bit key is at least $2^{-n}$, which is strictly positive. In this review, we restrict our attention to the asymptotic case, not because finite-key corrections are negligible --- quite the opposite seems to be true\footnote{For instance, in the only experiment analyzed with finite-key formalism to date \cite{has07}, the authors extracted $r\approx 2\%$, whereas, for the observed error rate, the asymptotic bound would have yielded $r\gtrsim 40\%$!} --- but because their estimate is still the object of on-going research (see \ref{sssfinite} for the state-of-the-art).

\subsection{Notions of Security}

\subsubsection{Unconditional security, and its conditions}
\label{sssuncond}

The appeal of QKD comes mainly from the fact that, in principle, it can achieve \textit{unconditional security}. This technical term means that security can be proved without imposing any restriction on the computational resources or the manipulation techniques that are available to the eavesdropper acting on the signal. The possibility of achieving unconditional security in QKD is deeply rooted in quantum physics. To learn something about the key, Eve must interact with the quantum system; now, if the coding uses randomly chosen non-orthogonal states, Eve's intervention necessarily modifies the state on average, and this modification can be observed by the parties. As we discussed in Sec. \ref{ssbasic}, there are many equivalent formulations of this basic principle. However formulated, it must be stressed that this criterion can be made \textit{quantitative}: the observed perturbations in the quantum channel allow computing a bound on the information that Eve might have obtained.

Like many other technical terms, the wording ``unconditional security'' has to be used in its precise meaning given above, and not as a synonym of ``absolute security'' --- something that does not exist. As a matter of fact, unconditional security of QKD holds under some conditions. First of all, there are some \textit{compulsory requirements}:

\begin{enumerate}

\item Eve cannot intrude Alice's and Bob's devices to access either the emerging key or their choices of settings (we shall see in Sec. \ref{ssstrojan} how complex it is to check this point thoroughly).

\item Alice and Bob must trust the random number generators that select the state to be sent or the measurement to be performed.

\item The classical channel is authenticated with unconditionally secure protocols, which exist \cite{carter79a,wegman81a,sti95}.

\item Eve is limited by the laws of physics. This requirement can be sharpened: in particular, one can ask whether security can be based on a restricted set of laws\footnote{As we have seen (\ref{sssorigin}), intuition suggests that the security of QKD can be traced back to a few specific principles or laws like ``no-cloning'' or ``non-locality without signaling''. One may ask whether this intuition may be made fully rigorous. Concretely, since any theory that does not allow signaling and is non-local exhibits a no-cloning theorem \cite{mas06,bar06}, and since non-locality itself can be checked, one may hope to derive security only from the physical law of \textit{no-signaling}. In this framework, as of today, unconditional security has been proved only in the case of strictly error-free channels and for a key of vanishing length \cite{bar05}. Only limited security has been proved in more realistic cases \cite{aci06,sca06}. Recently, Masanes showed that unconditional composable security can be proved if no-signaling is assumed not only between Alice and Bob, but also among the systems that are measured by each partner \cite{mas08}.}. In this review, as in the whole field of practical QKD, we assume that Eve has to obey the whole of quantum physics.

\end{enumerate} We shall take these requirements, the failure of which would obviously compromise any security, as granted. Even so, many other issues have to be settled, before unconditional security is claimed for a given protocol: for instance, the theoretical description of the quantum states must match the signals that are really exchanged; the implementations must be proved free of unwanted information leakage through side-channels or back-doors, against which no theoretical protection can be invoked.

\subsubsection{Definition of security}
\label{ssssecu}


The security of a key ${\cal K}$ can be parametrized by its deviation $\varepsilon$ from a perfect key, which is defined as a list of perfectly correlated symbols shared between Alice and Bob, on which Eve has no information (in particular, all the possible lists must be equally probable \textit{a priori}). A \textit{definition of security} is a choice of the quantity that is required to be bounded by $\varepsilon$; a key that deviates by $\varepsilon$ from a perfect key is called $\varepsilon$-secure. The main property that a definition of security must fulfill is \textit{composability}, meaning that the security of the key is guaranteed whatever its application may be --- more precisely: if an $\varepsilon$-secure key is used in an $\varepsilon'$-secure task\footnote{For instance, the One-Time Pad is a $0$-secure task; while any implementation of channel authentication, for which a part of the key is used (\ref{introchannels}), must allow for a non-zero $\varepsilon'$.}, composability ensures that the whole procedure is at least $(\varepsilon+\varepsilon')$-secure. 

A composable definition of security is the one based on the trace-norm \cite{ben05,ren05c}: $\frac{1}{2}\| \rho_{{\cal K} E} - \tau_{\cal K} \otimes \rho_E \|_1 \leq \varepsilon$, where $\rho_{{\cal K} E}$ is the actual state containing some correlations between the final key and Eve, $\tau_{\cal K}$ is the completely mixed state on the set ${\cal K}$ of possible final keys and $\rho_E$ is any state of Eve. In this definition, the parameter $\varepsilon$ has a clear interpretation as the maximum failure probability of the process of key extraction.
As the dates of the references show, the issue of composability was raised rather late in the development of QKD. Most, if not all, of the early security studies had adopted a definition of security that is \textit{not} composable, but the asymptotic bounds that were derived can be ``redeemed'' using a composable definition\footnote{The early proofs defined security by analogy with the classical definition: Eve, who holds a quantum state $\rho_E$, performs the measurement ${\cal M}$ which maximizes her mutual information with the key ${\cal K}$. This defines the so-called \textit{accessible information} $I_{acc}({\cal K}:\rho_E)=\max_{E={\cal M}(\rho_E)}I({\cal K}:E)$, and the security criterion reads $I_{acc}({\cal K}:\rho_E)\leq \varepsilon$. As for the history of claims, it is quite intricate. Accessible information was first claimed to provide composable security \cite{ben05}. The proof is correct, but composability follows from the use of two-universal hashing in the privacy amplification step (see \ref{12way}), rather than from the properties of accessible information itself. Indeed, shortly later, an explicit counterexample showed that accessible information is in general \textit{not} composable for any reasonable choice of the security parameter $\varepsilon$ \cite{koe05}. The reason why accessible information is not composable can be explained qualitatively: this criterion supposes that Eve performs a measurement to guess the key at the end of the key exchange. But Eve may prefer not to measure her systems until the key is actually used in a further protocol: for instance, if a plaintext attack can reveal some information, Eve has certainly better adapt her measurement to this additional knowledge. The counterexample also implies that the classical results on privacy amplification by two-universal hashing \cite{ben95} do not apply and have to be replaced by a quantum version of the statement \cite{ren05c}.}.

\subsubsection{Security proofs}
\label{sssproofs}

Once the security criterion is defined, one can derive a full \textit{security proof}, leading to an explicit (and hopefully computable) expression for the length of the extractable secret key rate. Several techniques have been used:
\begin{itemize}
\item The very first proofs by Mayers were somehow based on the \textit{uncertainty principle} \cite{may96,may01}. This approach has been revived recently by Koashi \cite{koa05,koa07}.
\item Most of the subsequent security proofs have been based on the correspondence between \textit{entanglement distillation} and classical post-processing, generalizing the techniques of Shor and Preskill \cite{sho00}. For instance, the most developed security proofs for imperfect devices follow this pattern \cite{got04}.
\item The most recent techniques use rather information-theoretical notions \cite{ben02,ren05a,ren05b,kra05}.

\end{itemize}

A detailed description on how a security proof is built goes beyond the scope of this review. The core lies in how to relate the security requirement $\frac{1}{2}\| \rho_{{\cal K} E} - \tau_{\cal K} \otimes \rho_E \|_1 \leq \varepsilon$ to a statement about the length $\ell$ of the secret key that can be extracted. This step is achieved using inequalities that can be seen as a generalization of the Chernoff bound. In other words, one must use or prove an inequality of the form
\ba
\mbox{Prob}\left[\| \rho_{{\cal K} E} - \tau_{\cal K} \otimes \rho_E \|_1 > 2\varepsilon\right]&\lesssim & e^{\ell-F(\rho_{{\cal K} E},\varepsilon)}
\ea
where we omitted constant factors. From such an inequality, one immediately reads that the security requirement will fail with exponentially small probability provided $\ell\lesssim F(\rho_{{\cal K} E},\varepsilon)$. Explicit security bounds will be provided below (Sec. \ref{sssf}) for the asymptotic limit of infinitely long keys --- note that in this limit one can take $\varepsilon\rightarrow 0$, whence no explicit dependence on $\varepsilon$ is manifest in those expressions.

\subsection{Explicit Protocols}
\label{protos}

\subsubsection{Three families}

The number of explicit QKD protocols is virtually infinite: after all, Bennett has proved that security can be obtained when coding a bit in just two non-orthogonal quantum states \cite{ben92}. But as a matter of fact, this possible variety has crystallized into three main families: discrete-variable coding (\ref{sssdiscr}), continuous-variable coding (\ref{ssscv}), and more recently distributed-phase-reference coding (\ref{sssmixed}). The crucial difference is the \textit{detection scheme}: discrete-variable coding and distributed-phase-reference coding use photon counting and post-select the events in which a detection has effectively taken place, while continuous-variable coding is defined by the use of homodyne detection (detection techniques are reviewed in Sec. \ref{ssdet}).

\textit{Discrete-variable} coding is the original one. Its main advantage is that protocols can be designed in such a way that, in the absence of errors, Alice and Bob would share immediately a perfect secret key. They are still the most implemented QKD protocols. Any discrete quantum degree of freedom can be chosen in principle, but the most frequent ones are \textit{polarization for free-space implementations} and \textit{phase-coding in fiber-based implementations}\footnote{Other degrees of freedom have been explored, for instance coding in sidebands of phase-modulated light \cite{mer99} and time-coding \cite{bou05}. Energy-time entanglement gives also rise to a peculiar form of coding \cite{tit00}.}.
The case for \textit{continuous-variable} coding stems from the observation that photon counters normally feature low quantum efficiencies, high dark count rates, and rather long dead times; while these inconveniences can be overcome by using homodyne detection. The price to pay is that the protocol provides Alice and Bob with correlated but rather noisy realization of a continuous random variable, because losses translate into noise (see \ref{sslight}): as a consequence, a significant amount of error correction procedures must be used. In short, the issue is, whether it is better to build up slowly a noiseless raw key, or rapidly a noisy one. As for \textit{distributed-phase-reference} coding, its origin lies in the effort of some experimental groups toward a more and more practical implementation. From the point of view of detection, these protocols produce a discrete-valued result; but the nature of the quantum signals is very different from the case of discrete-variable coding, and this motivates a separate treatment.

Despite the differences originating from the use of a different detection device, 
there is a strong conceptual unity underlying discrete- and continuous-variable QKD. To take just
one example, in both cases the ability to distribute a quantum key
is closely related to the ability to distribute entanglement, regardless of the detection scheme used and even if no actual entanglement is present.
These similarities are not very surprising since it has long been known that the quantum features of light
may be revealed either via photon counting (e.g., antibunching or anticorrelation experiments) or via 
homodyne detection (e.g., squeezing experiments). Being a technique that exploits these quantum features
of light, QKD has thus no reason to be restricted to the photon-counting regime. Surprisingly,
just like antibunching (or a single-photon source) is not even needed in photon-counting based QKD, 
we shall see that squeezing is not needed in homodyne-detection based QKD. The only quantum feature
that happens to be needed is the non-orthogonality of light states.

\subsubsection{Discrete-variable Protocols}
\label{sssdiscr}

\paragraph{BB84-BBM.} The best known discrete-variable protocol is of course BB84
\cite{bb84}, that we introduced in Sec. \ref{ssbasic}. The corresponding EB protocol is known as BBM \cite{ben92a}; the E91 protocol \cite{eke91} is equivalent to it when implemented with qubits. Alice prepares a single particle in one of the four states:
\ba
\begin{array}{lcl}\ket{+x},\ket{-x}\,,&\mbox{ eigenstates of
}&\sigma_x\\
\ket{+y},\ket{-y}\,,&\mbox{ eigenstates of
}&\sigma_y\end{array}\label{4states}\ea where the $\sigma$'s are Pauli operators. The states with ``$+$''
code for the bit value 0, the states with ``$-$'' for the bit
value 1. Bob measures either $\sigma_x$ or $\sigma_y$. In the
absence of errors, measurement in the correct basis reveals the bit-value encoded by Alice. The protocol includes a sifting phase: Alice reveals the
basis, $X$ or $Y$, of each of her signals; Bob accepts the values
for which he has used the same basis and discards the others\footnote{In
the original version of BB84, both bases are used with the same
probability, so that the sifting factor is $p_{sift}=\demi$, i.e.
only half of the detected bits will be kept in the raw key. But the protocol can be made asymmetric without changing
the security \cite{lo05b}: Alice and Bob can agree on using one
basis with probability $1-\epsilon$ where $\epsilon$ can be taken
as small as one wants, so as to have $p_{sift}\approx
1$ (recall that we are considering only asymptotic bounds; in the finite key regime, the optimal value of $\epsilon$ can be computed \cite{sca07}).}. 

Unconditional security of BB84-BBM has been proved with many different techniques \cite{may96,may01,lo99b,sho00,kra05}. The same coding can be implemented with other sources, leading to a family of BB84-like protocols. We review them at length in Sec. \ref{ssbb84}.

\paragraph{SARG04.} The SARG04 protocol \cite{aci04a,sca04} uses the same four
states (\ref{4states}) and the same measurements on Bob's side as
BB84, but the bit is coded in the basis rather than in the state
(basis $X$ codes for 0 and basis $Y$ codes for 1). Bob has to
choose his bases with probability $\demi$. The creation of the raw key is slightly more complicated than in BB84. Suppose for definiteness that Alice sends $\ket{+x}$: in
the absence of errors, if Bob measures $X$ he gets $s_b=+$; if he measures $Y$, he may
get both $s_b=+/-$ with equal probability. In the sifting phase, Bob reveals $s_b$; Alice tells him to accept if she had prepared a state with $s_a\neq s_b$, in which case Bob accepts the bit
corresponding to the basis he has {\em not} used. The reason is clear in the example above: in the absence of errors, $s_b=-$ singles out the wrong basis \footnote{In an alternative version of the sifting, Alice
reveals that the state she sent belongs to one of the
two sets $\{\ket{s_ax},\ket{s_ay}\}$, and Bob accepts if he has
detected a state $s_b\neq s_a$. This is a simplified version with respect to the
original proposal, where Alice could declare any of the four sets
of two non-orthogonal states. The fact, that the two versions are
equivalent in terms of security, was not clear when the first
rigorous bounds were derived \cite{bra05}, but was verified
later.}.

SARG04 was invented for implementations with attenuated laser sources, because it is more robust than BB84 against the PNS attacks. Unconditional security has been proved, we shall review the main results in Sec. \ref{sssarg04}.

\paragraph{Other discrete-variable protocols.} A large number of other discrete-variable protocols have been proposed; all of them have features that makes them less interesting for practical QKD than BB84 or SARG04.

The {\em six-state protocol} \cite{ben84b,bru98,bec99} follows the same
structure as BB84, to which it adds the third mutually unbiased
basis $Z$ defined by the Pauli matrix $\sigma_z$. Its unconditional security has been proved quite early \cite{lo01}. The interest of this protocol lies in the fact that the channel estimation becomes
``tomographically complete'', that is, the measured parameters
completely characterize the channel. As a consequence, more noise can be tolerated with respect to BB84 or SARG04. However, noise is quite low in optical setups, while losses are a greater concern (see \ref{sschannels}). Under this respect, six-state perform worse, because it requires additional lossy optical components. Similar considerations apply to the \textit{six-state version of the SARG04 coding} \cite{tam06} and to the {\em
Singapore protocol} \cite{eng04}.

The coding of BB84 and six-state has been generalized to {\em larger
dimensional quantum systems} \cite{bec00a,bec00b}. For any $d$, protocols that use
either two or $d+1$ mutually unbiased bases have been defined
\cite{cer02}. Unconditional security was not studied; for restricted attacks, the robustness to noise increases with $d$. Time-bin coding allows producing $d$-dimensional quantum states of light in
a rather natural way \cite{the04,der04}. However, the production and detection of
these states requires $d$-arm interferometers with couplers or
switches, that must moreover be kept stable. Thus again, the
possible advantages are overcome by the practical issues of losses
and stability.

Finally, we have to mention the {\em B92 protocol}
\cite{ben92}, which uses only two non-orthogonal states, each one coding for one bit-value. In terms of encoding, this is obviously the most economic possibility. Unfortunately, B92 is a rather sensitive protocol: as noticed already in the original paper, this protocol is secure only if some other
signal (e.g. a strong reference pulse) is present along with the two states that code the bit. Unconditional security has been proved for single-photon implementations \cite{tam03,tam04} and for some implementations with a strong reference pulse \cite{koa04,tam07}. Incidentally, SARG04 may be seen as a modified B92, in which a second set of non-orthogonal states is added --- actually, an almost forgotten protocol served as a link between the two \cite{hut95}.

\subsubsection{Continuous-variable Protocols}
\label{ssscv}

Discrete-variable coding can be implemented with several sources, but requires photon-counting techniques. An alternative approach to QKD has been suggested, in which the photon counters 
are replaced by standard telecom PIN photodiodes, which are faster (GHz instead of MHz) and more efficient
(typically 80$\%$ instead of 10$\%$). The corresponding schemes are then based on homodyne detection (\ref{ssshomodet}) and involve measuring data that are \textit{real amplitudes instead of discrete events}; hence these schemes are named continuous-variable (CV) QKD.

The first proposals suggesting the use of homodyne detection in QKD are due to \cite{ralph,hillery,reid}.
In particular, a squeezed-state version of BB84 was proposed in \cite{hillery}, where Alice's basis choice consists 
of selecting whether the state of light sent to Bob is squeezed in either quadrature $q=x$ or $q=p$. Next, this $q$-squeezed state is displaced in $q$ either by $+c$ or $-c$ depending on a random bit chosen by Alice, where $c$ is an appropriately chosen constant. Bob's random basis choice defines whether it is the $x$ or $p$ quadrature that is measured. The sifting simply consists in keeping only the instances where Alice and Bob's chosen quadratures coincide. In this case, the value measured by Bob is distributed according to a Gaussian distribution centered on the value ($+c$ or $-c$) sent by Alice. In some sense, this protocol can be viewed as ``hybrid'' because Alice's data are binary while Bob's data are real (Gaussian distributed).

These early proposals and their direct generalization are called \textit{CV protocols with discrete modulation}; at the same time, another class of CV protocols was proposed that rather use a continuous modulation, in particular a \textit{Gaussian modulation}. Although CV protocols are much more recent than discrete-variable protocols, 
their security proofs have been progressing steadily over the last years, 
and are now close to reach a comparable status: see a thorough discussion in Sec. \ref{statuscv}.

\paragraph{Gaussian protocols.} The first proposed Gaussian QKD protocol was based on squeezed states of light, which are modulated with a Gaussian distribution in the $x$ or $p$ quadrature by Alice, and are measured via homodyne detection by Bob \cite{Cerf01}. This protocol can be viewed as the proper continuous-variable counterpart of BB84 in the sense that the average state sent by Alice is the same regardless of the chosen basis (it is a thermal state, replacing the maximally-mixed qubit state in BB84). The security of this protocol can be analyzed using the connection with continuous-variable cloning \cite{CVcloning}; using a connection with quantum error-correcting codes, unconditional security was proved when the squeezing exceeds 2.51~dB \cite{GP01}. The main drawback of this protocol is the need for a source of squeezed light.

A second Gaussian QKD protocol was therefore devised, in which Alice generates coherent states of light,
which are then Gaussian modulated both in $x$ and $p$, while Bob still performs homodyne detection \cite{GG02}. A first proof-of-principle experiment, supplemented with the technique of reverse reconciliation\footnote{In all Gaussian QKD protocols, reversing the one-way reconciliation procedure (i.e., using Bob's measured data instead of Alice's sent data as the raw key) is beneficial in terms of attainable range, provided that the noise is not too large. We will come back to this point in Section \ref{seccv}.}, was run with bulk optical elements on an optics table \cite{Nature03}. Subsequent experiments have used optical fibers and telecom wavelengths. The scheme was thus implemented over distances up to 14~km using a Plug\&Play configuration \cite{leg06}, then up to 25~km by time-multiplexing the local oscillator pulses with the signal pulses in the same optical fiber and using an improved classical post-processing \cite{lodewyck-1,lodewyck-2}. Another fiber-based implementation over 5~km has been reported \cite{qi07b}.

Note that, in these two first protocols, Bob randomly chooses to homodyning one quadrature, either $x$ or $p$. In the squeezed-state protocol, this implies the need for sifting. Bob indeed needs to reject the instances where he measured the other quadrature than the one modulated by Alice, which results in a decrease of the key rate by a factor of 2 (this factor may actually be reduced arbitrarily close to 1 by making an asymmetric choice between $x$ and $p$, provided that the key length is sufficiently large) \cite{lo05b}.
In the coherent-state protocol, Alice simply forgets the quadrature that is not measured by Bob, so that all
pulses do carry useful information that is exploited to establish the final secret key. 

The fact that Alice, in this second protocol, discards half of her data may look like a loss of efficiency since 
some information is transmitted and then lost.
A third Gaussian QKD protocol was therefore proposed \cite{WL04}, in which Alice
still transmits doubly-modulated coherent states drawn from a bivariate Gaussian distribution, 
but Bob performs heterodyne instead of homodyne measurements\footnote{This possibility was also suggested for postselection-based protocols in \cite{lor04}, and the experiment has been performed \cite{lorenz06a}.}, that is, he measures both $x$ and $p$ quadratures simultaneously. At first sight, this seems to imply that the rate is doubled since Bob then acquires a pair of quadratures ($x,p$). Actually, since heterodyne measurement effects one additional unit of vacuum noise on the measured quadratures, the two quadratures received by Bob are noisier than the single quadrature in the homodyne-based protocol. The net effect, however, is often an increase of the key rate when the two quadratures are measured simultaneously. In addition, a technological advantage of this heterodyne-based coherent-state protocol is that there is no need to choose a random quadrature at Bob's side (that is, no active basis choice is needed). The experiment has been realized \cite{lan05}.

Finally, a fourth Gaussian QKD protocol was introduced recently \cite{RaulPhD}, which completes this family of Gaussian QKD protocols. Here, Alice sends again squeezed states, as in the protocol of \cite{Cerf01}, but Bob performs heterodyne measurements, as in the protocol of \cite{WL04}. This protocol is associated with the highest rate and range among all Gaussian QKD protocols, but requires a source of squeezed light.

As seen in the discussion about BB84 and SARG04 above, it turns out also for the CV QKD protocols that the classical processing is an essential element of the protocol. As will be discussed later (\ref{statuscv}), the performance of CV-QKD protocols depends crucially on the exact protocol that extracts the secret key from the experimental data. Two important tools here are reverse reconciliation \cite{GG02} and post-selection \cite{silberhorn}. As shown in \cite{heid07}, the combination of both will lead to the optimal key rate.

\paragraph{Discrete-modulation protocols.} On the side of practical implementation, it is desirable to keep the number of signals as low as possible, and also to minimize the number of parameters in the detection process that needs to be monitored. The deep reason behind this is that in practical implementation at some stage one has to consider finite size effects in the statistics and also in the security proof stage. For a continuous family of signals, it will be intuitively harder to get hold of these finite size effects and to include statistical fluctuations of observations into a full security proof.

For this reason, it becomes interesting to have a look at QKD systems that combine a finite number of signals with the continuous variable detection schemes: discrete-modulation protocols have been devised following this proposal, some based on coherent states instead of squeezed states \cite{silberhorn}. The signals consist here of a weak coherent state together with a strong phase reference. The signal is imprinted onto the weak coherent state by setting the relative optical phase between weak coherent state and reference pulse either to $0$ or $\pi$.  Schematically, the strong phase reference could be represented by two local oscillators, e.g. phase-locked lasers at the sending and receiving station. These type of signals have been used already in the original B92 protocol \cite{ben92}. The receiver then uses the local oscillator in the homodyne or heterodyne measurement. The security of this protocol is still based on the fact that the weak signal pulses represent non-orthogonal signal states. 

On the receiver side, homodyne detection is performed by choosing at random one of the two relevant quadrature measurement (one quadrature serves the purpose of being able to measure the bit values, the other one serves the purpose to monitor the channel to limit possible eavesdropping attacks). Alternatively, a heterodyne measurement can, in a way, monitor both quadratures. Consider for definiteness a simple detection scheme, in which bit-values are assigned by the sign of the detection signal, $+$ or $-$, with respect to the half-planes in the quantum optical phase space in which the two signals reside. As a result, both sender and receiver have binary data at hand. As in the case of Gaussian modulation, they can now perform post-selection of data, and use error-correction and privacy amplification to extract secret keys from these data.

\subsubsection{Distributed-phase-reference Protocols}
\label{sssmixed}

\begin{figure}[ht]
\includegraphics[scale=0.6]{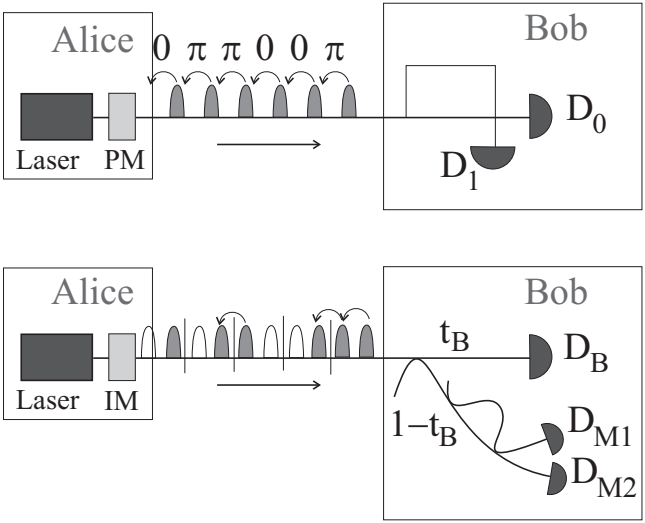}
\caption{The two distributed-phase reference protocol: differential phase shift (DPS, top) and coherent one-way (COW, bottom). Legend: PM: phase modulator; IM: intensity modulator. See text for description.}\label{fig:dpscow}
\end{figure}

Both discrete- and continuous-variable protocols have been invented by theorists. Some experimental groups, in their developments toward practical QKD systems, have conceived new protocols, which do not fit in the categories above. In these, like in discrete-variable protocols, the raw keys are made of realizations
of a discrete variable (a bit) and are already perfectly correlated in the absence of errors.
However, the quantum channel is monitored using the properties of coherent
states --- more specifically, by observing the phase coherence of subsequent pulses; whence the name \textit{distributed-phase-reference protocols}.

The first such protocol has been called {\em
Differential-Phase-Shift (DPS)} \cite{ino02,ino03}. Alice produces a
sequence of coherent states of same intensity \ba \ket{\Psi({\cal
S}_n)}&=&...\ket{e^{i\varphi_{k-1}}\sqrt{\mu}}
\ket{e^{i\varphi_{k}}\sqrt{\mu}}
\ket{e^{i\varphi_{k+1}}\sqrt{\mu}}...\label{statedps}\ea where each phase can be
set at $\varphi=0$ or $\varphi=\pi$ (Fig.~\ref{fig:dpscow}). The bits are coded in the
difference between two successive phases: $b_k=0$ if
$e^{i\varphi_{k}}=e^{i\varphi_{k+1}}$ and $b_k=1$ otherwise. This
can be unambiguously discriminated using an unbalanced
interferometer. The complexity in the analysis of this protocol
lies in the fact that $\ket{\Psi({\cal
S}_n)}\neq\ket{\psi(b_1)}\otimes...\otimes \ket{\psi(b_n)}$: the
$k$-th pulse contributes to both the $k$-th and the $(k+1)$-st
bit. The DPS protocol has been already the object of several experimental demonstrations \cite{tak05,dia06,tak07}.

In the protocol called {\em Coherent-One-Way (COW)} \cite{gis04,stu05},
each bit is coded in a sequence of one non-empty and one empty
pulse: \ba
\ket{\mathbf{0}}_k=\ket{\sqrt{\mu}}_{2k-1}\ket{0}_{2k}&,&
\ket{\mathbf{1}}_k=\ket{0}_{2k-1} \ket{\sqrt{\mu}}_{2k}\,.\ea
These two states can be unambiguously discriminated in an optimal
way by just measuring the time of arrival (Fig.~\ref{fig:dpscow}). For the channel
estimation, one checks the coherence between two successive
non-empty pulses; these can be produced on purpose as a ``decoy
sequence'' $\ket{\sqrt{\mu}}_{2k-1}\ket{\sqrt{\mu}}_{2k}$, or can
happen as $\ket{\sqrt{\mu}}_{2k}\ket{\sqrt{\mu}}_{2k+1}$ across a
bit separation, when a sequence
$\ket{\mathbf{1}}_k\ket{\mathbf{0}}_{k+1}$ is coded. This last
check, important to detect PNS attacks,
implies that the phase between any two successive pulses must be
controlled; therefore, as it happened for DPS, the whole sequence must be considered as a
single signal. A prototype of a full QKD system based on COW has been reported recently \cite{stu08}.

Both DPS and COW are P\&M schemes, tailored for laser sources. It has not yet been possible to derive a bound for unconditional security, because the existing techniques apply only when $\ket{\Psi({\cal S}_n)}$ can be decomposed in independent signals. We shall review the status of partial security proofs in Sec. \ref{secmixed}.

\subsection{Sources}

\subsubsection{Lasers}
\label{ssslaser}

Lasers are the most
practical and versatile light sources available today. For this
reason, they are chosen by the vast majority of groups working in
the field. Of course, all implementations in which the source is a laser are
P\&M schemes. For the purposes of this review, we don't have to delve deep into
laser physics. The output of a
laser in a given mode is described by a coherent state of the field \ba
\ket{\sqrt{\mu}\,e^{i\theta}}\,\equiv\,
\ket{\alpha}&=&e^{-\mu/2}\sum_{n=0}^{\infty}
\frac{\alpha^n}{\sqrt{n!}}\ket{n}\ea where $\mu=|\alpha^2|$ is the
average photon number (also called intensity). The phase factor
$e^{i\theta}$ is accessible if a reference for the phase is
available; if not, the emitted state is rather described by the
mixture \ba
\rho&=&\int_{0}^{2\pi}\frac{d\theta}{2\pi}\ket{\alpha}\bra{\alpha}\,=\,
\sum_n P(n|\mu)\ket{n}\bra{n} \label{rhopoiss}\ea with \ba
P(n|\mu)&=&e^{-\mu}\,\frac{\mu^n}{n!}\,. \ea Since
two equivalent decompositions of the same density matrix cannot be
distinguished, one may say as well that, in the absence of a phase
reference, the laser produces a {\em Poissonian mixture} of number
states.

The randomization of $\theta$ generalizes to multimode coherent
states \cite{mol97,van02}. Consider for instance the two-mode
coherent state $\ket{\sqrt{\mu}\,e^{i(\theta+\varphi)}}
\ket{\sqrt{\mu'}\,e^{i\theta}}$ that may describe for instance a
weak pulse and a reference beam. The phase $\varphi$ is the
relative phase between the two modes and is well-defined, but the
common phase $\theta$ is random. One can then carry out the same
integral as before; the resulting $\rho$ is the Poissonian mixture
with average photon number $\mu+\mu'$ and the number states
generated in the mode described by the creation operator $A^{\dagger}=\big(e^{i\varphi}\sqrt{\mu}
a_1^{\dagger} + \sqrt{\mu'}a_2^{\dagger}\big)/\sqrt{\mu+\mu'}$.

Let us turn now to QKD. The existence of a reference for the phase is essential in both continuous-variable and distributed-phase-reference protocols: after all, these protocols have been designed having specifically in mind the laser as a source. On the contrary, when attenuated lasers are used to approximate qubits in discrete protocols, the phase reference does not play any role. In this implementations, $\rho$ given in (\ref{rhopoiss}) is generically\footnote{One must be careful though: the fact that the phase reference is not used in the protocol does not necessarily mean that such a reference is physically not available. In particular, such reference \textit{is} available for some source, e.g. when a mode-locked laser is used to produce pulses. In such cases, even though Alice and Bob don't use the phase coherence in the protocol, the signal is no longer correctly described by (\ref{rhopoiss}), and Eve can in principle take advantage of the existing coherence to obtain more information \cite{lo06}. Therefore it is necessary to implement active randomization \cite{gis06,zha07}.} an accurate description of the quantum signal outside Alice's lab. Since $\rho$ commutes with the measurement of the
number of photons, this opens the possibility of the
photon-number-splitting (PNS) attacks \cite{ben92,bra00,lut00}, a major concern in practical QKD that will be addressed in Sec. \ref{sssside}.

\subsubsection{Sub-Poissonian Sources}

Sub-Poissonian sources (sometimes called ``single-photon sources'') come closer to a
single-photon source than an attenuated laser, in the sense that the probability of
emitting two photons is smaller. The quantum signal in each mode
is taken to be a photon-number diagonal mixture with a very small
contribution of the multi-photon terms. The quality
of a sub-Poissonian source is usually measured through the second order correlation function
\ba g_2(\tau)&=&\frac{\moy{:I(t)I(t+\tau):}}{\moy{I(t)}^2} \ea where $I(t)$ is the signal intensity emitted by the source and $:-:$ denotes normal ordering of the creation and annihilation operators. In particular, $g_2(0)\approx 2p(2)/p(1)^2$, while $p(n)$ is the probability that the source emits $n$ photons. For Poissonian sources, $g_2(0)=1$; the smaller $g_2(0)$, the closer the
source is to an ideal single-photon source. It has been noticed that the knowledge of the
efficiency and of $g_2$ is enough to characterize the performance
of such a source in an implementation of BB84 \cite{wak02a}.

Sub-Poissonian sources have been, and still are, the object of intensive research; recent reviews cover the most meaningful developments \cite{lou05,shi07}. In the context of QKD, the discovery of PNS attacks triggered a lot of interest in sub-Poissonian sources, because they would reach much higher secret fractions. QKD experiments have been performed with such sources \cite{bev02,wak02,all04}, also in fibers \cite{int07} thanks to the development of sources at telecom wavelengths \cite{war05,sai06,zin06}. At the moment of writing, this interest has significantly dropped, as it was shown that the same rate can be achieved with lasers by using decoy states, see \ref{sssdecoy} and \ref{ssestimatepm}. But the tide may turn again in the near future, for applications in QKD with quantum repeaters \cite{san07}.

\subsubsection{Sources of Entangled Photons}
\label{ssspdc}

Entangled photon pairs suitable for entanglement-based protocols
or for heralded sub-Poissonian sources are mostly generated by
spontaneous parametric down conversion (SPDC) \cite{man95}. In
this process some photons from a pump laser beam are converted due
to the non-linear interaction in an optical
crystal\footnote{Crystals like KNbO${}_{3}$, LiIO${}_{3}$,
LiNbO${}_{3}$, $\beta$-BaB${}_{2}$O${}_{4}$, etc. Very promising
are periodically-poled nonlinear materials \cite{tan01}. Besides
the spontaneous parametric down conversion, new sources of
entangled photons based on quantum dots are tested in laboratories
\cite{you06}. But these sources are still at an early stage of development.
Their main drawback is the need of cryogenic environment.} into
pairs of photons with lower energies. The total energy and
momentum are conserved. In QKD devices, cw-pumped sources are
predominantly used.

In the approximation of two output modes,
the state behind the crystal can be described as follows
\ba
 \ket{\psi}_{PDC}&=&\sqrt{1-\lambda^2} \sum_{n=0}^\infty \lambda^n \ket{n_A,n_B},
 \label{squeezed_vac}
\ea where $\lambda=\tanh\xi$ with $\xi$ proportional to the
pump amplitude, and where $\ket{n_A,n_B}$ denotes the state with $n$
photons in the mode destined to Alice and $n$ photons in the other
mode aiming to Bob. This is the so called two-mode squeezed
vacuum.

The photons are entangled in time and in frequencies
(energies); one can also prepare pairs of photons correlated in
other degrees of freedom: polarization \cite{kwi95,kwi99}, time
bins \cite{bre99,tit00}, momenta (directions), or orbital angular momenta
\cite{mai00}.

The state (\ref{squeezed_vac}) can be directly utilized in
continuous-variable protocols. In the case of discrete-variable
protocols, one would prefer only single pair of photons per
signal; however, SPDC always produces \textit{multi-pair components}, whose presence must be taken into account.
Let us describe this in the four-mode approximation, which is sufficient
for the description of fs-pulse pumped SPDC \cite{li05}. An ideal two-photon maximally entangled
state reads $\ket{\Psi_2}=\frac{1}{\sqrt{2}} \left( \ket{1,0}_A\ket{1,0}_B +
\ket{0,1}_A\ket{0,1}_B\right)$ where each photon can be in two different modes (orthogonal polarizations, different time-bins...). This state can be approximately achieved if $\lambda \ll 1$, i.e.~if the mean pair number per pulse $\mu = 2\lambda^2/(1-\lambda^2) \ll 1$. But there are \textit{multi-pair components}: in fact, again in the case of a four-mode approximation, the generated state reads
\ba \ket{\Psi} &\approx& \sqrt{p(0)}\,
\ket{0} + \sqrt{p(1)}\,\ket{\Psi_2}+ \sqrt{p(2)}\,\ket{\Psi_4}
\label{spdc_state} \ea
where $p(1) \approx \mu$
and $p(2) \approx \frac{3}{4} \mu^2$, $\ket{0}$ is the vacuum state, and the four-photon state is $\ket{\Psi_4}=\frac{1}{\sqrt{3}} \big(
\ket{0,2}\ket{0,2}+\ket{2,0}\ket{2,0}+\ket{1,1}\ket{1,1}\big)$.
We recall that this description is good for short pump pulses; when a cw-pumped source is used (or the pulse-pumped
source with the pulse duration much larger than the coherence time
$\tau$ of the down-converted photons) the four-mode approximation
is not applicable and a continuum of frequency modes must be taken
into account. The multiple excitations created during the
coherence time $\tau$ are coherent and partially correlated: in
this case, the four-photon state is a fully entangled state that
cannot be written as ``two pairs'' --- see $\ket{\Psi_4}$ above\footnote{Though a nuisance in qubit-based protocols, the existence of such four photon components can lead to new opportunities for QKD, as pointed out independently in \cite{bra00b} and \cite{dur02}.}. However, $\tau$ is usually much
shorter than the typical time $\Delta t$ that one can discriminate, this time being defined as the time resolution of the detectors for cw-pumped sources\footnote{However, a recent entanglement-swapping experiment combined fast detectors and narrow filters to achieve $\Delta t<\tau$ in cw-pumped SPDC \cite{hal07}.} or as the duration of a pulse for pulsed
sources. This implies that, when two photons arrive ``at the same time'', they may
actually arise from two incoherent processes, and in this case the
observed statistics corresponds to that of two independent pairs.
This physics has been the object of several studies
\cite{tap98,ou99,der04b,sca05,tsu04,eis04}.

What concerns us here
is the advantage that Eve may obtain, and in particular the
efficiency of PNS attacks. If the source is used in a P\&M scheme
as heralded single-photon source, then the PNS attack is effective
as usual, because all the photons that travel to Bob have been
actively prepared in the same state \cite{lut00}; ideas inspired from decoy
states can be used to detect it \cite{ada06,mau07}. In an EB scheme,
the PNS attack is effective on the fraction $\zeta\approx
\tau/\Delta t$ of coherent four-photon states; besides, all multi-pair contributions inevitably produce errors in the correlations Alice-Bob. We shall come back to these points in Sec. \ref{sssbb84eb}.

\subsection{Physical Channels}
\label{sschannels}

As far as the security is concerned, the quantum channel must be characterized only \textit{a posteriori}, because Eve has full freedom of acting on it. However, the knowledge of the \textit{a priori} expected behavior is obviously important at the moment of designing a setup. We review here the physics of the two main quantum channels used
for light, namely optical fibers and free space beams.

An important parameter of the quantum channel is the amount of \textit{losses}. Surely enough, a key can be built by post-selecting only those photons that have actually been detected. But, since quantum signals cannot be amplified, the raw key rate decreases with the distance as the transmission $t$ of
the channel; in addition, at some point the detection rate
reaches the level of the dark counts of the detectors, and this effectively limits the maximal achievable distance. Finally, in general the lost photons are
correlated to the signal and thus must be counted as information that leaked to Eve.

Concerning the interaction of photons with the environment in the channel, the
effect of {\em decoherence} depends strongly on the quantum degree
of freedom that is used; therefore, although weak in principle, it
cannot be fully neglected and may become critical in some
implementations.

\subsubsection{Fiber Links}

The physics of optical fibers has been explored in depth because
of its importance for communication \cite{agr97}. When we quote a
value, we refer to the specifications of the standard fiber
Corning SMF-28 (see e.g.
www.ee.byu.edu/photonics/connectors.parts/smf28.pdf); obviously,
the actual values must be measured in any experiment.

The {\em losses} are due to random scattering processes and depend
therefore exponentially on the length $\ell$: \ba
t&=&10^{-\alpha\,\ell/10}\,. \label{tfiber}\ea The value of $\alpha$ is strongly
dependent on the wavelength and is minimal in the two ``telecom
windows'' around 1330nm ($\alpha\simeq 0.34$dB/km) and 1550nm
($\alpha\simeq 0.2$dB/km).

The {\em decoherence} channels and their importance vary with the
coding of the information. Two main effects modify the state of
light in optical fibers. The first effect is {\em chromatic
dispersion}: different wavelengths travel at slightly different
velocities, thus leading to an incoherent temporal spread of a
light pulse. This may become problematic as soon as subsequent
pulses start to overlap. However, chromatic dispersion is a fixed
quantity for a given fiber, and can be compensated \cite{fas04b}.
The second effect is {\em polarization mode dispersion (PMD)}
\cite{gis92,gal05}. This is a birefringent effect, which defines a fast
and a slow polarization mode orthogonal to one another, so that
any pulse tends to split into two components. This induces a
depolarization of the pulse. Moreover, the direction of the birefringence may vary in time due to environmental factors: as such, it cannot be compensated statically. Birefringence effects induce decoherence in polarization coding, and may be problematic for all implementations that require a control on polarization. The importance of such effects depend on the fibers and on the sources; recent implementations can be made stable, even though they use a rather broadband source \cite{hub07}.

\subsubsection{Free Space Links}

A free space QKD link can be used in several very different scenarios, from 
short distance line-of-sight links with small telescopes mounted on 
rooftops in urban areas, to ground-space or even space-space links, involving the 
use of astronomical telescopes (see also \ref{sssexpfron}). Free-space QKD has been demonstrated in both the prepare-and-measure \cite{but98,rar01,hug02,kur02} and the entanglement-based configurations \cite{mar06,urs06,lin08,erv08}. 

The \textit{decoherence} of polarization or of any other degree of freedom is practically negligible. The \textit{losses} can roughly be divided into geometric and 
atmospheric. The \textit{geometric losses} are related with the apertures of receiving telescopes and with the 
effective aperture of the sending telescope (the one perceived by the receiving telescope, which is influenced by alignment, moving buildings, atmospheric turbulence etc.). The \textit{atmospheric losses} are due to scattering and to scintillation. Concerning scattering, within the 700-10.000nm  
wavelength range there are several 'atmospheric transmission windows', 
e.g. 780-850nm and 1520-1600nm, which have an attenuation 
$\alpha<$ 0.1dB/km in clear weather. Obviously, the weather conditions influence heavily such losses; numerical values are available, see e.g. \cite{kim01,blo03}. A simple model of the losses for a line-of-sight free space channel of length $\ell$ is therefore given by $t \approx\left(\frac{d_r}{d_s +D\,\ell}\right)^2\,10^{-\alpha\,\ell /10}$, where the first term is an estimate of the geometric losses ($d_s$ and $d_r$ are the apertures of the sending and receiving telescopes, $D$ is the divergence of the beam) and the second describes scattering ($\alpha$ is the atmospheric attenuation). We note that this formula does not account for scintillation, which is often the most critical factor in practice.

\subsection{Detectors}
\label{ssdet}

\subsubsection{Photon Counters}

Discrete-variable protocols use photon-counters as detectors. The main quantities characterizing photon-counters are the \textit{quantum efficiency} $\eta$ that represents the probability of a detector
click when the detector is hit by a photon, and the \textit{dark-count rate}
$p_d$ characterizing the noise of the detector -- dark counts are
events when a detector sends an impulse even if no photon has
entered it. An important parameter is also the \textit{dead time of the
detector}, i.e. the time it takes to reset the detector after a click. These three quantities are not independent. Most often, the overall \textit{repetition rate} at which the detector can be operated is determined by the dead time. For each of the detectors discussed below, the meaningful parameters are listed in Table \ref{tabledetectors}.

The most commonly used photon counters in discrete-variable
systems are \textit{avalanche photodiodes (APD)}. Specifically, for
wavelengths from the interval approximately 400--1000\,nm
Si APD can be used, for wavelengths from about 950\,nm to
1650\,nm, including telecom wavelengths, InGaAs/InP diodes
are most often applied. A whole \textit{savoir-faire} on the use of APDs has originated in the field of QKD \cite{gis02,cov04}. Because they can be operated with thermo-electric cooling, these detectors are an obvious choice for practical QKD, and in particular for commercial devices \cite{rib04,tri04}. Two recent developments are worth mentioning. First: instead of direct use of InGaAs APDs, one can detect signals at
telecom wavelengths (1310\,nm and 1550\,nm) by applying parametric
frequency up-conversion and then using efficient silicon APDs
\cite{dia05,the06}. Compared with InGaAs APDs, these \textit{up-conversion detectors} have
lower quantum efficiency but could in principle be operated in continuous mode thus
leading to repetition rates (GHz); however, as of today's
knowledge, they suffer from an intrinsic noise source that leads to high
dark count rates. Second: more recently, an improvement of the repetition rate and count rate by several orders of magnitude has been obtained by using a circuit that compares the output of the APD with that in the preceding clock cycle; such devices have been named \textit{self-differencing APDs} \cite{yua07b}.

Single-photon detectors other than APDs have been and are being developed. For instance, \textit{Visible Light Photon Counters} are semiconductor detectors that can also distinguish the number of impinging photons \cite{kim99,wak03,wak06b}. Other photon-counters are based on superconductors, for instance \emph{Superconducting Single Photon Detectors} \cite{ver02,ver04} and \textit{Transition Edge Sensors} \cite{mil03,ros05}; both types have been already used in QKD experiments \cite{had06,his06,ros07,ros08}. Each type has its own strong and weak features; in particular, all of them must be operated at cryogenic temperatures.

\begin{table}[t]
\begin{tabular}{|l|c|c|c|c|c|c|c|c|}\hline
Name & $\lambda$&$\eta$ & $p_d$ &Rep.&Count& Jitter& T &n \\
& [nm]& &  & [MHz]&[MHz]& [ps]& [K] &\\\hline
APDs:&& &  &&&&&\\
 Si & 600&50\% & 100Hz & cw& 15 &50-200 & 250 &N \\
InGaAs & 1550& 10\% & $10^{-5}$/g &10 &0.1 & 500 & 220 &N \\
Self-Diff.& &  &  &1250 &100 & 60 &  & \\ \hline Others:&& &  &&&&&\\
VLPC & 650 &58-85\% & 20kHz &cw & 0.015& N.A. & 6& Y\\ SSPD & 1550 &0.9\% & 100Hz & cw& N.A. &68 & 2.9 & N\\ TES & 1550&65\% & 10Hz &cw & 0.001 &9$\times 10^4$ & 0.1 &Y \\\hline
\end{tabular}
\caption{Overview of typical parameters of single-photon detectors: detected wavelength $\lambda$, quantum efficiency $\eta$, fraction of dark counts $p_d$ (g: gate), repetition rate (cw: continuous wave), maximum count rate, jitter, temperature of operation $T$; the last column refers to the possibility of distinguishing the photon numbers. For acronyms and references, refer to the main text.}\label{tabledetectors}
\end{table}

\subsubsection{Homodyne Detection}
\label{ssshomodet}

Continuous-variable QKD is based on the measurement of
quadrature components of light. This can conveniently be done
by means of \textit{optical homodyne detection}. This detection scheme uses two beams of the same frequency: the signal and the so-called local oscillator (much stronger and therefore often treated as classical). The beams are superimposed at a
balanced beam splitter. The intensity of light in each of the output modes is measured with proportional detectors, and the difference between the resulting photocurrents is recorded. If the
amplitude and the phase of the local oscillator are stable, the differential current
carries information about a quadrature component of the
input signal --- what quadrature component is actually
measured depends on the phase difference between the signal
and local oscillator. To keep this phase difference
constant, the signal and local oscillator are usually
derived from the same light source: the local oscillator beam needs to
be transmitted along with the signal from Alice to Bob; in practice, they are actually sent through the same channel, so that they experience the same phase noise and the relative phase remains unaltered --- note however that this practical change may render the scheme completely insecure, unless additional measurements are performed to verify the character of both the weak and the strong signal \cite{hae07}.

The intensities are measured by PIN diodes, which provide high detection
efficiency (typically 80$\%$) and relatively low noise. Therefore homodyne
detection could in principle operate at GHz repetition rates \cite{cam06} in contrast to photon counters based on APDs, whose detection rate is limited by the detector dead-time.

The use of such a high-rate homodyne detection technique
unfortunately comes with a price. Because of the uncertainty principle, the measurement of complementary quadratures is intrinsically noisy. The \textit{vacuum noise} (or \textit{intrinsic noise}) is the noise obtained when there is vacuum in the signal port (only the local oscillator is present). Now, the unavoidable transmission losses in the optical line, which simply cause ``missing clicks'' in photon-counting based schemes, result in a decrease of the signal-to-noise ratio in homodyne-detection based schemes. The vacuum noise is responsible for a rather significant added noise in continuous-variable QKD, which needs to be corrected during 
the classical post-processing stage: an additional computing effort in continuous-variable QKD.

In addition to the vacuum noise, an \textit{excess noise} is generated
mainly by detectors themselves and by the subsequent
electronics. In real systems, it is possible to reduce the
excess noise even 20\,dB below the shot noise; but this
ratio depends on the width of the spectral window, and
narrow spectral windows bound the modulation frequencies
(i.e. the repetition rates). 

\subsection{Synchronization and alignment}

\subsubsection{Generalities}

The problem of the \textit{synchronization} of two distant clocks, in itself, is a technical matter that has been solved efficiently in several different ways; basically, either one sends out a synchronization signal at regular intervals during the whole protocol, or one relies on an initial synchronization of two sufficiently stable clocks.
In the context of QKD, one has to consider possible hacking attacks that would exploit this channel (more in Sec. \ref{ssstrojan}).

The physical meaning of \textit{alignment} depends on the coding. For coding in polarization, it obviously means that Alice and Bob agree on the polarization directions. For phase coding, it refers rather to the stabilization of interferometers. Both procedures are most often performed by sending a servoing signal at a different frequency than the quantum signal, taking advantage of the bandwidth of the optical channel. Alternatively, self-stabilized setups have been proposed: this is the so-called Plug\&Play configuration, that we shall describe in the next paragraph in the context of phase-coding.

Before that, we have to mention that quantum mechanics allows also for a coding that does not require any alignment by exploiting the so-called ``decoherence-free subspaces'' \cite{zan97,boi04}. However, though demonstrated in some proof-of-principle experiments \cite{bou04,che06}, such coding is highly impractical, as it requires the preparation and measurement of complex multi-photon states; moreover, it is very sensitive to losses\footnote{The simplest example is the singlet state of two qubits: when both qubits are sent into the quantum channel, the state is robust against any misalignment $U$ since $U\otimes U\ket{\Psi^-}=\ket{\Psi^-}$. With \textit{four} physical qubits, there are two orthogonal states such that $U\otimes U\otimes U\otimes U\ket{\psi_{0,1}}=\ket{\psi_{0,1}}$; therefore, one can form an effective logical qubit $\ket{0}\equiv \ket{\psi_0}$ and $\ket{1}\equiv \ket{\psi_1}$ that is insensitive to misalignments. The states $\ket{\psi_{0,1}}$ are not easy to prepare and to detect. As a matter of fact, the available experiments did not produce those states: they produced a quite complex photonic state, that gives the required statistics conditioned on the observation of a specific detection pattern. In turn, this implies that all four photons must be transmitted and detected, therefore losses lead to a very fast decrease of the detection rate.}.

\subsubsection{Phase coding: two configurations}
\label{sspp}

\begin{figure}[ht]
\includegraphics[scale=0.6]{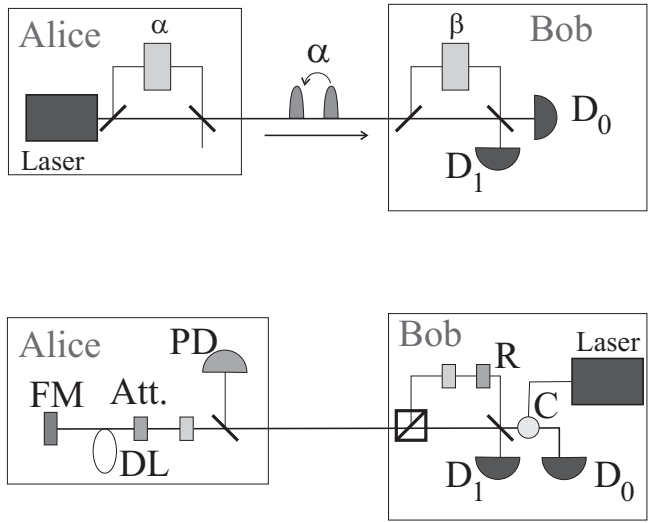}
\caption{Comparison of the one-way and two-way configurations for phase coding. The one-way configuration is called \textit{double Mach-Zehnder} (top). Alice splits each laser pulse into two pulses with relative phase $\alpha$; if Bob's phase is such that $\alpha-\beta =0$ modulo $\pi$, the outcome is deterministic in the absence of errors. In the two-way configuration, or \textit{Plug\&Play}(bottom), the source of light is on Bob's side. In detail: an intense laser pulse is sent through a circulator (C) into Bob's interferometer. The phase modulator is passive at this stage, but a polarization rotation (R) is implemented so that all the light finally couples in the fiber. On Alice's side, part of the light is deflected to a proportional detector (PD) that is used to monitor Trojan Horse attacks. The remaining light goes to a Faraday mirror (FM) that sends each polarization on the orthogonal one. On the way back, the pulses are attenuated down to the suitable level, then the coding is done as above. The role of the delay line (DL) is explained in the text.}\label{fig:mzpp}
\end{figure}

We consider P\&M schemes with phase coding. This coding has been the preferential choice in fiber implementations and has given rise to two possible configurations (Fig.~\ref{fig:mzpp}). In the configuration called \textit{one-way}, the laser is on Alice's side; it is typically realized with a double Mach-Zehnder interferometer \cite{ben92,tow93}. The other possible configuration has been called {\em Plug\&Play} configuration \cite{mul97,rib98}. As the name suggests, the goal of the Plug\&Play configuration is to achieve {\em self-alignment} of the system. Contrary to the one-way configuration, the Plug\&Play configuration puts the source of light on Bob's side: a strong laser pulse travels on the quantum channel from Bob to Alice. Alice attenuates this light to the suitable weak intensity (surely less than one photon per pulse in average, more precisions below and in Sec. \ref{ssestimatepm}), codes the information and sends the remaining light back to Bob, who detects. The coded signal goes as usual from Alice to Bob; but the same photons have first traveled through the line going from Bob to Alice. This way, interferometers become self-stabilized because the light passes twice through them; if the reflection on Alice's side is done with a Faraday mirror, polarization effects in the channel are compensated as well. These two configurations have shaped the beginning of practical QKD; we refer to a previous review \cite{gis02} for a thorough discussion.

It is useful here to address some problems that are specific for the Plug\&Play configuration, since they illustrate the subtleties of practical QKD. The system has an intrinsic duty cycle, which limits the rate at long distances: Bob
must wait a go-and-return cycle before sending other strong signals, otherwise the weak signal coded by Alice will be overwhelmed by the backscattered photons of the new strong ones\footnote{As a matter of fact, the back-scattering and the corresponding duty cycle could be avoided, but at the price of attenuating the pulses already at Bob's side. In turn, this implies that (i) a different channel should be used for synchronization, and (ii) the maximal operating distance is reduced in practice, especially if one takes Trojan Horse attacks into account,see below. Such a setup has been demonstrated \cite{bet00}.}. The nuisance has been reduced by having Bob send, not just one pulse, but a train of pulses; on Alice's side, a sufficiently long delay line must be added: all the pulses must have passed the phase modulator before the first one comes back and is coded. Still, this duty cycle is a serious bottleneck compared to one-way configurations. 

Also, two specific security concerns arise for the Plug\&Play configuration.
First concern: in full generality, there is no reason to assume that Eve
interacts only with the signal going from Alice to Bob: she might
as well modify the signal going from Bob to Alice. A
simple argument suggests that this is not helpful for Eve: Alice attenuates the light strongly and should actively randomize the global phase; then, whatever the state of the incoming light, the outgoing coded light
consists of weak signals with almost exact Poissonian statistics \cite{gis06}. Indeed, the rigorous analysis shows that unconditional security can be proved if the global phase is actively randomize, and that the resulting secret fractions are only slightly lower than those achievable with the one-way configuration \cite{zha08}. Second concern: since Alice's box must allow two-way transit of light, Trojan Horse attacks (see \ref{ssstrojan}) must be monitored actively, whereas in one-way setups they can be avoided by passive optical isolators. In practice, this may decrease the limiting distance\footnote{The argument goes as follows: upon receiving Bob's pulse, Alice attenuates it down to the desired intensity $\mu$. Now, it turns out that a simple error by a factor of 2, i.e. sending out $2\mu$ instead of $\mu$, would spoil all security (see \ref{ssestimatepm}). This implies that the intensity of the input pulse must be monitored to a precision far better than this factor 2. This precision may be hard to achieve at long distances, when Bob's pulse has already been significantly attenuated by transmission.}.

It is not obvious what the future perspectives of the Plug\&Play configuration will be: recently, stabilized one-way configurations have been demonstrated, which can also reach optical visibilities larger than 99\% and have a less constraining duty cycle \cite{gob04}. Still, the Plug\&Play configuration is an important milestone of practical QKD: in particular, the first commercial QKD systems are based on it\footnote{The configuration has been used also for continuous-variable coding \cite{leg06}, for a distributed-phase-reference protocol \cite{zho03} and for non-cryptographic quantum information tasks \cite{bra03}.}.

\section{Secret Key Rate}
\label{secssk}

We have seen in Sec. \ref{ssssecret} that the secret key rate $K$ is
the product of two terms (\ref{eqK}), the raw key rate $R$ and the
secret fraction $r$. This section is devoted to a detailed study
of these two factors. Clearly, the latter is by far the more
complex one, and most security studies are devoted only to it;
however the raw key rate is crucial as well in practice and its proper description involves some
subtleties as well. We will therefore start from this description.

\subsection{Raw key rate}
\label{ssraw}

The raw key rate reads \ba R&=&\nu_S\,\prob(\mbox{Bob accepts})
\label{rawgeneral}\ea The second factor depends both on the
protocol and on the hardware (losses, detectors) and will be
studied for each specific case. The factor $\nu_S$ is the
repetition rate.

In the case of \emph{pulsed sources} $\nu_S$ is the repetition
rate of the source of pulses. Of course, $\nu_S\leq \nu_S^{max}$,
the maximal repetition rate allowed by the source itself; but two
other limitations may become important in limiting cases, so that
the correct expression reads \ba
\nu_S^{\mathrm{pulse}}&=&\min\left(\nu_S^{max},\frac{1}{\tau_d\,\mu
t \,t_B \eta},\frac{1}{T_{dc}}\right)\,.\label{nus} \ea We explain
now what the two last terms mean.

The first limitation is due to the {\em dead-time of the
detectors} $\tau_d$. In fact, it is useless to send more light
than can actually be detected (worse, an excess of light may even
give an advantage to Eve). One can require that at most one
photon is detected in an interval of time $\tau_d$; the detection
probability is $\prob(\mbox{Bob detects})\approx \mu\, t\, t_B
\eta $ with $\mu=\moy{n}\lesssim 1$ the average number of photons
produced by the source, $t$ the transmittivity of the quantum channel, $t_B$ the losses in Bob's device and $\eta$ the efficiency of the detector.
Therefore, $\nu_S\lesssim \left(\tau_d\,\mu \,t \,t_B
\eta\right)^{-1}$. It is clear that this limitation plays a role
only at short distances: as soon as there are enough losses in the
channel, fewer photon will arrive to Bob than can actually be
detected.

The second limitation is associated to the existence of a {\em
duty cycle}: two pulses cannot be sent at a time interval smaller
than a time $T_{dc}$ determined by the setup. The expression for
$T_{dc}$ depends on the details of the setup. In Plug\&Play
configurations for instance, one cannot send the next train of bright pulses
before the weak signal of the earlier train has come back (\ref{sspp}): the effect becomes important at long distance.
Another example of a duty cycle is the one introduced by a
stabilization scheme for one-way configurations, in which each
coded signal is preceded by a strong reference signal
\cite{yua05}. Note finally that in any implementation with
time-bin coding, the advanced component of the next signal must
not overlap with the delayed component of the previous one.

In the case of heralded photon sources or
entanglement-based schemes working in a
\emph{continuous-wave (cw) regime} it is reasonable to define
$\nu_S$ as an average rate of Alice's detections,
thus\footnote{The source is assumed to be safe at Alice's side. It
is supposed that Alice's detectors are still ``open'' (not gated).
Dark counts and multi-pair contributions were neglected in the
estimation of $\nu_S^{\mathrm{cw}}$.} \ba
\nu_S^{\mathrm{cw}}&=&\min\left( \eta_A t_A \mu',
\frac{1}{\tau_d^A}, \frac{1}{\tau_d \, t \, t_B \eta},
\frac{1}{\Delta t} \right)\,. \label{nuscw} \ea Here $\eta_A t_A
\mu'$ is the trigger rate, with which Alice announces the pair
creations to Bob, with $\mu'$ being the pair-generation rate of
the source, $t_A$ is the overall transmittance of Alice's part of
the apparatus, and $\eta_A$ is the efficiency of Alice's
detectors. Of course, in practice this rate is limited by the dead
time of Alice's detectors $\tau_d^A$. The whole repetition rate is
limited by Bob's detector dead time $\tau_d$ and by the width of
coincidence window $\Delta t$ (usually $\Delta t \ll \tau_d$).

\subsection{Secret fraction}
\label{sssf}

\subsubsection{Classical information post-processing}
\label{12way}

To extract a short secret key from the raw key, classical
post-processing is required. This is the object of this paragraph, for more details see e.g.~\cite{ren05a,van06}. The security bounds for the secret fraction crucially depend on how this step is performed.

\paragraph{One-way post-processing.} These are the most studied and best known
procedures. One of the partners, the one who is chosen to hold the
reference raw key, sends classical information through the public
channel to the other one, who acts according to the established
procedure on his data but never gives a feedback. If the sender in
this procedure is the same as the sender of the quantum states
(Alice with our convention), one speaks of {\em direct
reconciliation}; in the other case, of {\em reverse
reconciliation}. The optimal one-way post-processing has been characterized and
consists of two steps.

The first step is {\em
error correction} (EC), also called \textit{information reconciliation}, at the end of which the lists of symbols
of Alice and Bob have become shorter but perfectly correlated. As proved by Shannon, the fraction of perfectly correlated symbols that can be extracted from a list of partially correlated symbols is bounded by the {\em mutual information} $I(A:B)=H(A)+H(B)-H(AB)$ where $H$ is the entropy of the probability distribution. In the context of one-way procedures with a sender S and a receiver R, it is natural to write $I(A:B)$ in the apparently asymmetric form $H(S)-H(S|R)$. This formula has an intuitive interpretation, if one remembers that the
entropy is a measure of uncertainty: the sender must reveal an
amount of information at least as large as the uncertainty the
receiver has on the reference raw key.

The second step is {\em privacy amplification} (PA). This
procedure is aimed at destroying Eve's knowledge on the reference raw key. Of course,
Alice and Bob will have chosen as a reference raw key the one on
which Eve has the smallest information: here is where the choice between direct and reverse reconciliation becomes meaningful\footnote{Note that, $I(A:B)$ being symmetric, there is no difference between direct and reverse reconciliation at the level of EC, as expected from the nature of the task.}. The fraction to be further
removed can therefore be written $\min\left(I_{EA},I_{EB}\right)$,
where $I_{E\cdot}$ is Eve's information on the raw key of Alice or Bob, that
will be defined more precisely in the next paragraph
\ref{sssattacks}. PA was first mentioned in \cite{ben88}, then established in \cite{ben95}. This reference has been considered as valid for one decade but, after the notion of universally composable security was introduced (see \ref{ssssecu}), it had to be replaced by a generalized version \cite{ren05c}. At the moment of writing, the only PA procedure that works in a provable way is the one based on two-universal hash functions\footnote{A set $\cal F$ of functions $f:X\rightarrow Z$ is called two-universal if $\mbox{Pr}[f(x)=f(x')]\leq \frac{1}{|Z|}$ for $x\neq x'$ and $f$ chosen at random with uniform probability. It is instructive to see why this definition is meaningful for privacy amplification. After EC, Alice and Bob share the same list of bits $x$; Eve has an estimate $x'$ of this list. For PA, Alice chooses $f$ from the two-universal set and announces it publicly to Bob. Both Alice and Bob end up with the shorter key $z=f(x)$; but the probability that Eve's estimate $z'=f(x')$ coincides with $z$ is roughly $1/|Z|$: Eve might as well choose randomly out of the set $Z$ of possible final keys.\\ Two-universal hash-functions, e.g. in the form of matrix multiplication, can be implemented efficiently \cite{carter79a,wegman81a}. The size of the matrices is proportional to the length $N$ of the raw key. Against a classical adversary, other \textit{extractors} exist whose size grows only like $\log N$; but at the moment of writing, it is not known whether a similar construction exists in the case where the adversary is quantum \cite{koe07}.}. Also, for
composability, the protocol must be symmetric under permutations:
in particular, the pairs for the parameter estimation must be
chosen at random, and the hash function has to be symmetric (as it
is usually).

In summary, the expression for the secret fraction extractable using one-way classical post-processing reads \ba r&=&I(A:B) - \min\left(I_{EA},I_{EB}\right)\,.
\label{genbound}\ea

\paragraph{Remarks on practical EC.} As mentioned above, the performance of EC codes is bounded by Shannon's mutual information. Practical EC codes however do not reach up to the Shannon bound. For \textit{a priori} theoretical estimates, it is fair to increase the number of bits to be removed by 10-20\%; more precise estimates are available \cite{lut99} but ultimately the performance must be evaluated on each code. We shall take this correction explicitly into account in Sections \ref{secdiscr}-\ref{seccompare}.

In addition, most of the efficient EC codes that are actually implemented, e.g. Cascade \cite{brassard93a}, use \textit{two-way communication}. To fit these two-way EC codes in the framework of one-way post-processing, one can give the position of the errors to Eve and treat all communication as one-way communication \cite{lut99}. Alternatively, one can use encryption of the EC data, as suggested in \cite{lut99} and formally proved in \cite{lo03}.

Note finally that it is not necessary to estimate the error rate with a small sample of the data: instead, the parties learn naturally the precise number of errors during the EC procedure.

\paragraph{Other forms of post-processing.} Bounds can be improved by {\em
two-way post-processing}, one refers to any possible procedure in which both
partners are allowed to send information. Since its first appearance in QKD
\cite{gis99,cha02,gotlo}, this possibility has been the object of several
studies\footnote{We note that some of the security claims in the first paper dealing with advantage distillation \cite{gis99} were imprecise. These works have also had an intriguing
offspring, the conjecture of the existence of ``bound
information'' \cite{gis00}, later proved for three-partite
distributions \cite{aci04}.}. Contrary to the one-way case, the
optimal procedure is still not known, basically because of the
complexity of taking feedback into account.

More recently, a further trick to improve bounds
was found, called {\em pre-processing}: before post-processing,
the sender (for one-way) or both partners (for two-way) can add
locally some randomness to their data. Of course, this decreases
the correlations between them, but it decreases Eve's information
as well, and remarkably the overall effect may be positive
\cite{kra05,ren05b}.

Both pre-processing and two-way post-processing are easy to implement and allow extracting a secret key in a parameter
region where one-way post-processing would fail; in particular, the critical tolerable error rate is pushed much higher\footnote{The order of magnitude of the
improvements is roughly the same for all examples that have been
studied. Consider e.g.~BB84 in a single-photon implementation, and
security against the most general attacks: the critical QBER for
one-way post-processing without pre-processing is $11\%$
\cite{sho00}; bitwise pre-processing brings this value up to
$12.4\%$ \cite{kra05}, more complex pre-processing up to $12.9\%$
\cite{smi06}; two-way post-processing can increase it
significantly further, at least up to $20.0\%$, but at the
expenses of drastically reduced key rate \cite{gotlo,cha02,bae07}. In
weak coherent pulses implementations, pre-processing increases the
critical distance of BB84 and of SARG04 by a few kilometers, both
for security against individual \cite{bra05} and most general attacks \cite{kra07}.}. To our knowledge though, they have been implemented only once in real systems \cite{ma06}. The reason is that, in terms of secret key rate, an improvement can be appreciated only when the dark counts become dominant\footnote{Recall that optical error is routinely kept far below $5\%$; therefore, the total error rate exceeds $\sim 10\%$ when the error is largely due to the dark counts.}, a regime in which few systems tend to operate --- see however \cite{ros08,tan08,yua08}. Therefore, in what follows, we shall present only bounds for one-way classical post-processing without pre-processing.

\subsubsection{Individual, Collective and Coherent Attacks}
\label{sssattacks}

As stressed from the beginning (\ref{sssuncond}), one aims ultimately at proving unconditional security, i.e. security bounds in the case where Eve's attack on the quantum channel is not restricted. Such a \textit{lower bound} for security has been elusive for many years (\ref{ssmilestones}); it has nowadays been proved for many protocols, but is still missing for others. In order to provide an ordered view of the past, as well as to keep ideas that may also be useful in the future, we discuss now several levels of security.

\paragraph{Individual (or incoherent) attacks.} This family describes the most constrained attacks that have been studied. They are characterized by the following properties:

\begin{itemize}

\item[(I1)] Eve attacks each of the systems flying from Alice to
Bob independently from all the other, and using the same strategy\footnote{We note here that this ``same strategy'' may be probabilistic
(with probability $p_1$, Eve does something; with probability
$p_2$, something else; etc), provided the probabilities are fixed
during the whole key exchange. Strange as it may seem from
the standpoint of practical QKD, an attack, in which Eve would
simply stop attacking for a while, belongs to the family of the
most general attacks!}. This property is easily formalized in
the EB scheme: the state of $n$ symbols for Alice and Bob has the
form $\rho_{\mathbf{AB}}^n= \left(\rho_{AB}\right)^{\otimes n}$.

\item[(I2)] Eve must measure her ancillae before the classical
post-processing. This means that, at the beginning of the
classical post-processing, Alice, Bob and Eve share a product
probability distribution of classical symbols.

\end{itemize}

In this case, the security bound for one-way post-processing is
the {\em Csisz\'ar-K\"orner bound}, given by (\ref{genbound}) with
\ba I_{AE}=\max_{Eve}I(A:E)&\quad& \mbox{(individual
attacks)}\label{ckbound}\ea and of course similarly for $I_{BE}$
\cite{csi78}. Here, $I(A:E)$ is the mutual information between the
classical symbols; the notation $\max_{Eve}$ recalls that one must
maximize this mutual information over Eve's strategies. There is actually an ambiguity in the literature, about the moment where Eve is forced to perform her measurement: namely, whether she is forced to measure immediately after the interaction \cite{lut96,cur05,bec06} or whether she can keep the signals in a quantum memory until the end of the sifting and error correction phase \cite{fuc97,slu98,bru98,bec99,lut99,bra00,cer02,her08}. The first case is associated to the hardware assumption that Eve is restricted not to have a quantum memory\footnote{Generalizing
\cite{wan01}, it is conjectured that individual attacks should be
optimal under the weaker assumption of a quantum memory that would
be bounded, either in capacity or in lifetime; but only rougher bounds have been derived so far \cite{dam05,koe06,dam07}.}. The second case is associated to the hardware assumption that Eve cannot perform arbitrary coherent measurements and can be useful as a step on the way to unconditional security proofs. However, we stress that the bound for collective attacks can nowadays be calculated more easily and gives more powerful results\footnote{At the moment of writing, there is still something that is known only for individual attacks, and this is Eve's full strategy; the optimal procedures been found both for the scenario without quantum memory \cite{lut96} and with it \cite{lut99,her08}. On the contrary, the bound for collective and coherent attacks is computed by optimizing the Holevo bound over all possible interactions between the signal and Eve's ancillae (see below): one implicitly assumes that suitable measurements and data processing exist, which will allow Eve to extract that amount of information. It would be surely interesting to exhibit explicit procedures also for more general attacks.}.

An important sub-family of individual attacks are the
{\em intercept-resend (IR) attacks}. As the name indicates, Eve
intercepts the quantum signal flying from Alice to Bob, performs a
measurement on it, and conditioned on the result she obtains she
prepares a new quantum signal that she sends to Bob. If performed
identically on all items, this is an individual attack. Moreover,
it obviously realizes an entanglement-breaking channel between Alice and Bob, thus providing an easily computed upper bound on the security of a protocol \cite{cur05,bec06}.

\paragraph{Collective attacks.} This notion was first proposed by Biham, Mor
and coworkers, who proved the security of BB84 against them and
conjectured that the same bound would hold for the most general attacks
\cite{bih97,bih02}. Collective attacks are defined as follows:

\begin{itemize}

\item[(C1)] The same as (I1).

\item[(C2)] Eve can keep her ancillae in a quantum memory until
the end of the classical post-processing, and more generally until any later time convenient to her (for instance: if the key is used to encode a message, part of which is vulnerable to plaintext attack, Eve may delay her measurement until she obtains the information coming from this attack). She can then perform the best measurement compatible with what she knows. In general, this
will be a collective measurement.

\end{itemize}

Only (C1) is an assumption on Eve's power. The generic bound
for the secret key fraction achievable using one-way
post-processing ({\em Devetak-Winter bound}) is given by (\ref{genbound}) with
\ba I_{AE}=\max_{Eve}\chi(A:E)&\quad& \mbox{(collective
attacks)}\label{dwbound}\ea and $I_{BE}$ defined in the analog
way \cite{dev05}. Here, $\chi(A:E)$
is the so-called {\em Holevo quantity} \cite{hol73} \ba
\chi(A:E)&=&S(\rho_E)-\sum_{a}p(a) S(\rho_{E|a})
\label{chiholevo}\ea where $S$ is von Neumann entropy, $a$ is a
symbol of Alice's classical alphabet distributed with probability
$p(a)$, $\rho_{E|a}$ is the corresponding state of Eve's ancilla
and $\rho_E= \sum_{a}p(a) \rho_{E|a}$ is Eve's partial state. The
Holevo quantity bounds the capacity of a channel, in which a
classical value (here $a$) is encoded into a family of quantum
states (here, the $\rho_{E|a}$): in this sense, it is the natural
generalization of the mutual information.

As mentioned, it is actually easier to compute (\ref{dwbound})
than (\ref{ckbound}). The reason lies in the optimization of Eve's
strategy. In fact, the Holevo quantity depends only on Eve's
states $\rho_{E|a}$, that is, on the unitary operation with which
she couples her ancilla to the system flying to Bob. In contrast to that, the mutual information depends both on Eve's states {\em
and} on the best measurement that Eve can perform to discriminate
them, which can be constructed only for very
specific examples of the set of states \cite{hel76}.

\paragraph{General (or coherent) attacks.} Eve's most
general strategy includes so many possible variations (she may
entangle several systems flying from Alice to Bob, she may modify
her attack according to the result of an intermediate
measurement...) that it cannot be efficiently parametrized. A
brute force optimization is therefore impossible. Nevertheless, as mentioned several times already, bounds for unconditional security have been found in many cases.  In all these cases, it turns out \textit{the bound is the same as for collective attacks}. This remarkable result calls for several comments.

First remark: this result has an intuitive justification. If the state $\ket{\Psi({\cal S}_n)}$ that codes the
sequence ${\cal S}_n$ has the tensor product form
$\ket{\psi(s_1)}\otimes...\otimes \ket{\psi(s_n)}$, then the
states flying from Alice to Bob are uncorrelated in the quantum
channel; therefore Eve does not seem to have any advantage in
introducing artificial correlations at this point\footnote{Of course, one is not saying that Eve \textit{does} fulfill (I1): Eve can do whatever she wants; but there exist an attack that fulfills (I1) and that performs as well as the best possible attack.}. However, correlations do appear later, during the classical post-processing of the raw
key; such that in fact, the final key is determined by the {\em
relations} between the symbols of the raw key, rather than by
those symbols themselves. Thus, Eve must not try and
guess the value of each symbol of the raw key, but rather some
relation between them --- and this is typically a situation in
which entanglement is powerful. This vision also clarifies why unconditional security is still elusive for those protocols, for which $\ket{\Psi({\cal S}_n)}$ is not of the tensor product form (see \ref{statusdpr}).

Second remark: for BB84, six-state and other protocols, assuming the squashing property of detectors (see \ref{sssquashing}), this result is a consequence of the internal symmetries \cite{kra05,ren05b}. The explicit calculations are given in Appendix \ref{appshor}. In a more general framework, the same conclusion can be reached by invoking the \textit{exponential De Finetti theorem} \cite{ren05a,ren07}. This theorem says that, after some suitable symmetrization, the statistics of the raw key are never significantly different from those that would be obtained under constraint (I1). This is a very powerful result, but again does not solve all the issues: for instance, because the actual exponential bound depends on the dimension of the Hilbert space of the quantum signals, it cannot be applied to continuous-variable QKD (see however the \textit{Note added in proof} at the end of this paper). Also recall that we consider only the asymptotic bound: the finite-key bounds obtained by invoking the De Finetti theorem are over-pessimistic \cite{sca07}.

\subsubsection{Quantum side channels and zero-error attacks}
\label{sssside}

The possibility of zero-error attacks seems to be at odds with the
fundamental tenet of QKD, namely that Eve must introduce
modifications in the state as soon as she obtains some
information. However, there is no contradiction: for instance, in the presence of \textit{losses} the quantum signal is also changed between the source and the receiver. Even if in most protocols (see discussion in Sec. \ref{sslight}) losses do not lead to errors in the raw key, some information about the value of the coded symbol
may have leaked to Eve.

Losses are the most universal example of leakage of information in a {\em quantum side-channel}, i.e. in some degree of freedom
other than the one which is monitored. We stress that the
existence of side-channels does not compromise the security,
provided the corresponding attacks are taken into
account in the privacy amplification.

The {\em beam-splitting (BS) attack} translates the fact that all
the light that is lost in the channel must be given to Eve:
specifically, Eve could be simulating the losses by putting a beam-splitter
just outside Alice's laboratory, and then forwarding the remaining
photons to Bob on a lossless line. The BS attack does not modify the optical mode that Bob receives:
it is therefore always possible for lossy channels and does not introduce any
error\footnote{For some sources, this attack simply does not give Eve any information:
for a perfect single-photon source, if the photon goes to
Eve, nothing goes to Bob, and viceversa.}. For an explicit computation of BS attacks, see \ref{ssmixedatt}.

When the signal can consist of more than one photon, Eve can
count the number of photons in each signal and act differently
according to the result $n$ of this measurement. Such attacks are
called {\em photon-number splitting (PNS) attacks}
\cite{ben92,dus99,lut00,bra00} and can be much more powerful than the BS
attack. They were discovered as zero-error attacks against BB84 implemented with weak laser pulses; in the typical parameter regime of QKD, even the Poissonian photon number distribution can be preserved \cite{lut02}, so that the PNS attack cannot be detected even in principle as long as one known signal intensity is used. To use different intensities in order to detect PNS attacks is the idea behind the {\em decoy states} method \cite{hwa03,lo05,wan05}. Also the distributed-phase-reference protocols detect the PNS attacks \cite{ino05,stu05}.

Finally, we mention the possibility of attacks based on {\em
unambiguous state discrimination (USD)} followed by resend of a
signal \cite{dus00}. These can be part of a PNS attack
\cite{sca04} or define an attack of its own \cite{bra07,cur07}; they are
clearly zero-error attacks and modify the
photon-number statistics in general.

Of course, a quantum side-channel may hide in any imperfect component of
the device (e.g., a polarizer which would also distort the wave
function according to the chosen polarization). The list of the possibilities is unbounded, whence the need for careful testing\footnote{Some very specific protocols and the corresponding security proofs can be made robust against such imperfections \cite{aci07}.}.

\subsubsection{Hacking on Practical QKD}
\label{ssstrojan}

In practical QKD, the security concerns are not limited to the
computation of security bounds for Eve's action on the quantum channel. Any specific implementation must
be checked against hacking attacks and classical leakage of information.

{\em Hacking attacks} are related to the weaknesses of an implementation. A first common
feature of hacking attacks is that they are feasible, or almost
feasible, with present-day technology. The best-known example is
the family of {\em Trojan Horse Attacks}, in which Eve probes the
settings of Alice's and/or Bob's devices by sending some light
into them and collecting the reflected signal \cite{vak01}.
Actually, the first kind of hacking attack that was considered is
a form of Trojan Horse that would come for free: it was in fact
noticed that some photon counters (silicon-based avalanche
photo-diodes) emit some light at various wavelengths when they
detect a photon \cite{kur01}. If this light carries some
information about which detector has fired, it must be prevented
to propagate out, where Eve could detect it. On these two
examples, one sees also the second common feature of all hacking
attacks, namely, that once they have been noticed, they can be
countered by adding some component. In all setups where light goes
only one way (out of Alice's lab and into Bob's lab), the solution
against Trojan Horse attacks consists in simply putting an optical
isolator; in implementations where light must go both ways
(typically, the Plug \& Play setups), the solution is provided by
an additional monitoring detector \cite{gis06}.

Apart from Trojan Horses, other hacking attacks have been invented to exploit
potential weaknesses of specific implementations, e.g. faked state
attacks \cite{mak05,mak06,mak07}, phase-remapping attacks \cite{fun06}, time-shift attacks \cite{qi07,zha07b}. It has also been noticed that a too precise timing disclosed in the Alice-Bob synchronization protocol may disclose information about which detector actually fired \cite{lam07}.

\subsubsection{A crutch: the ``uncalibrated-device scenario''}
\label{ssstrusted}

As stressed, all the errors and losses in the quantum channel must be attributed to Eve's intervention. But in a real experiment, there are errors and losses also inside the devices of the authorized partners. In particular, the detectors have finite efficiency (losses) and dark counts (errors); these values are known to the authorized partners, through calibration of their devices. A security proof should take this fact into account.

The task of integrating this knowledge into security proofs, however, has proved harder than one might think. In general, the naive approach, consisting in taking an attack and removing the device imperfections from the parameters used in privacy amplification, gives only an upper bound, even at the level of individual attacks\footnote{Consider a PNS attack (\ref{sssside}) on BB84 implemented with weak coherent pulses, and focus on the pulses for which Eve has found $n=2$ photons. The obvious PNS attack consists in Eve keeping one photon in a quantum memory and sending the other one to Bob, because in this case she obtains full information and introduces no error. But there is no information on non-detected photons: in particular, if Eve cannot control the losses in Bob's apparatus $t_B$ and the detector efficiency $\eta$, her information rate on such events will be $I_{2\rightarrow 1+1}=t_B\eta$. Now, consider another strategy: Eve applies a quantum cloner $2\rightarrow 3$, keeps one photon and sends the other two to Bob. Since no perfect cloning is possible, this introduces an error $\varepsilon_2$ on Bob's side and Eve's information on each detected bit is $I(\varepsilon_2)<1$. But Eve's information rate is $I_{2\rightarrow 2+1} =[1-(1-t_B\eta)^2] I(\varepsilon_2)\approx 2t_B\eta I(\varepsilon_2)$ and can therefore become larger than $I_{2\rightarrow 1+1}$. The full analysis must be done carefully, taking into account the observed total error rate; in the family of individual attacks, the cloning strategy performs indeed better than the ``obvious'' one for typical values of $t_B\eta$ \cite{cur04,nie05}. Note that there is no claim of optimality in this example: another attack may be found that performs still better.}. In particular, unconditional security proofs, whenever available, have been provided only under the assumption that \textit{all the losses and all the errors are attributed to Eve} and must therefore be taken into account in privacy amplification. We refer to this assumption as to the \textit{uncalibrated-device scenario}, because it all happens as if Alice and Bob would have no means of distinguishing the losses and errors of their devices from those originating in the channel\footnote{The name ``uncalibrated-device scenario'' is proposed here for the first time. In the literature, the assumption used to be named ``untrusted-device scenario''; but this name is clearly inadequate (see \ref{sssuncond} for the elements that must be always trusted in a QKD setup, and \ref{sssdevindep} for those may not be trusted in some very specific protocols).}. These issues have been raised in a non-uniform way in the literature. Most of the discussions have taken place for discrete-variable protocols; the security studies of distributed-phase-reference protocols are in a too early stage, but will surely have to address the question. The case of CV QKD may prove different because of the difference in the detection process (homodyne detection instead of photon counting).

At the moment of writing, the uncalibrated-device scenario is still a necessary condition to derive lower bounds. In the following sections, we shall work with this scenario. In \ref{ssupptrust} and \ref{ssstrustplots}, we shall compare the best available lower bounds with the upper bounds obtained with a naive approach to calibrated devices: we shall show (for the first time explicitly, to our knowledge) that in some cases the two bounds coincide for every practical purpose. In \ref{sssunctrust}, we summarize the status of this open problem.

\section{Discrete-variable protocols}
\label{secdiscr}

\subsection{Generic Assumptions and Tools}

As argued in Sec. \ref{ssstrusted}, in order to present lower bounds as they are available today, we work systematically in the uncalibrated-device scenario; paragraph \ref{ssupptrust} will present how to derive an upper bound for calibrated devices.

\subsubsection{Photon-number statistics}
\label{sssstatn}

We suppose that each signal is represented by a diagonal state in
the photon-number basis, or in other words, that there is no phase
reference available and no coherence between successive
signals\footnote{In some cases like Plug\&Play implementations,
the randomization of the phase should in principle be ensured
actively \cite{gis06,zha08}.}. Thus, \textit{Alice's source} can be described as
sending out a pulse that contains $n$ photons with probability
$p_A(n)$; Eve can learn $n$ without modifying the state, so this
step is indeed part of the optimal collective attack (Eve may
always choose not to take advantage of this information).

The statistical parameters that describe a key
exchange are basically detection rates and error rates\footnote{We
assume that these parameters are independent of Bob's
measurements, either because they are really measured to be the
same for all bases (a reasonable case in practice), or because,
after the sifting procedure, Alice and Bob forget from which
measurement each bit was derived and work with average values.}. Here are the main notations:
\begin{itemize}
\item $R$: total detection rate;
\item $R_n$: detection rate for the events when Alice sent $n$ photons ($\sum_nR_n=R$);
\item $Y_n=R_n/R$ a convenient notation ($\sum_nY_n=1$);
\item $R_n^w$: wrong counts among the $R_n$;
\item $\varepsilon_n=R_n^w/R_n$ the error rate on the $n$ photon signals;
\item $Q=\sum_nY_n\varepsilon_n$ the total error rate (QBER).
\end{itemize}
Concerning photon statistics on Bob's side, it is important to notice the following. If the channel introduces random losses, the photons that enter Bob's device are distributed according to $p_{B}^{\,t}(k)= \sum_{n\geq k}
p_A(n)\,C_n^k t^k(1-t)^{n-k}$ where $C_n^k=\frac{k!}{n!(n-k)!}$
is the binomial factor; one could compute $R_n$ from this value and the details of the protocol. However, Eve can adapt her strategy to the value of $n$, so the photon-number statistics $p_B(k)$ on Bob's
side may be \textit{completely different} from $p_{B}^{\,t}(k)$ \cite{lut02}.

\subsubsection{Qubits and Modes}
\label{sssquashing}
Many, though not all, security proofs can be obtained by finding qubit protocols in the optical implementations that work with optical modes.

\paragraph{Sources: Tagging.} On the source side, this can be done by 'tagging', by assuming that all multi-photon signals (with respect to the total signal)  becoming fully known to an eavesdropper. This leaves us effectively with qubits, using the single photons and the coding degree of freedom, for example polarization or relative phase between two modes. This method has been used in \cite{lut00,ina07}, but the term tagging has been introduced only in \cite{got04}. Note that security proofs can be done without this assumptions, e.g. in the case of the SARG protocol. 

\paragraph{Detectors: Squashing.} Detectors act on optical modes, and typically threshold detectors are used that cannot resolve the incoming photon number. Some security proofs \cite{may96,may01,koa05} can directly deal with this situation. In other security proofs one has either to search through all possible photon number of arriving signals to prove that it is Eve's optimal strategy to send preferentially single photons to Bob \cite{lut99}. It was there realized  that double clicks in detection devices, resulting from multi-photon signals or dark counts, cannot be simply ignored, as a security loophole would open up. \footnote{A simple attack exploiting this loophole goes as follows:  Eve performs an intercept/resend attack and resend a pulse containing a large number of photons in the detected polarization. If Bob measures in the same basis as Eve, he will receive a single detector click, about which Eve has full information. If Bob measures in a different basis, he will receive almost always double clicks, which he would discard. Therefore Eve has perfect information about all signals retained by Eve, allowing her to break the QKD scheme.} As a countermeasure, in \cite{lut99,lut00} it was introduced to assign double clicks at random to the values corresponding to single click events.

The concept of squashing, originally introduced in a continuous variable context \cite{GP01}, has been coined in \cite{got04}, where it is assumed that the detection device can be described by a two-step process: in a first step, the optical signal is mapped (squashed) into a single photon (qubit), and then the ideal measurement in the qubit description is performed. Only recently, it has been shown that a squashing model actually exists for the BB84 protocol \cite{tsu08b,bea08} with the given assignment of double clicks to random single detector clicks. Actually, in \cite{bea08}, a framework has been developed to find squashing maps for different detector set-ups, including the implementation of passive basis choice in the BB84 protocols via a beamsplitter. Note that the existence of a squashing model should not be taken for granted, as for example the six-state protocol does not admit a squashing model. However, a six-state protocol measurement with a passive basis choice via a linear optics array admits a squashing model for suitable assignment of multi-clicks. \cite{bea08b}.

Note again that it is not necessary to find a squashing model to prove security, but it is certainly an elegant short cut, as now the combination of tagging in the source and squashing in the detector allows to reduce the security analysis of QKD to qubit protocols. For the remainder of this review, however, we adopt the squashing model view.

\subsubsection{Secret key rate}

The bound for the secret fraction is (\ref{genbound}). In the case of the protocols under study, $H(A)=H(B)=1$ and $H(A|B)=H(B|A)=h(Q)$, where $h$ is binary entropy and $Q$ is the QBER. Therefore $I(A:B)=1-h(Q)$. However, we want to provide formulas that take imperfect error correction into account. Therefore we shall use
\ba
K&=&R\,\left[1-\mbox{leak}_{EC}(Q)-I_E\right]\label{Kdiscrete}
\ea
with $\mbox{leak}_{EC}(Q)\geq h(Q)$ and $I_E=\min\left(I_{AE},I_{BE}\right)$. Let us study this last term. Eve gains information only on the non-empty pulses, and provided
Bob detects the photon she has forwarded. Since, due to the squashing model, the exponential De
Finetti theorem applies to discrete-variable protocols (see discussion in Sec. \ref{sssattacks}), and
since the optimal collective attack includes the measurement of
the number of photons, the generic structure for the Eve's
information reads\footnote{More explicitly, this formula should read $I_E=\min\left(I_{AE},I_{BE}\right)$ with $I_{AE}=\max_{Eve}\,\sum_n Y_n\,I_{AE,n}$ and similarly for $I_{BE}$. In the development of QKD, this formula
was derived first for BB84 \cite{got04}, then for SARG04
\cite{fun06a}, then generalized to all discrete-variable protocols
\cite{kra07}.} \ba 
I_{E}&=&\max_{Eve}\,\sum_n Y_n\,I_{E,n}
\label{eqidiscr}\ea where, as usual, the maximum is to be taken on
all Eve's attacks compatible with the measured parameters.

\subsection{BB84 coding: lower bounds}
\label{ssbb84}

In the BB84 coding, the probability that Bob accepts an item depends only on
the fact that he has used the same basis as Alice, which happens with probability $p_{sift}$. Therefore, writing $\tilde{\nu}_S=\nu_S\,p_{sift}$, we have \ba R_{n} &=&\tilde{\nu}_S \,p_A(n)\,f_n\ea where $f_n$ is the probability that Eve forwards some signal to Bob for $n$-photon pulses. Eve's attack must be optimized over the possible $\{f_n\}_{n\geq 0}$ compatible with $\sum_{n}R_n=R$. Now we consider different implementations of this coding.

\subsubsection{Prepare-and-Measure: Generalities}
\label{ssspmgen}

In P\&M BB84, $I_{AE}=I_{BE}$. On the events when Alice sends no photons ($n=0$) but Bob has a detection, the intuitive result $I_{E,0}=0$ \cite{lo05x} has indeed been proved \cite{koa06}. On the single-photon pulses, Eve
can gain information only at the expense of introducing an error
$\varepsilon_1$; the maximal information that she can obtain this
way is $I_{E,1}=h(\varepsilon_1)$ where $h$ is binary
entropy \cite{sho00}. A possible demonstration of this well-known result is given in Appendix \ref{appshor}.
For multi-photon pulses, the best attack
is the PNS attack in which Eve forwards one photon to Bob and keeps the others: i.e. for $n\geq 2$,
$\varepsilon_n=0$ and $I_{E,n}=1$ \cite{got04,fun06a,kra07}. Therefore (\ref{eqidiscr}) becomes
\ba I_E &=&\max_{Eve}
\left[Y_1h(\varepsilon_1)+\big(1-Y_0-Y_1\big)\right]\nonumber\\&=&1-\min_{Eve}\,\left\{Y_0+Y_1[1-h(\varepsilon_1)]\right\}\,.
\label{iebb84}\ea

\subsubsection{P\&M without decoy states}

In P\&M schemes without decoy states, the only measured parameters
are $R$ and $Q$. We have to assume $\varepsilon_{n\geq 2}=0$; therefore we obtain $\varepsilon_1=Q/Y_1$. From this and (\ref{iebb84}), we see\footnote{First proved in \cite{ina07} in the context of unconditional security.} that Eve's optimal attack compatible with the measured parameters is the one which
minimizes $Y_1$, a situation which is obviously achieved by
setting $f_0=0$ and $f_{n\geq 2}=1$. One finds then \ba Y_1&=&
1-(\tilde{\nu}_S/R)\,p_A(n\geq 2)\,. \label{y1bb84}\ea
As a conclusion, for BB84 in a P\&M scheme without decoy states,
the quantity to be subtracted in PA is \ba
I_E&=&1-Y_1[1-h(Q/Y_1)]\,; \ea the corresponding
achievable secret key rate (\ref{Kdiscrete}) is \ba
K&=&R\,\left[Y_1\left(1-h(Q/Y_1)\right)-\mbox{leak}_{EC}(Q)\right] \label{kbb84}\ea where $Y_1$
is given in (\ref{y1bb84}). As expected,
$K$ contains only quantities that are known either from
calibration or from the parameter estimation of the protocol ($R$,
$Q$).

\subsubsection{P\&M with decoy states}
\label{sssdecoy}

The idea of {\em decoy states} is simple and deep. Alice changes the nature of the quantum signal at random during the protocol; at the end of the exchange of quantum signals, she will reveal which state she sent in each run. This way, Eve cannot adapt her attack to Alice's state, but in the post-processing Alice and Bob can estimate their parameters conditioned to that knowledge. The first proposal using one- and two-photon signals \cite{hwa03} was rapidly modified to the more realistic implementation in which Alice modulates the intensity of the laser
\cite{lo05,wan05}. As we mentioned, several experiments have already been performed \cite{zha06, ma06,pen07,ros07,yua07}, more recently even including finite-key effects \cite{has07}.

Let $\xi$ be some tunable parameter(s) in the
source, the typical example being $\xi=\mu$ the intensity (mean
photon-number) of a laser. Alice changes the value of $\xi$
randomly from one pulse to the other; at the end of the exchange
of quantum signals, Alice reveals the list of values of $\xi\in{\cal X}$, and
the data are sorted in order to estimate the parameters separately
for each value. With this simple method, Alice and Bob measure $2|{\cal X}|$ parameters, namely the $R^{\xi}$ and the $Q^{\xi}$.

The set ${\cal X}$ is publicly known as part of the protocol; but if $|{\cal X}|>1$,
Eve cannot adapt her strategy to the actual value of $\xi$ in each
pulse, because she does not know it. Therefore, $f_n$ and $\varepsilon_n$
are independent of $\xi$; in particular, $R_n^\xi=\tilde{\nu}_S \,p_A(n|\xi)\,f_n$. The measured parameters \ba R^\xi=\sum_{n\geq 0}R_n^\xi\,&\mbox{ and
}&\, Q^\xi=\sum_{n\geq 0}\frac{R_n^\xi}{R^\xi}\varepsilon_n \label{measdecoy}\ea define a
linear system with $2|{\cal X}|$ equations for the $f_n$ and the
$\varepsilon_n$. The optimization in (\ref{iebb84}) must then be performed using the lower bounds for $Y_1^\xi$ and
the upper bound for $\varepsilon_1$ as obtained from the
measured quantities $\{R^\xi,Q^\xi\}_{\xi\in{\cal X}}$ \cite{tsu08}. In practice, the
meaningful contributions are typically the $n=0,1,2$ terms, and a decoy-state protocol with $|{\cal X}|=3$ reaches very
close an exact determination \cite{hay07b}. For simplicity, here we suppose that all the 
$f_n$ and $\varepsilon_n$ have been determined exactly\footnote{As a side remark: one might find $\varepsilon_{n\geq 2}>0$, but this does not modify the discussion in Sec. \ref{ssspmgen} about the optimal attack. Indeed, Eve might have performed the attack that gives $\varepsilon_{n\geq 2}=0$, then added some errors ``for free''.}. Also, we consider a protocol in which the
classical post-processing that extracts a key is done separately
on the data that correspond to different $\xi$. For each $\xi$,
the quantity to be subtracted in PA is\footnote{Note the presence of $Y^\xi_0$ in the next two equations.} \ba
I_E^\xi&=&1-Y^\xi_0-Y^\xi_1[1-h(\varepsilon_1)] \ea with $Y^\xi_{0,1}=R^\xi_{0,1}/R^\xi$ and the
corresponding achievable secret key rate is \ba
K^\xi&=&R^\xi\,\left[Y^\xi_0+Y_1^\xi\left(1-h(\varepsilon_1)\right)-\mbox{leak}_{EC}(Q^\xi)\right]\,.
\label{kbb842}\ea The total
secret key rate is $K=\sum_\xi'K^\xi$, where the sum is taken on
all the values of $\xi$ such that $K^\xi\geq 0$. If the classical
post-processing were done on the whole raw key, the total secret
key rate would read $K=R[1-\mbox{leak}_{EC}(Q)]-\sum_\xi R^\xi I_E^\xi$. The
two expressions coincide if there exists a $\xi$ which is used
almost always.

\subsubsection{P\&M: analytical estimates}
\label{ssestimatepm}

Alice and Bob can optimize $K$ by playing with the parameters of
the source, typically the intensity. A rigorous optimization can be done only numerically. In this paragraph, we re-derive some often-quoted results for P\&M implementations of BB84. For this \textit{a priori} estimate, one has to assume that some ``typical'' values for the $R_n$ and the $Q_n$ will indeed be observed. As stressed above, security must be based on the actually measured values: what follows provides only guidelines to start working with the correct orders of magnitude. Here, we chose to work in a regime in which the rate of detection of true photons is much larger than the dark count rate. For simplicity, we also assume optimal error correction, so that $\mbox{leak}_{EC}(Q)=h(Q)$. 

The reference case is the case of single-photon sources, for which the meaningful scheme is P\&M without decoy states. For this source, $p_A(1)=1$ therefore $Y_1=1$; the expected detection rate is $R=\tilde{\nu}_S t\,t_B\eta$, and Eq.~(\ref{kbb84}) yields immediately
\ba
K&\approx&\tilde{\nu}_S t\,t_B\eta\,[1-2h(Q)]\quad\mbox{(single photon)}\,.
\ea
As expected, $K$ scales linearly with the losses in the line and the efficiency of the detector.

The most widespread source in P\&M schemes are attenuated lasers. The estimate can be made by considering only the single-photon and the two-photon emissions: $p_A(1)=\mu e^{-\mu}$, $p_A(2)=\mu^2e^{-\mu}/2$. The expected detection rate is $R=\tilde{\nu}_S \mu t\,t_B\eta$. The important feature, which is absent in the study of single-photon sources, is the existence of an {\em optimal value for the intensity} $\mu$, a compromise between a large $R$ and a small $p_A(2)$. We focus first on implementations without decoy states. We can set $p_A(1)\approx\mu$ and $p_A(2)\approx\mu^2/2$, but still, the optimal value of $\mu$ cannot be estimated exactly in general, because $Y_1=1-\frac{\mu}{2t\,t_B\eta}$ depends on $\mu$ and appears in a non-algebraic function. Let us then consider first the limiting case $Q=0$: Eq.~(\ref{kbb84}) becomes $K/\tilde{\nu}_S\approx \mu t\,t_B\eta-\mu^2/2$, whose maximal value is $\demi(t\,t_B\eta)^2$ obtained for $\mu_0=t\,t_B\eta$ \cite{lut00}. To obtain estimates for the $Q>0$ case, we can make the approximation of using $\mu_0$ to compute $Y_1$, i.e. to set $Y_1=\demi$. Then, the optimization of Eq.~(\ref{kbb84}) is also immediate: writing $F(Q)=1-h(2Q)-h(Q)$, the highest achievable secret key rate is
\ba
\frac{K}{\tilde{\nu}_St\,t_B\eta}&\approx &\demi\,\mu_{opt}\,F(Q) \quad\mbox{(laser, no decoy)}
\ea
obtained for the optimal mean photon number
\ba
\mu_{opt}&\approx& t\,\,t_B\eta\frac{F(Q)}{1-h(2Q)}\,.\ea

Let us now perform the estimate for an implementation using decoy states. The decoy consists in varying the intensity of the laser from one pulse to the other, so that the general parameter $\xi$ is in fact $\mu$. We suppose that a given value $\mu$ is used almost always (and this one we want to optimize), while sufficiently many decoy values are used in order to provide a full parameter estimation. The expected values are $R^{\mu}=\tilde{\nu}_S\mu t\,t_B\eta$, $R_1^{\mu}=\tilde{\nu}_S\mu e^{-\mu} t\,t_B\eta$ and $\varepsilon_1=Q$. Inserted into Eq.~(\ref{kbb842}), one obtains $K\approx \tilde{\nu}_S\mu t\,t_B\eta[e^{-\mu}(1-h(Q))-h(Q)]$; using $e^{-\mu}\approx 1-\mu$, this expression reaches the maximal value
\ba
\frac{K}{\tilde{\nu}_St\,t_B\eta}&\approx &\demi\,\mu_{opt}\,[1-2h(Q)]  \quad\mbox{(laser, decoy)}
\ea
for the optimal mean photon number
\ba
\mu_{opt}&\approx& \demi\left[1-\frac{h(Q)}{1-h(Q)}\right]\,.
\ea
Let us summarize. Without decoy states, $\mu_{opt}\sim t$ and consequently $K\propto t^2$: the larger the losses, the more attenuated must the laser be. The reason are PNS attacks: Alice must ensure that Eve cannot reproduce the detection rate at Bob's by using only photons that come from 2-photon pulses (on which she has full information). With decoy states, one can {\em determine} the fraction of detections that involve photons coming from 2-photon pulses; if this fraction is as low as expected, one can exclude a PNS attack by Eve --- as a benefit, the linear scaling $K\propto t$ is recovered. This is the same scaling obtained with single-photon sources, with the obvious benefit that lasers are much more versatile and well-developed than strongly sub-Poissonian sources. Another interesting remark is that, both with and without decoy states, $\mu_{opt}\approx\demi\mu_{crit}$, where the critical value $\mu_{crit}$ is defined as the one for which $K\approx 0$. In other words, \textit{an intensity double than the optimal one is already enough to spoil all security}. In implementations without decoy states, where $\mu$ decreases with $t$, this calibration may be critical at long distances.

\subsubsection{Entanglement-Based}
\label{sssbb84eb}

If \textit{Alice holds
the down-conversion source}, as is the case in almost all the EB QKD experiments performed to date\footnote{We are aware of a single case, in which the source was in the middle \cite{erv08}. As we shall discuss below in this paragraph, security proofs have been provided also for this situation.}, an EB scheme is
equivalent to a P\&M one (see \ref{sssprotoq}) so the
corresponding security proofs could be applied. The only specific
difference to address concerns the events in which more than one
pair is produced inside a coincidence window. As described in
Sec.~\ref{ssspdc}, two kinds of such contributions exist and Eve
is able to distinguish between them:
\begin{itemize}

\item A fraction of the multi-pair events contain partial
correlations in the degrees of freedom used for symbol encoding;
thus, Eve can get information on the key bit by some form of PNS
attacks. This situation is similar to the multi-photon case in
P\&M schemes, although here it is difficult to determine exactly
the amount of information that leaks out. To be on the safe side
we will suppose that Eve can obtain full information without
introducing any errors.

\item The other, usually much larger fraction of
multi-pair events consists of independent uncorrelated
pairs. In this case Eve cannot obtain any information on Bob's
symbol using the PNS attack. She can only apply ``standard''
single particle attack. We suppose that Eve can somehow find
out which one of multiple pairs were selected by Alice's detector,
so we treat all such multi-pair contributions as if they
were single pairs.
\end{itemize}
Therefore Eq.~(\ref{iebb84}) is replaced by \ba I_E \le Y'_m +
Y'_1\,h\left(\frac{Q}{Y'_1}\right), \ea where $Y'_1$ is the
fraction of single-pair plus uncorrelated multi-pair events and
$Y'_m$ is the fraction of multi-pair events which are (partially)
correlated in the degree of freedom the information is encoded in.
Explicitly, \ba Y'_m=p_A(n\ge2) \frac{\tilde{\nu}_S}{R} \,\zeta
\label{eqzeta}\ea with $\zeta$ being the ratio of the number of
partially correlated multi-pair contributions to all multi-pair
contributions (see Sec.~\ref{ssspdc}). In total $Y'_m+Y'_1=1$.
Finally, the achievable secret-key rate reads \ba K =
R\,\left[Y'_1\left(1-h(Q/Y'_1)\right)-\mbox{leak}_{EC}(Q)\right]\,.
\label{kbb843} \ea  Recall that these formulas apply to
implementations, in which the source is safe on Alice's side.
Notice also that two different sorts of multi-pair contributions
are considered and for each of them different eavesdropping
strategy is assumed. However, in reality there is a smooth
transition between correlated and uncorrelated pairs. All
multi-pair events which exhibit non-negligible correlations must
be counted as correlated.

Recently security has been demonstrated also for EB systems, in
which \textit{the source is under Eve's control} \cite{ma07}. The
authors describe the conditions, under which the whole object
``Eve's state preparation and Alice's measurement'' behaves like
an uncharacterized source in the sense of Koashi and Preskill
\cite{koa03}. Alice has a box where she can dial a basis and gets
an information bit from her box indicating which signal (0 or 1)
was sent. Whatever state Eve prepares, when she gives one part
into Alice's box and Alice chooses a measurement, then the average
density matrix outside this box is independent of this choice
(assuming that the no-click event probability is basis
independent).\footnote{This is clearly true for an active basis
choice. In case of the passive basis selection some additional
assumptions on the detection may be necessary.} On Alice's side no
Hilbert space argument is needed, but on Bob's side the squshing property of the detection is required (see \ref{sssquashing}). The formula for the achievable secret-key rate then reads
\ba
K = R\,\left[1-h(Q)-\mbox{leak}_{EC}(Q)\right]\,.
\label{eb-ma}
\ea
Formally, this is the same as obtained in a P\&M
scheme using single photons [Eq.~(\ref{kbb84}) with $Y_1=1$]. As
such, it is a remarkable result: it states that, under the
assumptions listed above, all the deviations from a perfect
two-photon source --- in particular, the presence of multi-photon
components --- are taken care of by measuring the error rate $Q$
\cite{koa03}. Besides, it has been found that the EB QKD can
tolerate higher losses if the source is placed in the middle
between Alice and Bob rather than if it is in Alice's side
\cite{wak02b,ma07}.

Finally, we note that very recently another proof of the security of entanglement-based systems with real detectors was proposed, that does not rely on the squashing property but rather on the measurement of the double-click rate \cite{koa08}.

\subsection{BB84 coding: upper bounds incorporating the calibration of the devices}
\label{ssupptrust}

As explained in Sec. \ref{ssstrusted}, the bounds for unconditional security are always found for the uncalibrated-device scenario, which is over-pessimistic. It is instructive to present some upper bounds that take the calibration of the devices into account: the comparison between these and the lower bounds will determine the ``realm of hope'', i.e. the range in which improvements on $K$ may yet be found. Clearly, the contribution $\mbox{leak}_{EC}(Q)$ of error correction is independent of the scenario: one has to correct for all the errors, whatever their origin. The difference appears in the quantity to be removed in privacy amplification.

\subsubsection{Statistical parameters}

In order to single out the parameters of the devices, one has first to recast the general notations (\ref{sssstatn}) in a more elaborated form. The detection rates must be explicitly written as
\ba
R_n&=&R_{n,p}+R_{n,d}
\ea
where $R_{n,p}$ is the contribution of detections and $R_{n,d}$ is the contribution of dark counts. Since Eve can act only on the first part, it is convenient to redefine $Y_n=R_{n,p}/R$, so that $\sum_nY_n\equiv Y<1$. The errors on the line $\varepsilon_n$ are introduced only on the photon contribution, while the dark counts always give an error rate of $\demi$; therefore the total error is
\ba
Q&=&Y\varepsilon+\delta
\ea where $\varepsilon = \sum_{n\geq 1}\frac{Y_n}{Y}\varepsilon_n$ and $\delta=\frac{1-Y}{2}$.  

Note that the content of this paragraph is not specific to BB84; but all that follows is.

\subsubsection{Upper bounds}

To derive an upper bound, we use a simple recipe, which consists in following closely the calculations of the previous subsection \ref{ssbb84} and just making the necessary modifications, although this is known to be sub-optimal and no squashing model is known in this situation to justify the assumption. In particular, Eve is still supposed to forward to Bob at most one photon, although this is known to be sub-optimal. Therefore \ba R_{n,p}&=&\tilde{\nu}_Sp_A(n)f_n\,t_B\eta\\R_{n,d}&=&\tilde{\nu}_Sp_A(n)(1-f_n\,t_B\eta)\,2p_d\ea where $p_d$ is the dark count rate. Note the presence of $\,t_B\eta$ in these formulas: the detector efficiency has \textit{not} been incorporated into $f_n$. Extracting $f_n\,t_B\eta$ from these equations, one finds
\ba
Y&=& \left(1-2p_d\tilde{\nu}_S/R\right)/(1-2p_d)
\ea
which means that the ratio between detections and dark counts depends only on the total detection rate $R$. Also, for our simple recipe, it is immediate that the modification of the general expression (\ref{iebb84}) reads
\ba I_E &=&\max_{Eve}
\left[Y_1h(\varepsilon_1)+\big(Y-Y_1\big)\right]\nonumber\\&=&Y-\min_{Eve}\,Y_1[1-h(\varepsilon_1)]\,.
\label{iebb84trust}\ea 

We restrict now to the P\&M schemes. In the \textit{implementation with decoy states}, the $Y_n$ and the $\varepsilon_n$ are known, so the only difference with the uncalibrated-device formula (\ref{kbb842}) is the role of dark counts:
\ba
K^\xi&=&R^\xi\,\left[Y_1^\xi\left(1-h(\varepsilon_1)\right)+2\delta^{\xi}-\mbox{leak}_{EC}(Q^\xi)\right]
\label{kbb842trust}\ea where $Y_0$ is replaced by the very slightly larger term\footnote{In the notation of this paragraph, the previous $Y_0$ would read $R_0/R=R_{0,d}/R$; while $2\delta=\sum_{n\geq 0}R_{n,d}/R$. Note that, strictly speaking, $R_0=R_{0,d}$ is an assumption: a priori, one can imagine that Eve creates some photons to send to Bob also when Alice is sending no photons --- but we don't consider here such a highly artificial situation.} $2\delta^{\xi}=1-Y^{\xi}$. Things are different for the \textit{implementation without decoy states}, because now $Y_1$ and $\varepsilon_1$ are not directly measured, only $R$ and $Q$ are. Since we are supposing that the optimal strategy is still such that $\varepsilon_{n\geq 2}=0$ and $f_{n\geq 2}=1$, we have \ba Y_1=
Y-\,t_B\eta\frac{\tilde{\nu}_S}{R}\,p_A(n\geq 2)&\mbox{ and }& \varepsilon_1=\frac{Q-\delta}{Y_1}\,. \label{y1bb84trust}\ea Note that $Y_1$ can be significantly larger than in the uncalibrated-device scenario, eq.~(\ref{y1bb84}): in fact, although $Y$ is slightly smaller than one, the term to be subtracted is multiplied by $\,t_B\eta$. This difference is specifically due to the fact that Eve is not supposed to influence the efficiency of the detector. Finally, one obtains
\ba
K&=&R\,\left[Y_1\left(1-h(\varepsilon_1)\right)+2\delta-\mbox{leak}_{EC}(Q)\right]
\label{kbb841trust}\ea with the expressions (\ref{y1bb84trust}) and with $2\delta=1-Y$.

\subsection{Bounds for the SARG04 coding}
\label{sssarg04}

We sketch here the analysis of SARG04 because it contains a certain number of instructive
differences with respect to BB84. Here we note
$\tilde{\nu}_S=\nu_S/2$ because Bob must always choose the bases
with probability $\demi$, even if Alice would almost always use
the same set of states. The raw key rates are different from those
of BB84. For definiteness, suppose that Alice send $\ket{+x}$, so
the bit is accepted if Bob finds ``$-$''. If Bob measures $X$, he accepts the bit only if he obtains ``$-$'', but this can only be due to an error. We write $R_{n}^w=\tilde{\nu}_Sp_A(n)\,f_n\,\tilde{\varepsilon}_n$ where the relation of $\tilde{\varepsilon}_n$ to the induced error rate ${\varepsilon}_n$ will be computed just below. If Bob measures $Y$, he gets ``$-$'' in half of the
cases\footnote{As such, this statement contains an assumption on
Eve's attack, namely $\Tr[\si_y \rho(\pm x)]=0$ where $\rho(\pm
x)$ is the state received by Bob after Eve's intervention, when
Alice has sent $\ket{\pm x}$. But the result holds in general for
the average detection rate, if Alice prepares all four states with
equal probability.} and the bit value is correct. So \ba
R_{n}&=&
\tilde{\nu}_Sp_A(n)\,f_n\Big(\demi+\tilde{\varepsilon}_n\Big)\,.\ea We see
that the detection rate increases in the presence of errors, contrary to BB84 where the detection rate is determined only by $p_{sift}$. The error rate is \ba
\varepsilon_n&=&\frac{\tilde{\varepsilon}_n}{\demi+\tilde{\varepsilon}_n}\,:
\label{epssarg}\ea for a given perturbation $\tilde{\varepsilon}_n$ in the quantum channel, the error introduced in
SARG04 is roughly twice the error $\varepsilon_n
=\tilde{\varepsilon}_n$ which would be introduced in BB84.

The protocol can be analyzed following the same pattern as the one presented for BB84. Here we just review the main results:
\begin{itemize}

\item SARG04 was invented as a method to reduce the effect of PNS attacks, taking advantage of the fact that Eve cannot extract full information from the 2-photon pulses \cite{aci04a,sca04}. This initial intuition has been confirmed by all subsequent, more rigorous studies. In particular, it was proved that a fraction of fully secure secret key can be extracted from the 2-photon pulses \cite{tam06}, and that in implementations using weak coherent lasers and without decoy states, for small error rate SARG04 performs indeed better than BB84 and shows a scaling $\sim t^{3/2}$ as a function of the distance \cite{koa05b,bra05,kra07}.

\item In the literature one finds the claim that, when implemented with decoy states, SARG04 performs worse than BB84 \cite{fun06a,kra07}. This must be properly understood: decoy states are a method to gain additional knowledge on Eve's attack. If this method does not reveal any PNS attack (as it will be the case in most experiments, because losses appear random and therefore Eve is acting as a beam-splitter), indeed the BB84 rate is better than the one of SARG04. However, if one would find that Eve is actually performing a PNS attack, SARG04 would of course be more robust, consistently with what we wrote in the previous item.  

\item An interesting case arises if one considers implementations with single-photon sources. The first unconditional security bound yielded that SARG04 tolerates a smaller QBER than BB84 \cite{tam06}. But this bound was improved shortly later: the optimal $I_{E,1}$, which is not known analytically but can easily be computed numerically, goes to zero for $\varepsilon_1\approx 11.67\%$ \cite{kra07}. This improved value is slightly better than the corresponding value for BB84, $\varepsilon_1\approx 11.0\%$: it seems therefore that SARG04 would perform better than BB84 also in a single-photon implementation. The picture is however different if one relates the error rate to the parameters of the channel, typically the visibility of interference fringes: this parameter is related to the ones introduced here through $\tilde{\varepsilon_1}=\frac{1-V}{2}$. For BB84, $\tilde{\varepsilon_1}=\varepsilon_1$ and consequently the critical visibility is $V\approx 78\%$; while for SARG04, because of (\ref{epssarg}), the critical visibility is worse, namely $V\approx 87\%$.

\end{itemize}

\section{Continuous-variable protocols}
\label{seccv}

\subsection {Status of security proofs}
\label{statuscv}

In the case of \textit{Gaussian modulation}, security has been proved against \textit{collective attacks} \cite{nav06,gar06}. We shall present this bound below (\ref{sscvbounds}) and use it for the comparison with the other platforms (\ref{seccompare}). There is some hope that the same bound would hold also for the most general attack, as it is the case for discrete-variable systems: in particular, we note that the ``intuitive'' reason behind that equivalence (\ref{sssattacks}) would apply also to CV protocols. Unfortunately, the exponential de Finetti bound \cite{ren07} does not help because it explicitely depends on the dimension of the quantum signals. On this issue, see \textit{Note add in proof} at the end of this paper.

In the case of \textit{discrete modulation}, the security status is even less advanced. Technically, the difficulty lies in the fact that the raw key is made of discrete variables for Alice, while Bob has a string of real numbers.
A full analysis has been possible only in the case where the quantum channel does not add excess noise to the signal, so that the observed conditional variances still describe minimum uncertainty states. In this case, the eavesdropper's attack is always describable as a generalized beam-splitting attack, simulating the observed loss. The corresponding key rates depend on the classical communication protocols chosen (with or without post-selection of data, in reverse or direct reconciliation); the best known protocol involves a combination of post-selection and reverse reconciliation, especially when the error correction algorithms work away from the asymptotic Shannon efficiency \cite{heid06a}. In the presence of excess noise, the formula for the key rate is the object of ongoing research; it has at least been possible to derive entanglement witnesses \cite{rigas06a}. Entanglement verification has been performed and has shown that excess noise in typical installations does not wipe out the quantum correlation within the experimentally accessible domain \cite{lorenz06a}.

Finally, in all works on CV QKD with no exception, it has been assumed that Eve does not act on the local oscillator\footnote{This amounts at viewing the local oscillator as an \textit{authenticated} channel, building on the closeness to classical signals. In an alternative set-up, this problem can be circumvented by Bob measuring the phase of the local oscillator, followed by the recreation within Bob's detector of a local oscillator with the measured phase \cite{koa04}.} --- of course, she is allowed to have access to it in order to measure quadratures. Since the local oscillator travels through Eve's domain, this assumption opens a security loophole\footnote{For the setups as they have been implemented, all observed correlations are compatible with an intercept/resend attack involving both the signal and the local oscillator. Security against this specific attack can be easily recovered by simple modifications of the setups, for example the independent measurement of the intensity of the phase reference pulse and the signal pulse \cite{hae07}.}. Note that a similar situation burdened until very recently the security of Plug\&Play configurations, for which finally unconditional security could be proved (see \ref{sspp}); it is not clear however that the same approach will work here, since the strong pulses have very different roles in the two schemes. In any case, the open issue just discussed, together with the fact that the existing exponential de Finetti theorem does not apply to infinitely-dimensional systems, are the main reasons unconditional security proofs are not available yet for CV QKD.

As mentioned earlier (\ref{ssscv}), continuous variable protocols show interesting features also on the classical part. In contrast to typical discrete variable protocols, where losses simply reduce the number of detected signals, continuous variable protocols will always detect a result, so that loss corresponds now to increased noise in the signal. Two main methods have been formulated to deal with this situation at the protocol level: reverse reconciliation \cite{GG02} and post-selection \cite{silberhorn}. The first method can be realized using one-way EC schemes, but turns out to be sensitive to the efficiency of those very schemes. Its main advantage is that 
its security can be rigorously assessed versus general collective attacks (and has been conjectured to hold even for coherent attacks)
In contrast, the second method can use both one-way and two-way EC schemes, and is fairly stable even if those schemes do not perform at the Shannon limit. However, its security can be analyzed only 
by making assumptions on Eve's interception (see below). The status
of its security is not clear even for general individual attacks.
Note that for close-to-perfect EC, reverse reconciliation outperforms post-selection. While progress is being made in the efficiency of EC schemes, it turns out that a combination of post-selection and reverse reconciliation provides a practical solution to obtain reasonable rates with current technology, both for discrete-modulation \cite{heid06a} and for Gaussian-modulation protocols \cite{heid07}.

\subsection{Bounds for Gaussian protocols}
\label{sscvbounds}

\subsubsection{Generalities}

As announced, we provide an explicit security bound for coherent-state homodyne-detection protocol of \cite{GG02}. Like all Gaussian protocols, this prepare-and-measure protocol can be shown to be equivalent to an entanglement-based scheme \cite{GC03}. In such a scheme, Alice prepares an EPR state --- more precisely, the two-mode squeezed vacuum state (\ref{squeezed_vac}). By applying an heterodyne measurement on mode $A$, she prepares in the second mode of the EPR pair a coherent state, whose displacement vector is Gaussian distributed in $x$ and $p$. Then, Bob applies a homodyne measurement on mode $B$, measuring quadrature $x$ or $p$. It can be shown that \textit{reverse reconciliation} is always favorable for Alice and Bob, so we have to compute Eq.~(\ref{genbound}) with $I_{EB}$ on the right hand side.

It has been proved that Eve's optimal attack is \textit{Gaussian} for both individual \cite{GC04,RaulPhD,ng} and collective attacks \cite{nav06,gar06}. We can therefore assume that Eve effects 
a Gaussian channel, so that the quantum state $\rho_{AB}$ just before Alice and Bob's measurements 
can be assumed to be a Gaussian two-mode state with zero mean value and covariance matrix $\gamma_{AB}$.

The Gaussian channel is characterized by two parameters: the transmittance, which here, since we work in the uncalibrated-device scenario, is $t\eta$ with $\eta$ the efficiency of the detectors; and the noise $\delta$ referred to the input of the channel\footnote{The observed noise in channels such as optical fibers is typically symmetric and uncorrelated in both quadratures $x$ and $p$ (there is no preferred phase), so we restrict to this case here.}. Since the two-mode squeezed state (\ref{squeezed_vac}) is also symmetric and has no correlations between $x$ and $p$, the resulting covariance matrix of modes A and B can be written in a block-diagonal form,
\begin{equation}
\gamma_{AB}= 
\left(
\begin{array}{cc}
\gamma_{AB}^{x} & 0 \\
0& \gamma_{AB}^p
\end{array}
\right)	
\label{eq:EBCovMat}	
\end{equation}
with 
\begin{equation}
\gamma_{AB}^{x(p)}=
\left(
\begin{array}{cc}
v & \pm\sqrt{t\eta(v^2-1)} \\
\pm\sqrt{t\eta(v^2-1)} & t\eta(v+\delta)
\end{array}
\right)
\end{equation}
where the signs $+$ and $-$ correspond to $\gamma_{AB}^{x}$ and $\gamma_{AB}^{p}$, respectively. 
Here, $v$ is the variance of both quadratures of Alice's output thermal state expressed in shot-noise units,
that is, $v=v_A+1$, $v_A$ being the variance of Alice's Gaussian modulation.

For what follows, it is convenient to define $v_{X|Y}$, the conditional variance that quantifies
the remaining uncertainty on $X$ after the measurement of $Y$:
\begin{equation}
v_{X|Y}=\langle x^2\rangle-\langle xy\rangle^2/\langle y^2\rangle\,,
\label{eq:CondVar}
\end{equation}
expressed in shot-noise units.

\subsubsection{Modeling the noise}

The noise $\delta$ is the total noise of the channel Alice-Bob. 
It can be modeled as the sum of three terms:
\ba
\delta&=&\frac {1-t}{t}\,+\,\frac{\delta_h}{t}\,+\,\epsilon\,.
\label{defchi}
\ea
The first term $(1-t)/t$ stands for the loss-induced vacuum noise (referred to the input); this term
is at the origin of the higher sensitivity to losses of continuous-variable QKD. The second term stands for the noise added by the imperfection of the homodyne detection. This is modeled by assuming that the signal reaching Bob's station is attenuated 
by a factor $\eta$ (detection efficiency) and mixed 
with some thermal noise $v_{el}$ (electronic noise of the detector),
giving\footnote{Replacing the expression for $\delta_h$ into (\ref{defchi}), one obtains $\delta=(1-t\eta+v_{el})/(t\eta)+\varepsilon$, which depends only on $t\eta$ as it should in the uncalibrated-device scenario.}
\ba\delta_{h}=\frac{1+v_{el}}{\eta}-1\,.
\ea
The third term $\epsilon$ is the excess noise (referred to the input) that is not due to line losses nor detector imperfections. For a perfect detector, it can be viewed as the continuous-variable counterpart of the QBER in discrete-variable QKD; it is zero for a lossy but noiseless line.

\subsubsection{Information Alice-Bob}
In the EB version of the coherent-state
protocol considered here \cite{GG02}, Alice performs heterodyne detection, 
so her uncertainty on Bob's quadratures is expressed as 
\begin{equation}
v_{B|A_M}=t\eta(\delta+1)\,.
\label{V_BAM}
\end{equation}
The mutual information between Alice
and Bob is therefore given by \ba
I(A:B) &=&\frac{1}{2} \log_2 \left[ \frac{v_B}{v_{B|A_M}}  \right]\,=\, \frac{1}{2} \log_2 \left[ \frac{\delta+v}{\delta+1}  \right]\,.\label{iabcv}
\ea As mentioned above, the main bottleneck of continuous-variable QKD schemes comes from
the heavy post-processing that is needed in order to correct the errors due to the vacuum noise
that is induced by the line losses. In practice, the amount of information left after error correction will be a fraction $\beta$ of $I(A:B)$. This value has an important effect on the achievable secret key rate and the limiting distance (as we shall discuss below, for $\beta=1$ a secure key can in principle be extracted for arbitrarily large distances). This provides a strong incentive
for developing better reconciliation algorithms. The first technique that was proposed to perform
continuous-variable error correction relied
on a so-called ``sliced reconciliation"
method \cite{gilles}, and gave an efficiency $\beta\approx 80\%$. These algorithms have been improved by using turbo-codes \cite{kim} and low-density 
parity codes (LDPC) \cite{mathieu}, which both allow to work with
noisy data, hence longer distances. More recently, multi-dimensional
reconciliation algorithms have been introduced, which allow to deal with even noisier data while keeping similar or higher reconciliation efficiencies \cite{lev07}.

\subsubsection {Individual attacks}

To become familiar with the security analysis, we first present individual attacks. In order to address the security of the protocol, we assume as usual that Eve 
holds the purification of $\rho_{AB}$. Then, by measuring their systems, 
Alice and Eve project Bob's share of the joint pure state $|\Psi_{ABE}\rangle$ 
onto another pure state (we may assume without loss of generality that 
Eve's projection results from a rank-one POVM). Applying the Heisenberg
uncertainty relation on the pure state held by Bob conditionally
on Alice and Eve's measurements, we have
\begin{equation}
v_{X_B|E}\,v_{P_B|A}\geq 1, \qquad v_{P_B|E}\,v_{X_B|A}\geq 1,
\label{eq:HeisenbergRR}
\end{equation}
where $X_B$ and $P_B$ are the canonically conjugate quadratures 
of Bob's mode. 
Equation (\ref{eq:HeisenbergRR}) can be written as a single uncertainty relation
\begin{equation}
v_{B|E}\,v_{B|A}\geq 1
\label{eq:HeisenbergRRf}
\end{equation} 
where $B$ stands for any quadrature of Bob's mode. This inequality
can be used to put a lower bound on the uncertainty of Eve's estimate 
of the key in reverse reconciliation, that is, 
when the key is made out of Bob's data while Alice and Eve compete 
to estimate it.

Now, $v_{B|A}$ is not necessarily given by (\ref{V_BAM}): Eve's attack cannot depend on how the mixed state sent by Alice (i.e., the thermal state) has been prepared, since all possible ensembles are indistinguishable. An acceptable possibility is Alice performing homodyne measurement, or, equivalently, preparing 
squeezed states just as in the protocol of \cite{Cerf01}; in which case we obtain
\begin{equation}
v_{B|A}=t\eta(\delta+1/v)\,.
\label{V_BA}
\end{equation} It can be shown that this is the lowest possible value of $v_{B|A}$, hence from (\ref{eq:HeisenbergRRf})
\begin{equation}
v_{B|E} \ge \frac{1}{t\eta(\delta+1/v)}\,.
\label{V_BE}
\end{equation} This gives a bound for $I(B:E)$, so the extractable secret key rate under the assumption of individual attacks becomes\ba 
r&=&I(A:B)-I(E:B)\,=\,\frac{1}{2} \log_2 \left[ \frac{v_{B|E}}{v_{B|A_M}}  \right] \nonumber  \\
&\ge& \frac{1}{2} \log_2 \left[ \frac{1}{(t\eta)^2 (\delta+1/v)(\delta+1)}  \right]
\ea
as shown in \cite{Nature03}. Note that the scheme that implements the optimal
attack (saturating this bound) is the entanglement cloner defined in \cite{entcloner}. Using Eq.~(\ref{defchi}), it appears that in the case of high losses ($t\eta\to 0$) and large modulation ($v\to \infty$), the secret key rate $r$
remains non-zero provided that the excess noise satisfies $\epsilon<1/2$. This is a remarkable result, due to reverse reconciliation: for direct reconciliation, obviously there can be no security when Eve has as much light as Bob, i.e. for $t\eta\leq \demi$.

A similar reasoning can be followed to derive the security of all Gaussian QKD protocols
against individual attacks \cite{RaulPhD}. The only special case concerns
the coherent-state heterodyne-detection protocol, whose security study 
against individual attacks is more involved \cite{hetero1,hetero2}.

\subsubsection {Collective attacks}

The security of the coherent-state homodyne-detection scheme against the class of collective attacks has been fully studied. The corresponding rates were first provided assuming
that Eve's collective attack is Gaussian \cite{coll1,coll2}. Later on, it was proved that this choice is actually optimal \cite{nav06,gar06}. This implies that
it remains sufficient to assess the security against Gaussian collective attacks,
which are completely characterized by the covariance matrix $\gamma_{AB}$ 
estimated by Alice and Bob. A long but straightforward calculation shows that
\ba
\chi(B:E)&=&
g(\tilde{\lambda}_1)+g(\tilde{\lambda}_2)-g(\tilde{\lambda}_3)
\label{chibecv}
\ea
where $g(x)=(x+1)\log_2 (x+1)-x\log_2 x$ is the entropy of a thermal state with 
a mean photon number of $x$ and $\tilde{\lambda}_k=\frac{\lambda_k -1}{2}$ where
\ba
\lambda_{1,2}^2\,=\, \frac{1}{2}(A\pm\sqrt{A^2-4B}) &,&
\lambda_{3}^2\,=\,v \frac{1+v\delta}{v+\delta}
\ea
with $A=v^2(1-2t\eta)+2t\eta+[t\eta(v+\delta)]^2$ and $B=[t\eta(v \delta+1)]^2$.

In conclusion, the secret key rate achievable against collective attacks is obtained by inserting expressions (\ref{iabcv}) and (\ref{chibecv}) into
\ba
K&=&R\,\left[\beta\,I(A:B)\,-\,\chi(B:E)\right]\label{ratecv}\,.
\ea
Finally, we note that the optimality of Gaussian attacks is actually valid also for protocols that use heterodyne detection; a bound for security against Gaussian collective attacks in these protocols has been provided recently \cite{pir08}.

\subsubsection{Collective attacks and post-selection}
In the case where all observed data are Gaussian, including the observed noise, we can again provide a security proof which also allows to include post-selection of data in the procedure. The starting point of this security proof is the protocol with Gaussian distribution of the amplitude together with the heterodyne detection by Bob. In this case, in a collective attack scenario, we can assume a product structure of the subsequent signals, and the density matrix $\rho_{AB}$ of the joint state of Alice and Bob is completely determined due to the tomographic structure of the source replacement picture and the measurement. In this scenario, we can therefore determine the quantum states in the hand of the eavesdropper as Eve holds the system E of the purification $\ket{\Psi}_{ABE}$ of $\rho_{AB}$. 

Let us consider the situation where all observed data in this scenario are Gaussian distributions, which is the typical observation made in experiments. Note that this is an assumption that can be verified in each run of the QKD protocol! In principle, one can now just use the standard formula for the key rate in the collective scenario, Eq.~(\ref{ratecv}). However, we would like to introduce a post-selection procedure \cite{silberhorn} to improve the stability of the protocol against imperfections in the error correction protocol.

To facilitate the introduction of post-selection, we add further public announcements to the CV QKD protocol: Alice makes an announcement 'a' consistent of  the imaginary component $\alpha_y$ and the modulus of the real component $|\alpha_x|$ of the complex amplitude $\alpha$ of her signals. That leaves two possible signals state open. Similarly, Bob makes an announcement 'b' which contains again the complex component $\beta{y}$ and the modulus $|\beta_x|$ of the complex measurement result $\beta$ of her heterodyne measurement. That leaves, again, two possible measurements from Eve's point of view. For any announcement combination $(a,b)$ we have therefore an effective binary channel between Alice and Bob. As the purification of the total state $\rho_{AB}$ is known, we can calculate for each effective binary channel a key rate
\begin{equation}
\Delta I (a,b) = \max\left\{(1 - f(e^{a,b})h[e^{a,b}]- \chi^{a,b}),0\right\}\, .
\end{equation}
This expression contains the post-selection idea in the way that whenever $1-h[e^{a,b}]-\chi^{a,b}$ is negative, the data are discarded, leading to a zero contribution of the corresponding effective binary channel to the overall key rate. 
The expressions for $\chi^{a,b}$ have been calculated analytically in \cite{heid07}, which is possible since now the conditional states of Eve, as calculated from the purification of $\rho_{AB}$, are now at most of rank four. Several scenarios have been considered there, but the one that is of highest interest is the combination of post-selection with reverse reconciliation.  The explicit expressions are omitted here, as they do not give additional insight. The evaluations of the overall key rate 
\begin{equation}
K = R \int \mathrm{d} a \; \mathrm{d} b \; \Delta I (a,b)
\end{equation}
is then done numerically.

\section{Distributed-phase-reference protocols}
\label{secmixed}

\subsection{Status of security proofs}
\label{statusdpr}

As we said in Sec. \ref{sssmixed}, distributed-phase-reference protocols were invented by experimentalists, looking for practical solutions. Only later it was noticed that these protocols, in addition to be practical, may even yield better rates than the traditional discrete-variable protocols, i.e. rates comparable to those of decoy-states implementations. The reason is that the PNS attacks are no longer zero-error attacks both for DPS \cite{ino05} and for COW \cite{gis04,stu05}. In fact, the number of photons in a given pulse and the phase coherence between pulses are incompatible physical quantities. At the moment of writing, no lower bound is known for the unconditional security of DPS or COW, but several restricted attacks have been studied \cite{wak06,bra07,bra07b,cur07,tsu07,cur08,gom08}. In these studies, it has also been noticed that DPS and especially COW can be modified in a way that does not make them more complicated, but may make them more robust \cite{bra07b}. Since this point has not been fully developed though, we restrict our attention to the original version of these protocols.

\subsection{Bounds for DPS and COW}
\label{ssmixedatt}

\subsubsection{Collective beam-splitting attack}

We present the calculation of the simplest zero-error collective attack, namely the beam-splitting attack \cite{bra07b}. For both DPS and COW, Alice prepares a sequence of coherent states $\bigotimes_k\ket{\alpha(k)}$: each $\alpha(k)$ is chosen in $\{+\alpha,-\alpha\}$ for DPS, in $\{+\alpha,0\}$ for COW. Eve simulates the losses with a beam-splitter, keeps a fraction of the signal and sends the remaining fraction $\tau=t\,t_B\eta$ to Bob on a lossless line --- note that, although this security study does not provide a lower bound, we work in the uncalibrated-device scenario for the sake of comparison with the other protocols. Bob receives the state $\bigotimes_k\ket{\alpha(k)\sqrt{\tau}}$: in particular, Bob's optical mode is not modified, i.e. BSA introduces no error\footnote{Apart from BSA, other attacks exist that do not introduce errors: for instance, photon-number-splitting attacks over the whole key, preserving the coherence (these are hard to parametrize and have never been studied in detail). For COW, there exist also attacks based on unambiguous state discrimination \cite{bra07}.}. Eve's state is $\bigotimes_k\ket{\alpha(k)\sqrt{1-\tau}}$; let us introduce the notations $\alpha_E=\alpha\sqrt{1-\tau}$ and \ba \gamma&=&e^{-|\alpha_E|^2}\,=\,e^{-\mu(1-\tau)}\,.\ea When Bob announces a detection involving pulses $k-1$ and $k$, Eve tries to learn the value of his bit by looking at her systems. Assuming that each bit value is equally probable, Eve's information is given by  $I_{Eve}=S(\rho_E)-\demi S(\rho_{E|0})-\demi S(\rho_{E|1})$ with $\rho_E=\demi \rho_{E|0}+\demi\rho_{E|1}$. 

The information available to Eve differs for the two protocols, because of the different coding of the bits.
In DPS, the bit is $0$ when $\alpha(k-1)=\alpha(k)$ and is $1$ when $\alpha(k-1)=-\alpha(k)$. So, writing $P_{\psi}$ the projector on $\ket{\psi}$, the state of two consecutive pulses reads $\rho_{E|0}=\demi P_{+\alpha_E,+\alpha_E}+\demi P_{-\alpha_E,-\alpha_E}$ and $\rho_{E|1}=\demi P_{+\alpha_E,-\alpha_E}+\demi P_{-\alpha_E,+\alpha_E}$; therefore, noticing that $|\braket{+\alpha_E}{-\alpha_E}|=\gamma^2$, we obtain
\ba
I_{E,BS}^{DPS}(\mu)&=&2h[(1-\gamma^2)/2]-h[(1-\gamma^4)/2]
\ea
where $h$ is the binary entropy function, and
\ba
K(\mu)&=&\nu_S\,\left(1-e^{-\mu t \,t_B\eta}\right)\,\left[1-I_{E,BS}^{DPS}(\mu)\right]\,.\label{rate_DPS_BS}
\ea
In COW, the bit is $0$ when $\alpha(k-1)=\sqrt{\mu}\,,\alpha(k)=0$ and is $1$ when $\alpha(k-1)=0\,,\alpha(k)=\sqrt{\mu}$; so, with similar notations as above,
$\rho_{E|0}=P_{+\alpha_E,0}$ and $\rho_{E|1}=P_{0,+\alpha_E}$; therefore, noticing that $|\braket{+\alpha_E}{0}|=\gamma$, we obtain
\ba
I_{E,BS}^{COW}(\mu)&=&h[(1-\gamma)/2]\,.
\ea
The secret key rate is given by
\ba
K(\mu)&=&\tilde{\nu}_S\,\left(1-e^{-\mu t \,t_B\eta}\right)\,\left[1-I_{E,BS}^{COW}(\mu)\right] \label{rate_COW_BS}
\ea where $\tilde{\nu}_S=\nu_S\,\frac{1-f}{2}$ because the fraction $f$ of decoy sequences does not contribute to the raw key, and half of the remaining pulses are empty.

\subsubsection{More sophisticated attacks}
\label{sssattackscow}

For the purpose of comparison with other protocols later in this review, it is useful to move away from the strictly zero-error attacks. As mentioned above, several examples of more sophisticated attacks have indeed been found. Instead of looking for the exact optimum among those attacks, we prefer to keep the discussion simple, bearing in mind that all available bounds are to be replaced one day by unconditional security proofs.

We consider attacks in which Eve interacts coherently with \textit{pairs of pulses} \cite{bra07b}. Upper bounds have been provided in the limit $\mu t\ll 1$ of \textit{not-too-short distances}. Even within this family, a simple formula is available only for COW. For COW, there is no \textit{a priori} relation between the error on the key $\varepsilon$ and the visibility $V$ observed on the interferometer. If $e^{-\mu}\leq \xi\equiv 2\sqrt{V(1-V)}$, one finds $I_E^{COW}(\mu)=1$: $\mu$ is too large and no security is possible. If on the contrary $e^{-\mu}>\xi$, the best attack in the family yields
\ba
I_E^{COW}(\mu)&=&\varepsilon+(1-\varepsilon)h\left(\frac{1+F_V(\mu)}{2}\right)
\ea
with $F_V(\mu)=(2V-1)e^{-\mu}-\xi\sqrt{1-e^{-2\mu}}$. Therefore
\ba
K(\mu)&=&R\,\left[1-I_E^{COW}(\mu)-\mbox{leak}_{EC}(Q)\right] \label{rate_COW}
\ea
where the value of $R$ is constrained by the definition of the attack to be $\tilde{\nu}_S[\mu t \,t_B\eta+2p_d]$.

As for DPS, numerical estimates show that its robustness under the same family of attacks is very similar (slightly better) than the one of COW. Therefore, we shall use (\ref{rate_COW}) as an estimate of the performances of distributed-phase-reference protocols in the presence of errors; again, for the sake of comparison with the other protocols, we have adopted the uncalibrated-device scenario here\footnote{For the family of attacks under study, the rate scales linearly with the losses, therefore the difference between calibrated and uncalibrated devices is only due to the dark counts. We have to warn that the attacks based on unambiguous state discrimination, which have been studied explicitly for calibrated devices \cite{bra07}, are expected to become significantly more critical in the uncalibrated-device scenario. However, this more complex family of attacks can be further restricted by a careful statistical analysis of the data: we can therefore leave it out of our analysis, which is anyway very partial.}.

\section{Comparison of experimental platforms}
\label{seccompare}

\subsection{Generalities}
\label{ssgen}

After having presented the various forms that practical QKD can take, it is legitimate to try and draw some comparison. If one would dispose of unlimited financial means and manpower, then obviously the best platform would
just be the one that maximizes the secret key rate $K$ for the desired distance. A choice in the real world will obviously put other parameters in the balance, like simplicity, stability, cost... Some partial comparisons are available in the literature; but, to our knowledge, this is the first systematic attempt of comparing all the most meaningful platforms of practical QKD. Of course, any attempt of putting all platforms on equal footing contains elements of arbitrariness, which we shall discuss. Also, we are bounded by the state-of-the-art, both concerning the performance of the devices and the development of the security proofs, as largely discussed in the previous sections. We have chosen to compare the best available bounds, which however \textit{do not correspond to the same degree of security}: for the implementations of the BB84 coding, we have bounds for unconditional security; for continuous variable systems, we have security against collective attacks; for the new protocols like COW and DPS, we have security only against specific families of attacks. Also, one must be reminded that all security proofs hold under some assumptions: these have been discussed in Sections \ref{secdiscr}, \ref{seccv} and \ref{secmixed}; it is crucial to check if they apply correctly to any given implementation.

As stressed many times, the security of a given QKD realization must be assessed using \textit{measured} values. Here, we have to present some \textit{a priori} estimates: they necessarily involve choices, which have some degree of arbitrariness. The first step is to provide a \textit{model for the channel}: the one that we give (\ref{ssschannel}) corresponds well to what is observed in all experiments and is therefore rather universally accepted as an \textit{a priori} model. At the risk of being redundant, we stress that the actual realization of this specific channel is \textit{not} a condition for security: Eve might realize a completely different channel, and the general formulas for security apply to any case\footnote{The attacks we studied against DPS and COW, Section \ref{secmixed}, \textit{do} suppose a model of the channel. This is a signature of the incompleteness of such studies. Security can be guaranteed by adding that, if the channel deviates from the expected one, the protocol is aborted. A full assessment of the channel, of course, requires additional tests: the fact that data can be reproduced by a channel model does not imply that the channel model is correct (for instance, in weak coherent pulses implementations of BB84 without decoy states, the observed parameters are compatible both with a BS and a PNS attack).}. Once the model of the channel accepted, one still has to choose \textit{the numerical values for all the parameters}.

\subsubsection{Model for the source and channel}
\label{ssschannel}

We assume that the \textit{detection rates} are those that are
expected in the absence of Eve, given the source and the distance
between Alice and Bob. As for the \textit{error rates}, we
consider a depolarizing channel with visibility $V$. For an \textit{a priori} choice, the modeling of the channel just sketched is rather universally accepted. In detail, it gives
the following:

\textit{Discrete-variable protocols, P\&M.} We consider implementations
of the BB84 coding. The rate is estimated by
$R=\tilde{\nu}_S[{\cal{P}}+{\cal{P}}_d]$ with
${\cal{P}}=\sum_{n\geq 1}p_A(n)[1-(1-t\,t_B\eta)^n]$ and ${\cal{P}}_d=2p_d\sum_{n\geq 0}p_A(n)(1-t\,t_B\eta)^n$. The
error rate in the channel is $\varepsilon=(1-V)/2$, so the expected error rate is $Q=[\varepsilon{\cal{P}}+{\cal{P}}_d/2]/(R/\tilde{\nu}_S)$. For weak coherent
pulses without decoy states, $p_A(1)=e^{-\mu}\mu$, $p_A(n\geq
2)=1-e^{-\mu}(1+\mu)$, and we optimize $K$, given by (\ref{kbb84}),
over $\mu$. For weak coherent pulses with decoy states, we
consider an implementation in which one value of $\mu$ is used
almost always, while sufficiently many others are used, so that
all the parameters are exactly evaluated. The statistics of the source are as above; $Y_0$ is estimated by
$\tilde{\nu}_S\,2p_dp_A(0)/R$, $Y_1$ by
$\tilde{\nu}_Sp_A(1)t\,t_B\eta/R$, and we optimize $K$ given by
(\ref{kbb842}) over $\mu$. For perfect single-photon sources,
$p_A(1)=1$ and $p_A(n\geq 2)=0$; we just compute (\ref{kbb84}), as
there is nothing to optimize.

\textit{Discrete-variable protocols, EB.} Again, we consider implementations
of the BB84 coding. Since most of the experiments have been performed using cw-pumped sources, we shall restrict to this case\footnote{Pulsed sources can be treated in a similar way. For short pulse schemes, one would have $p_A(1) \approx \mu$ and $p_A(2)
\approx \frac{3}{4} \mu^2$ if $\mu \ll 1$; for long-pulse pumping, the statistics of pairs is
approximately Poissonian: $p_A(1) \approx \mu$ and $p_A(2) \approx
\mu^2/2$ if $\mu \ll 1$ and the most of the multi-pair events are
uncorrelated. In both cases, the intrinsic error rate due to double-pair events
is $\varepsilon' \approx \mu/2$ \cite{eis04,sca05}. Note that the parameter $\zeta$ may be different from 0 in the case of short pulse schemes.}. For such sources, the probability of having multiple pairs is $\zeta=0$ with good precision, therefore the bounds (\ref{kbb843}) and (\ref{eb-ma}) for $K$ are identical. $K$ will be optimized over $\mu'$, the mean pair-generation
rate of the source. Note that $\nu_S^{\mathrm{cw}}$ given by Eq.~(\ref{nuscw}) depends on $\mu'$; given this, one has
$p_A(1) \approx 1$ and $p_A(2) \approx \mu'\,\Delta t$ if
$\mu' \Delta t \ll 1$: indeed, neglecting dark counts, whenever any of Alice's
detectors fires there is at least one photon
going to Bob; and the probability that another pair appears during the
coincidence window $\Delta t$ is approximately $\mu' \Delta t$. The total expected error is $Q=[(\varepsilon+\varepsilon'){\cal{P}}+{\cal{P}}_d/2]/(R/\tilde{\nu}_S)$, where $\varepsilon=(1-V)/2$ as above and $\varepsilon' \approx \frac{\mu' \Delta
t}{2}$ is the error rate due to double-pair events.

\textit{Continuous-variable protocols.} We consider the protocol that uses coherent states with Gaussian modulation, and compute the best available bound (\ref{ratecv}), which give security against collective attacks. The reference beam is supposed to be so intense, that there is always a signal arriving at the homodyne detection, so $R=\tilde{\nu}_S$. The error is modeled by (\ref{defchi}). Now, just as for discrete variable protocols one can optimize $K$ over the mean number of photons (or of pairs) $\mu$ for each distance, here one can optimize $K$ over the variance $v$ of the modulation. Note that this optimization outputs rather demanding values, so that only recently it has become possible to implement them in practice, thanks to the latest developments in error correction codes \cite{lev07}.

\textit{Distributed-phase-reference protocols.} As mentioned, apart from the errorless case, a simple formula exists only for COW, which moreover is valid only at not too short distances. We use this bound to represent distributed-phase reference protocols in this comparison, keeping in mind that DPS performs slightly better, but that anyway only upper bounds are available. Specifically, we have $R\approx\tilde{\nu}_S[\mu t \,t_B\eta+2p_d]$; we optimize then $K(\mu)$ given by (\ref{rate_COW}) over $\mu$, and keep the value only if $\mu_{opt}t\leq 0.1$. The expected error rate is formally the same as for P\&M BB84; recall however that here the bit-error $\varepsilon$ is \textit{not} related to the visibility of the channel and must be chosen independently.

\subsubsection{Choice of the parameters}

\begin{table}[t]
\begin{center}
\begin{tabular}{|l|l|c|c|}\hline
Platform & Parameter & Set \#1 & Set \#2\\\hline
 & $\mu$ mean intensity & (opt.) & (opt.)\\
 & $V$ visibility: P\&M & 0.99 & 0.99\\
 & $V$ visibility: EB & 0.96 & 0.99\\
BB84, & $t_B$ transmission in Bob's device& 1 & 1\\
COW & $\eta$ det. efficiency& 0.1 & 0.2\\
 & $p_d$ dark counts & $10^{-5}$ & $10^{-6}$\\
 & $\varepsilon$ (COW) bit error & 0.03 & 0.01\\
 & $\zeta$ (EB) coherent 4 photons & 0 & 0\\
 & leak EC code & 1.2 & 1\\\hline
 & $v=v_A+1$ variance & (opt.) & (opt.)\\
 & $\varepsilon$ optical noise & 0.005 & 0.001\\
CV & $\eta$ det. efficiency& 0.6 & 0.85\\
 & $v_{el}$ electronic noise& 0.01 & 0\\
 & $\beta$ EC code& 0.9 & 0.9\\\hline
\end{tabular}
\caption{Parameters used for the \textit{a priori} plots in this Section. See main text for notations and comments. The caption (opt.) means that the parameter will be varied as a function of the distance in order to optimize $K$.}
\label{tab:params}
\end{center}
\end{table}

We shall use two sets of parameters (Table \ref{tab:params}): set \#1 corresponds to today's state-of-the-art, while set \#2 reflects a more optimistic but not unrealistic development. Moreover, we make the following choices:

\begin{itemize}

\item Unless specified otherwise (see \ref{ssstrustplots}), the plots use the formulas for the uncalibrated-device scenario. The reason for this choice is the same as discussed in Sec. \ref{ssstrusted}: unconditional security has been proved only in this over-pessimistic scenario.

\item Since we are using formulas that are valid only in the asymptotic regime of infinitely long keys, we remove the nuisance of sifting by allowing an asymmetric choice of bases or of quadratures. Specifically, this leads to $\tilde{\nu}_S=\nu_S$ for both BB84 and continuous-variables. Similarly, for COW we can set $f=0$, whence $\tilde{\nu}_S=\nu_S/2$.

\item For definiteness, we consider fiber-based implementations; in particular, the relation between distance and transmission will be (\ref{tfiber}) with $\alpha=0.2$dB/km; and the parameters for photon counters are given at telecom wavelengths (Table \ref{tab:params}). The reader must keep in mind that in free space implementations, where one can work with other frequencies, the rates and the achievable distance may be larger.

\end{itemize}

\subsection{Comparisons based on $K$}
\label{ssKplots}

\subsubsection{All platforms on a plot}
\label{sssnoiseplots}

\begin{figure}[ht]
\includegraphics[scale=0.55]{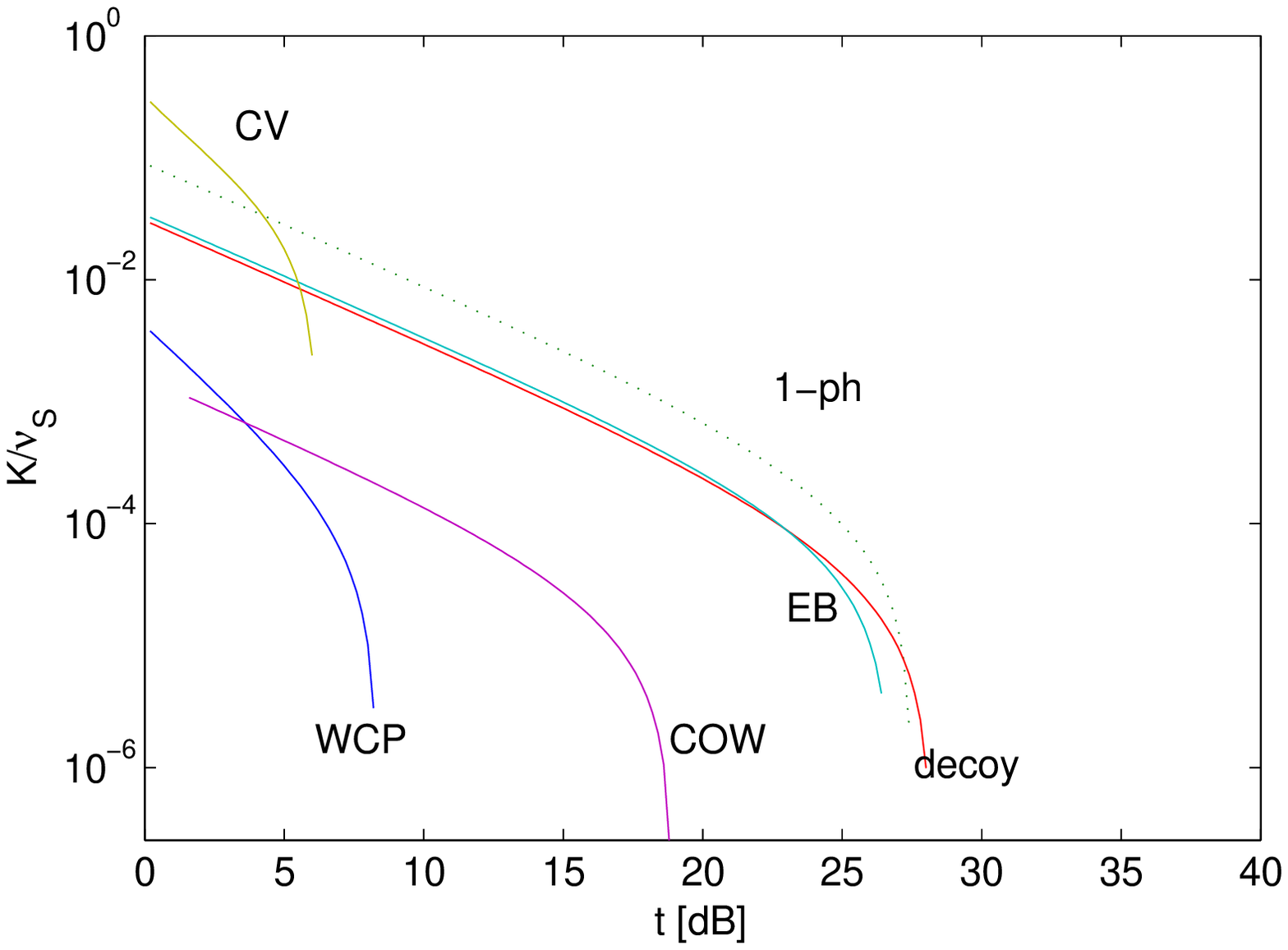}
\vspace{5mm}
\includegraphics[scale=0.55]{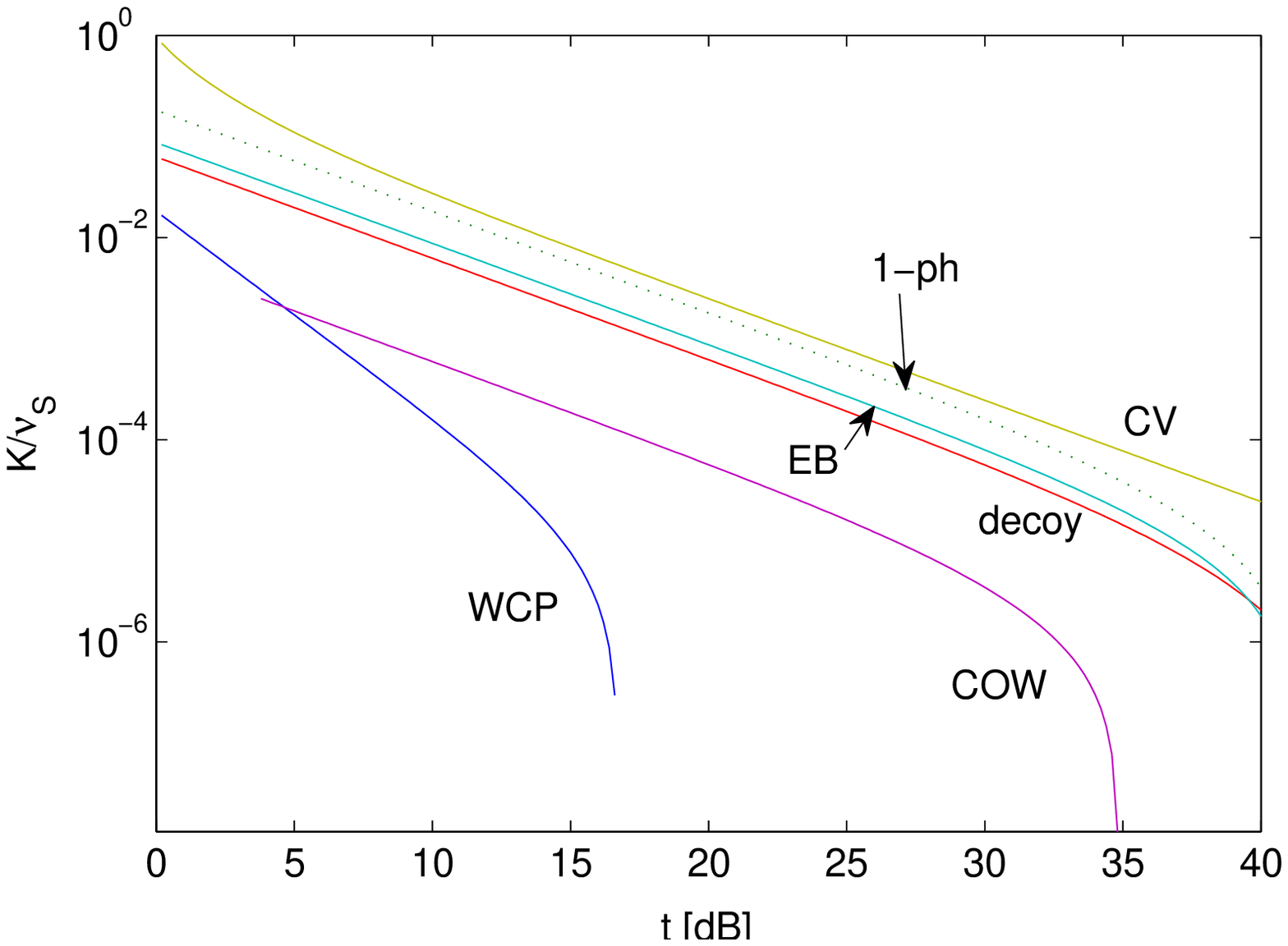}
\caption{(Color online). $K/\nu_S$ as a function of the transmittivity $t$, for all the platforms. \textit{Legend:} 1-ph: perfect single-photon source, unconditional; WCP: weak coherent pulses without decoy states, unconditional; decoy: weak coherent pulses with decoy states, unconditional; EB: entanglement-based, unconditional; CV: continuous-variables with Gaussian modulation, security against collective attacks; COW: Coherent-One-Way, security against the restricted family of attacks described in Sec. \ref{sssattackscow}. Parameters from Table \ref{tab:params}: set \#1 upper graph, set \#2 lower graph.}\label{fig:Ktall}
\end{figure}

As a first case study, we compare all the platforms on the basis by plotting $K/\nu_S$ as a function of the transmittivity $t$ of the channel. The result is shown in Fig.~\ref{fig:Ktall}. As promised, we have to stress the elements of arbitrariness in this comparison (in addition to the choices discussed above). First of all, we recall that the curves do not correspond to the same degree of security (see \ref{ssgen}). Second, we have considered ``steady-state'' key rates, because we have neglected the time needed for the classical post-processing; this supposes that the setup is stable enough to run in that regime (and it is fair to say that many of the existing platforms have not reached such a stage of stability yet). Third, the \textit{real} performance is of course $K$: in particular, if some implementations have bottlenecks at the level of $\nu_S$ (see \ref{ssraw}), the order of the curves may change significantly.

\subsubsection{Upper bound incorporating the calibration of the devices}
\label{ssstrustplots}

As a second case study, we show the difference between the lower bounds derived in the uncalibrated-device scenario, and some upper bounds that incorporate the calibration of the devices.

\begin{figure}[ht]
\includegraphics[scale=0.6]{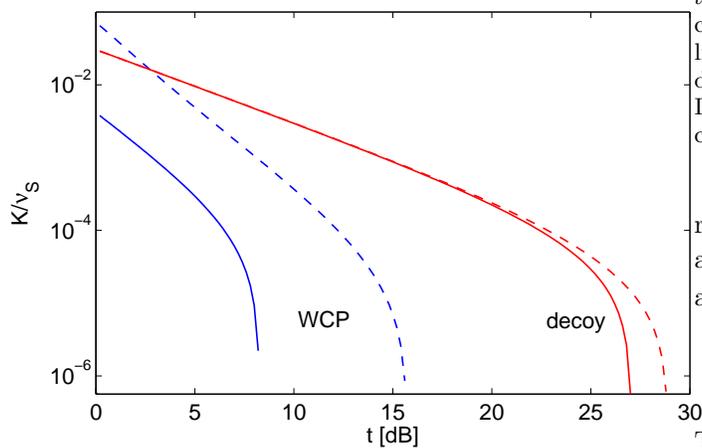}
\caption{(Color online). $K/\nu_S$ as a function of the transmission $t$ for the P\&M implementations of BB84 with weak coherent pulses: comparison between the lower bound (solid lines, same as in Fig.~\ref{fig:Ktall}, upper graph) and the upper bound for calibrated devices (dashed lines). Legend as in Fig.~\ref{fig:Ktall}. Parameters from Table \ref{tab:params}, set \#1.}\label{fig:trust}
\end{figure}

We focus first on \textit{BB84 implemented with weak coherent pulses}; the upper bounds under study have been derived in Sec. \ref{ssupptrust}. The plots in Fig.~\ref{fig:trust} show how much one can hope to improve the unconditional security bounds from their present status. As expected, the plot confirms that basically no improvement is expected for implementations with decoy states, because there only the treatment of dark counts is different; while the bound for implementations without decoy states may still be the object of significant improvement.

\begin{figure}[ht]
\includegraphics[scale=0.6]{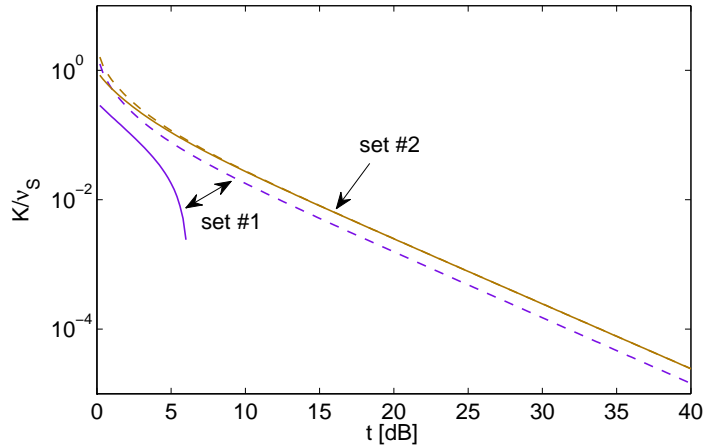}
\caption{(Color online). $K/\nu_S$ as a function of the transmission $t$ for CV QKD with Gaussian modulation, security against collective attacks, comparison between the lower bound (solid lines, same as in Fig.~\ref{fig:Ktall}) and the upper bound for calibrated devices (dashed lines) for both sets of parameters from Table \ref{tab:params}. Compared to Fig.~\ref{fig:Ktall}, the color of the lines of set \#1 was changed for clarity.}\label{fig:trustcv}
\end{figure}

We turn now to \textit{CV QKD with Gaussian modulation}. Bounds for the security against collective attacks assuming calibrated devices are given in Eqs (5)-(12) of \cite{lodewyck-2}. The plots are shown in Fig.~\ref{fig:trustcv}. One sees that the difference between the two scenarios is significant for set \#1 of parameters, but is negligible for the more optimistic set \#2. This is interesting, given that the efficiency $\eta$ of the detectors is ``only'' $85\%$ in set \#2.

\subsection{Comparison based on the ``cost of a linear network''}
\label{sscost}

\begin{figure}[ht]
\includegraphics[scale=0.55]{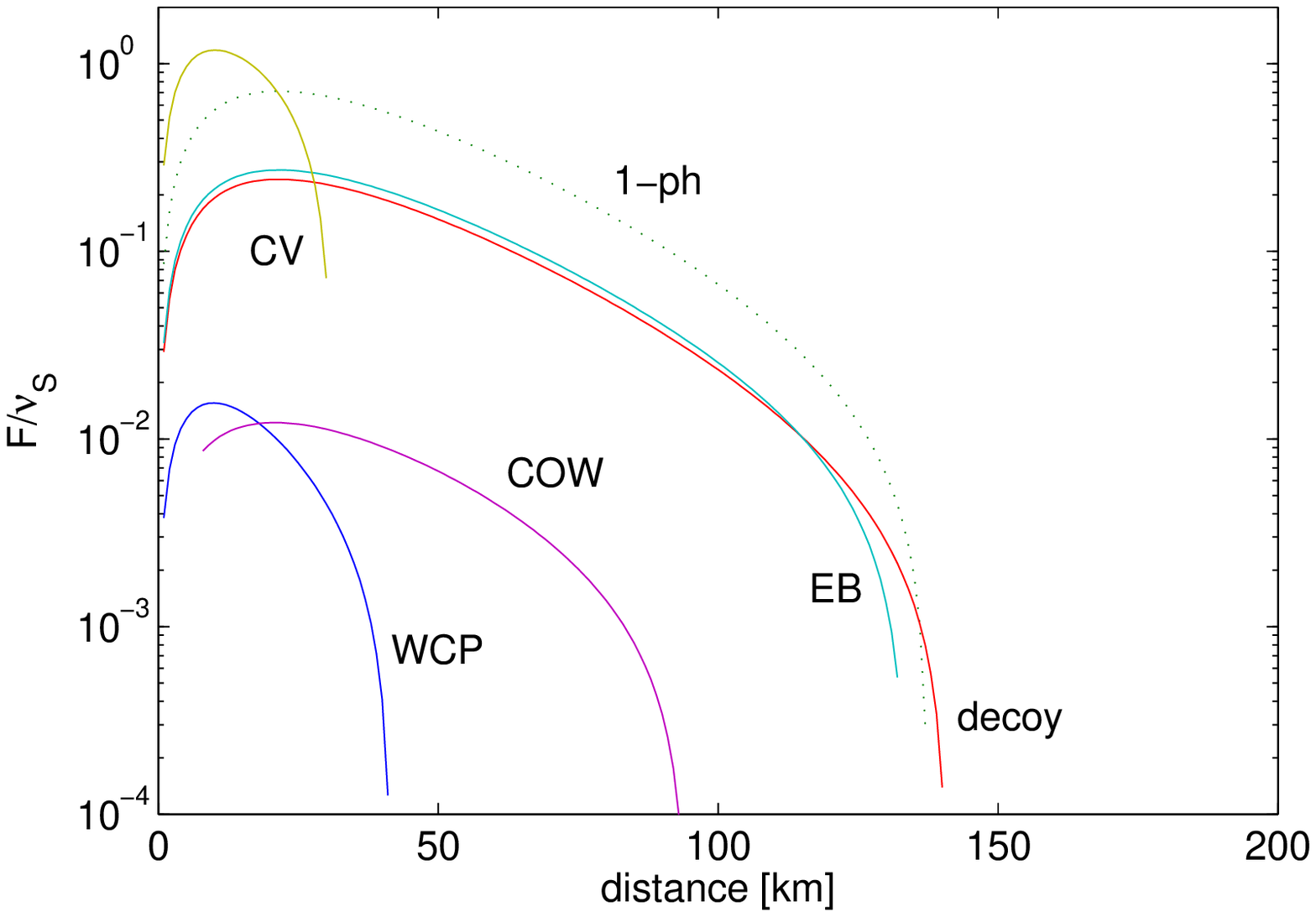}
\vspace{5mm}
\includegraphics[scale=0.55]{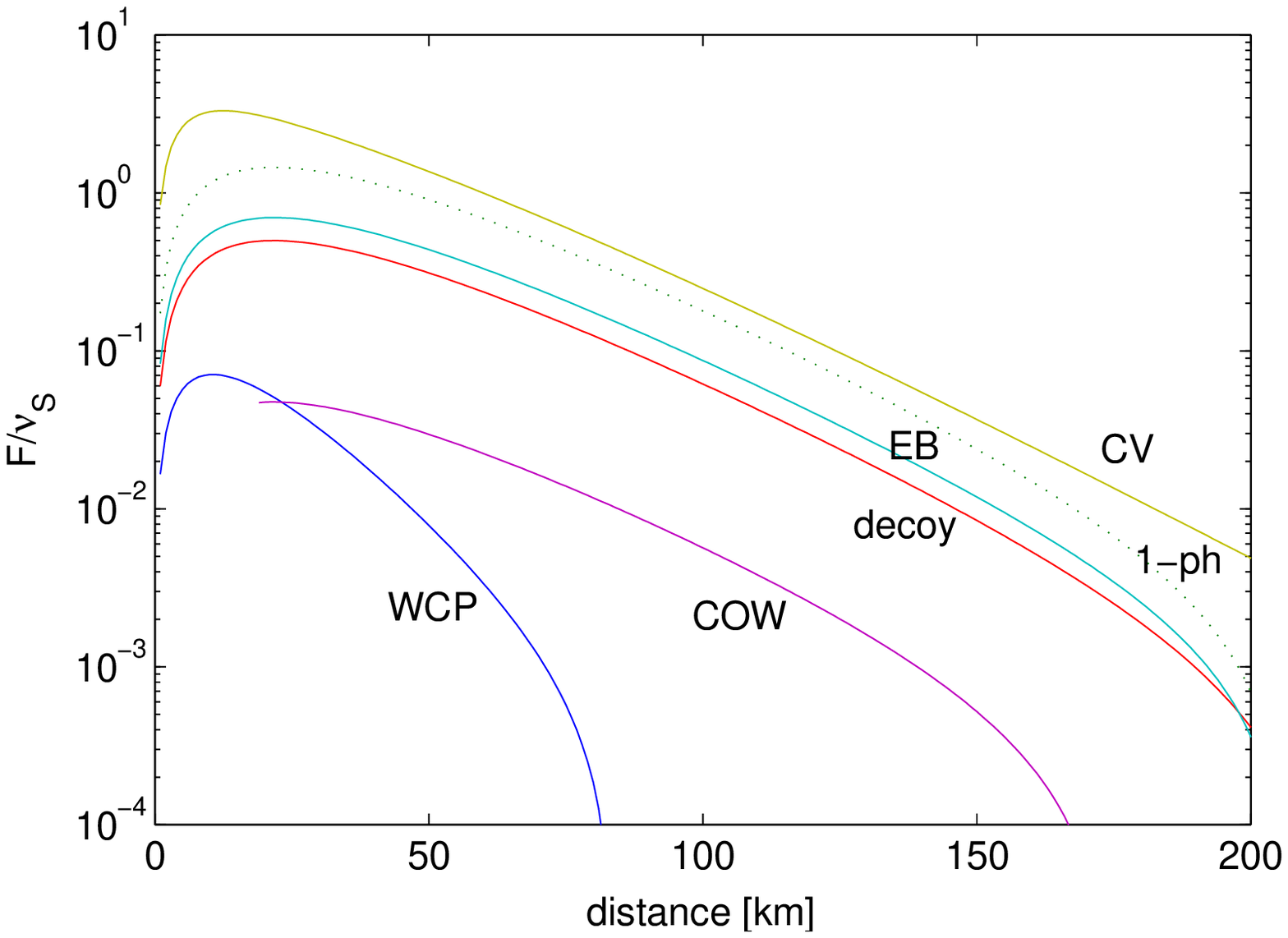}
\caption{(Color online). $F/\nu_S$ as a function of the distance $\ell$ for all the platforms. Legend as in Fig.~\ref{fig:Ktall}. Parameters from Table \ref{tab:params}: set \#1 upper graph, set \#2 lower graph.}\label{fig:Fdall}
\end{figure}

We consider a linear chain of QKD devices, aimed at achieving a secret key rate $K_{\rm target}$ over a distance $L$. Many devices can be put in parallel, and trusted repeater stations are built at the connecting points. Each individual QKD device is characterized by the point-to-point rate $K(\ell)$ it can achieve as a function of the distance $\ell$, and by its cost $C_1$. We need $N=\frac{L}{\ell} \frac{K_{\rm target}}{K(\ell)}$ devices to achieve the goal, so the \textit{cost of the network} is\footnote{In this first toy model, we neglect the cost of the trusted repeater stations; see \cite{all07b} for a more elaborated model.}
\ba
C_{\rm tot}[\ell]&=& C_1 \,\frac{L}{\ell} \frac{K_{\rm target}}{K(\ell)}\,.
\ea
The best platform is the one that minimizes this cost, i.e., the one that maximizes $F(\ell)=\ell K(\ell)$. This quantity, normalized to $\nu_S$, is plotted in Fig.~\ref{fig:Fdall} as a function of the distance for both sets of parameters defined in Table \ref{tab:params}. Of course, this comparison presents the same elements of arbitrariness as the previous one.

The optimal distances are quite short, and this can be understood from a simple analytical argument. Indeed, typical behaviors are $K(\ell)\propto t$ (single-photon sources, attenuated lasers with decoy states, strong reference pulses) and $K(\ell)\propto t^2$ (weak coherent pulses without decoy states). Using $t=10^{-\alpha\ell/10}$, it is easy to find $\ell_{\rm opt}$ which maximizes $F(\ell)$:
\ba
K(\ell)\propto t^k &\longrightarrow & \ell_{\rm opt}= 10/ (k\alpha \ln 10)\,.
\ea
In particular, for $\alpha \approx 0.2$dB/km, one has $\ell_{\rm opt} \approx 20$km for $k=1$ and $\ell_{\rm opt} \approx 10$km for $k=2$.

In conclusion, our toy model suggests that, in a network environment, one might not be interested in pushing the maximal distance of the devices; in particular, detector saturation (which we neglected in the plots above) may become the dominant problem instead of dark counts.

\section{Perspectives}
\label{secpersp}

\subsection{Perspectives within QKD}
\label{ssperspqkd}

\subsubsection{Finite-key analysis}
\label{sssfinite}

As stressed, all the security bounds presented in this review are valid only in the asymptotic limit of infinitely long keys. Proofs of security for \textit{finite-length keys} are obviously a crucial tool for practical QKD. The estimate of finite-key effects, unfortunately, has received very limited attention so far. The pioneering works \cite{may96,ina07}, as well as some subsequent ones \cite{wat04,hay06}, have used non-composable definitions of security (see \ref{ssssecu}). This is a problem because the security of a finite key is never perfect, so one needs to know how it composes with other tasks. Others studied a new formalism but failed to prove unconditional security \cite{mey06}. The most recent works comply with the requirements \cite{hay07a,sca07}; finite statistics have been incorporated in the analysis of an experiment \cite{has07}. Without going into details, all these works estimate that no key can be extracted if fewer than $N\approx 10^5$ signals are exchanged.

\subsubsection{Open issues in unconditional security}
\label{sssunctrust}

We have said above that, for CV QKD and distributed-phase reference protocols, no unconditional security proof is available yet. However, there is an important difference between these cases. In the existing CV QKD protocols, the information is coded in independent signals; as such, it is believed that unconditional security proofs can be built as generalizations of the existing ones (see also \textit{Note added in proof} below). On the contrary, the impossibility of identifying signals with qubits in distributed-phase reference protocols will require a completely different approach, which nobody has been able to devise at the moment of writing.

As explained in Sec. \ref{ssstrusted}, all unconditional security proofs have been derived under the over-conservative assumption of uncalibrated devices. Ideally, such an assumption should be removed: one should work out unconditional security proofs taking into account the knowledge about the detectors; this would lead to better rates. A possible solution consists in including the calibration of the devices in the protocol itself; the price to pay seems to be a complication of the setup \cite{qi06}. The idea is somehow similar to the one used in decoy states.
We also discussed how calibrated-device proofs may ultimately provide significant improvement only for some protocols (see \ref{ssstrustplots}). The difference between protocols can be understood from the fact that typically $K\sim t^\alpha$ where $t$ is the transmittance and $\alpha\geq 1$. When $\alpha=1$, then the only advantage of calibrating the devices can come from the dark count contribution. If on the contrary $\alpha>1$ (weak coherent pulses without decoy states: $\alpha=2$ for BB84, $\alpha=\frac{3}{2}$ for SARG04), then the difference is much larger, because it matters whether $\,t_B\eta$ is included in the losses or not. The urgency of this rather ungrateful\footnote{Here is an example of the complications that might appear. When taking the calibration into account, it is often assumed that the dark counts do not enter in Eve's information. Actually, things are more subtle. On the one hand, most of the dark counts will actually {\em decrease} Eve's information, because she does not know if a detection is due to the physical signal (on which she has gained some information) or is a completely random event. On the other hand, if a detection happens shortly after a previous one, Eve may guess that the second event is in fact a dark count triggered by an afterpulse, and therefore learn some correlations between the two results. Admittedly, these are fine-tuning corrections, and have never been fully discussed in the literature; but if one wants to prove unconditional security, also these marginal issues must be properly addressed.} task is therefore relative to the choice of a protocol.

\subsubsection{Black-box security proofs}
\label{sssdevindep}

The development of commercial QKD systems makes it natural to ask whether the ``quantumness'' of such devices can be proved in a black-box approach. Of course, the compulsory requirements (\ref{sssuncond}) must hold. For instance, the random number generator cannot be within the black box, because it must be trusted; one must also make sure that no output port is diffusing the keys on the internet; and so on. Remarkably though, all the quantum part can in principle be kept in a black-box. The idea is basically the one that triggered Ekert's discovery \cite{eke91}, although Ekert himself did not push it that far: the fact, that Alice and Bob observe correlations that violate a Bell inequality, is enough to guarantee entanglement, independent of the nature of the quantum signals and even of the measurements that are performed on them. This has been called ``device-independent security''; a quantitative bound was computed
for collective attacks on a modification of Ekert's protocol, the goal of proving unconditional security is still unattained \cite{aci07}. Device-independent security can be proved only for entanglement-based schemes: for this definition of security, the equivalence EB-P\&M presented in Sec. \ref{sssprotoq} does not hold. As long as the detection loophole is open, these security proofs cannot be applied to any system; but by re-introducing some knowledge of the devices, they might provide a good tool for disposing of all quantum side-channels (\ref{ssstrojan}).

\subsubsection{Toward longer distances: satellites and repeaters}
\label{sssexpfron}

The attempt of achieving efficient QKD over long distances is triggering the most ambitious experimental developments. Basically two solutions are being envisaged. The first is to use the techniques of free space quantum communication to realize \textit{ground-to-satellite links} \cite{but98,rar02,asp03}. The main challenges are technical: to adapt the existing optical tracking techniques to the needs of quantum communication, and to build devices that can operate in a satellite without need of maintenance.

The second solution are \textit{quantum repeaters} \cite{bri98,dur99}. The basic idea is the following: the link A-B is cut in segments A-C$_1$, C$_1$-C$_2$, ..., C$_n$-B. On each segment \textit{independently}, the two partners exchange pairs of entangled photons, which may of course be lost; but whenever both partners receive the photon, they store it in a quantum memory. As soon as there is an entangled pair on each link, the intermediate stations perform a Bell measurement, thus ultimately swapping all the entanglement into A-B. Actually, variations of this basic scheme may be more practical \cite{dua01}. Whatever the exact implementation, the advantage is clear: one does not have to ensure that all the links are active simultaneously; but the advantage can only be achieved if quantum memories are available. The experimental research in quantum memories has boosted over the last years, but the applications in practical QKD are still far away because the requirements are challenging (see Appendix \ref{appqmem}).

Teleportation-based links have been studied also in the absence of quantum memories (\textit{quantum relays}). They are rather inefficient, but allow to reduce the nuisance of the dark counts and therefore increase the limiting distance \cite{jac02,col03}; however, it seems simpler and more cost-effective to solve the same problem by using cryogenic detectors (see \ref{ssdet}).

\subsubsection{QKD in networks}

QKD is a point-to-point link between two users. But only a tiny fraction
of all communication is done in dedicated point-to-point links, most
communication takes place in networks, where many users are 
interconnected. Note that one-to-many connectivity between QKD devices can be obtained with optical switching \cite{tow94,ell02,ell05}.

In all models of QKD networks, the nodes are operated by authorized partners, while Eve can eavesdrop on all the links. If the network is built with quantum repeaters or quantum relays, no secret information is available to the nodes: indeed, the role of these nodes is to perform entanglement swapping, so that Alice and Bob end up with a maximally entangled --- therefore fully private --- state. Since quantum repeaters are still a challenge, \textit{trusted relays QKD networks} have been considered. In this case, the nodes learn secret information during the protocol. In the simplest model, a QKD key is created between two consecutive nodes and a message is encrypted and decrypted hop-by-hop. 
This model has been adopted by BBN Technologies and by the SECOQC QKD networks 
\cite{ell02,ell05,dia206,all07,dia208}. Alternatively, the trusted relays can perform an intercept-resend chain at the level of the quantum signal \cite{bec05}.

\subsection{QKD versus other solutions}

\label{sscompare}

Information-theoretically (unconditionally) secure key distribution (key agreement), is a cryptographic task that, as is well known, cannot be solved by public communication alone, i.e. without employing additional resources or relying on additional assumptions. Besides QKD, the additional resource in this case being the quantum channel, a number of alternative schemes to this end have been put forward \cite{Wyn75, csi78, Mau93, Ahls93}, to which one can also count the traditional "trusted courier" approach \cite{all07}. While the latter is still used in certain high security environments, QKD is the sole \textit{automatic}, practically \textit{feasible} and \textit{efficient} information-theoretically secure key agreement technology, whereby in the point-to-point setting, limitations of distance and related key rate apply. These limitations can be lifted by using QKD networks, see \ref{ssperspqkd}.

With this in mind, we address below typical secure communication solutions in order to relate this subsequently to the assets offered by QKD. Secure communication in general requires encrypted (and authentic) transition of communication data. In current standard cryptographic practice both the encryption schemes and the key agreement protocols used (whenever needed) are not unconditionally secure. While there is really a very broad range of possible alternatives and combinations, the most typical pattern for confidential communication is the following: public key exchange protocols are used to ensure agreement of two identical keys; the encryption itself is done using symmetric-key algorithms. In particular, most often some realization of the Diffie-Hellman algorithm \cite{diffie76new} is used in the key agreement phase. The symmetric-encryption algorithms most widely used today belong to the bloc-cipher class and are typically 3DES \cite{3DES}or AES \cite{AES}.

The security of the Diffie-Hellman algorithm is based on the assumption that the so called Diffie-Hellman problem is hard to solve, the complexity of this problem being ultimately related to the hardness of the discrete logarithm problem (see \cite{MW99, MW00} for a detailed discussion). It is widely believed, although it was never proven, that the discrete logarithm problem is classically hard to solve. This is not true in the quantum case, since a quantum computer, if available, can execute a corresponding efficient algorithm by Peter Shor \cite{sho94,sho97}, which is based on the same fundamental approach as is the Shor factoring algorithm, already mentioned in Sec. \ref{secintrocrypto}.

It should be further noted that that, similar to QKD, the Diffie-Hellman protocol can trivially be broken, if the authenticity of the communication channel is not ensured. There are many means to guarantee communication authenticity with different degrees of security but in any case additional resources are needed. In current common practice public key infrastructures are employed, which in turn rely on public-key cryptographic primitives (digital signatures), i.e. rely on similar assumptions as for the Diffie-Hellman protocol itself, \textit{and} on trust in external certifying entities.

Turning now to encryption it should be underlined that the security of a block-cipher algorithm is based on the assumption that it has no structural weaknesses, i.e. that only a brute force attack amounting to a thorough search of the key space (utilizing pairs of cipher texts and corresponding known or even chosen plain texts) can actually reveal the secret key. The cost of such an attack on a classical computer is $\mathcal{O}(N)$ operations, where $N$ is the dimension of the key space. The speed-up of a quantum computer in this case is moderate, the total number of operations to be performed being $\mathcal{O}(\sqrt{N})$ \cite{Gro96, Gro97}. The assumption on the lack of structural weaknesses itself is not related to any particular class of mathematical problems and in the end relies merely on the fact that such a weakness is not (yet) known. Cryptographic practice suggests that for a block-cipher algorithm such weaknesses are in fact discovered at the latest a few decades after its introduction\footnote{Vincent Rijmen, private communication.}.

Before turning to a direct comparison of the described class of secure communication schemes with QKD-based solutions, it should be explained why public-key based generation combined with symmetric-key encryption is actually the most proliferated solution. The reason is that currently AES or 3DES encryption, in contrast to direct public-key (asymmetric) encryption, can ensure a high encryption speed and appears optimal in this respect. Typically high speed is achieved by designing dedicated hardware devices, which can perform encryption at very high rate and ensure a secure throughput of up to 10Gb per second. Such devices are offered by an increasing number of producers (see e.g. ATMedia GmbH, www.atmedia.de) and it is beyond the scope of the current article to address these in any detail. We would like however to underline an important side-aspect. In general, security of encryption in the described scenario is increased by changing the key "often", the rate of change being proportional to the dimension of the key space. In practice, however, even in the high speed case, the key is changed at a rate lower than once per minute (often once per day or even more seldom). The reason for this is twofold: on the one hand public key agreement algorithms are generally slow and on the other, and more importantly, current design of the mentioned dedicated encryption devices is not compatible with a rapid key change.

The question now is how QKD compares with the standard practice as outlined above. It is often argued that QKD is too slow for practical uses and that the limited distance due to the losses is a limitation to the system as such. In order to allow for a correct comparison one has to define the relevant secure communication scenarios. There are two basic possibilities: (i) QKD is used in conjunction with One-Time Pad, (ii) QKD is used together with some high speed encryptor (we note in passing that the second scenario appears to be a main target for the few QKD producers).

The rate as a function of distance has been discussed in detail in the preceding sections. Here we shall consider an "average" modern QKD device operating in the range of 1 to 10kbps over 25 km; the maximal distance of operation at above 100 bps being around 100 km.

Case (i) obviously offers information-theoretic security of communication if the classical channel, both in the key generation and the encryption phase, is additionally authenticated with the same degree of security. As this overhead to this end is negligible the QKD generation rates as presented above are also the rates for secure communication. Obviously this is not sufficient for broad-band data transmission but pretty adequate for communicating very-highly sensitive data. Another advantage of this combination is the fact that keys can be stored for later use.

The security of the case (ii) is equivalent to the security of the high speed encryption, which we addressed above, while all treats related to the key generation-phase are eliminated. At 25 km the QKD speed would allow key refreshment (e.g. in the case of AES with 256 bit key length) of several times per second. This is remarkable for two reasons: first, this is on or rather beyond the key-exchange capacity of current high speed encryptors; second, it compares also to the performances of high level classical link encryptors, which refresh AES keys a few times per second using Diffie-Hellman elliptic curve cryptography for key generation.

So in the second scenario QKD over performs the standard solution at 25 km distance both in terms of speed and security.

Regarding the distance an interesting point is that classical high-end encryptors use direct dark fibers, not for reasons related to security but for achieving maximal speed, which also gives them a limitation in distance. However, classical key generation performed in software is naturally not bounded by the distance. In this sense standard public-key based key agreement appears superior. This is however a QKD limitation, which is typical for the point-to-point regime. As mentioned above, it is lifted in QKD networks.

\section*{Note added in proof} While this paper was being finalized, three groups have independently claimed to have solved one of the pending issues toward unconditional security proofs of CV QKD (see Sec.~\ref{statuscv}): namely, the fact that the security bound for collective and for general attacks should coincide asymptotically. On the one hand, a new exponential de Finetti theorem has been presented, which would apply to infinite-dimensional systems under some assumptions that are fulfilled in CV QKD \cite{ren08,zha08b}. A different argument reaches the same conclusion without any need for a de Finetti-type theorem altogether \cite{lev08}.

\section*{Acknowledgements} This paper has been written within the European Project SECOQC. The following members of the QIT sub-project have significantly contributed to the report that formed the starting point of the present review: Stefano Bettelli, Kamil Br\'adler, Cyril Branciard, Nicolas Gisin, Matthias Heid, Louis Salvail.

During the preparation of this review, we had further fruitful
exchanges with the above-mentioned colleagues, as well as with: Romain All\'eaume, Lucie Bart{\accent23 u}\v{s}kov\'a, Alexios Beveratos, Hugues De Riedmatten, Eleni Diamanti, Artur Ekert, Philippe Grangier, Fr\'ed\'eric Grosshans, Hannes Huebel, Micha\l{} Horodecki, Masato Koashi, Christian Kurtsiefer, Antia Lamas-Linares, Anthony Leverrier, Hoi-Kwong Lo, Chiara Macchiavello, Michele Mosca, Miguel Navascu\'es, Andrea Pasquinucci, Renato Renner, Andrew Shields, Christoph Simon, Kiyoshi Tamaki, Akihisa Tomita, Yasuhiro Tokura, Zhiliang Yuan, Hugo Zbinden.

\begin{appendix}

\section{Unconditional security bounds for BB84 and six-states, single-qubit signals}
\label{appshor}

In this Appendix, we present a derivation of the unconditional security bounds for the BB84 \cite{sho00} and the six-state protocol \cite{lo01} for the case where each quantum signal is a single qubit, or more generally when the quantum channel is a qubit channel followed by a qubit detection\footnote{For real optical channels, we assume therefore the tagging method for real sources and the squashing model for the detection, see \ref{sssquashing}.}.

As usual, the proof is done in the EB scheme, the application to the P\&M case following immediately as discussed in Sec. \ref{sssprotoq}. Alice produces the state $\ket{\Phi^+}=\frac{1}{\sqrt{2}}\left(\ket{00}+\ket{11}\right)$, she keeps the first qubit and sends the other one to Bob. This state is such that $\moy{\si_z\otimes\si_z}=\moy{\si_x\otimes\si_x}=+1$ (perfectly correlated outcomes) and $\moy{\si_y\otimes\si_y}=-1$ (perfectly anti-correlated outcomes); to have perfect correlation in all three bases, Bob flips his result when he measures $\sigma_y$. We suppose an \textit{asymmetric implementation} of the protocols: the key is extracted only from the measurements in the $Z$ basis, which is used almost always; the other measurements are used to estimate Eve's knowledge on the $Z$ basis, and will be used on a negligible sample (recall that we work in the asymptotic regime of infinitely long keys).

Now we follow the techniques of \cite{kra05,ren05b}. Without loss of generality, the symmetries of the BB84 and the six-state protocols\footnote{Actually, a lower bound can be computed in the same way for a very general class of protocols; but it may not be tight, as explicitly found in the case of SARG04 \cite{bra05,kra07}.} imply that one can compute the bound by restricting to collective attacks, and even further, to those collective attacks such that the final state of Alice and Bob is \textit{Bell-diagonal}:
\ba
\rho_{AB}&=&\lambda_1\ket{\Phi^+}\bra{\Phi^+}+\lambda_2\ket{\Phi^-}\bra{\Phi^-}\nonumber\\
&&+\lambda_3\ket{\Psi^+}\bra{\Psi^+}+\lambda_4\ket{\Psi^-}\bra{\Psi^-}
\ea
with $\sum_i\lambda_i=1$. Since $\ket{\Phi^{\pm}}$ give perfect correlations in the $Z$ basis, while $\ket{\Psi^{\pm}}$ give perfect anti-correlations, the QBER $\varepsilon_z$ is given by
\ba
\varepsilon_z&=&\lambda_3+\lambda_4\,.
\ea
The error rates in the other bases are
\ba
\varepsilon_x=\lambda_2+\lambda_4&\;,\;&\varepsilon_y=\lambda_2+\lambda_3\,.
\ea
Eve's information is given by the Holevo bound (\ref{chiholevo})
$I_{E}=S(\rho_E)-\demi S(\rho_{E|0})-\demi S(\rho_{E|1})$ since both values of the bit are equiprobable in this attack. Since Eve has a purification of $\rho_{AB}$, $S(\rho_E)=S(\rho_{AB})=H\left(\{\lambda_1,\lambda_2,\lambda_3,\lambda_4\}\right)\equiv H(\underline{\lambda})$ where $H$ is Shannon entropy. The computation of $\rho_{E|b}$ is made in two steps. First, one writes down explicitly the purification\footnote{All purifications are equivalent under a local unitary operation on Eve's system, so Eve's information does not change with the choice of the purification.} $\ket{\Psi}_{ABE}=\sum_i\sqrt{\lambda_i}\ket{\Phi_i}_{AB}\ket{e_i}_E$, where we used an obvious change of notation for the Bell states, and where $\braket{e_i}{e_j}=\delta_{ij}$. Then, one traces out Bob and projects Alice on $\ket{+z}$ for $b=0$, on $\ket{-z}$ for $b=1$. All calculations done, the result is $S(\rho_{E|0})=S(\rho_{E|1})=h(\varepsilon_z)$. So we have obtained
\ba
I_E(\underline{\lambda})&=&H(\underline{\lambda})-h(\varepsilon_z)\,.
\ea
Now we have to particularize to the two protocols under study.

Let's start with the \textit{six-state protocol}. In this case, both $\varepsilon_x$ and $\varepsilon_y$ are measured, so all the four $\lambda$'s are directly determined. After easy algebra, one finds
\ba
I_E(\underline{\varepsilon})&=&\varepsilon_z\, h\left[\frac{1+(\varepsilon_x-\varepsilon_y)/\varepsilon_z}{2}\right]\nonumber\\
&&+(1-\varepsilon_z)\,h\left[\frac{1-(\varepsilon_x+\varepsilon_y+\varepsilon_z)/2}{1-\varepsilon_z}\right]\,.
\ea 
Under the usual assumption of a depolarizing channel, $\varepsilon_x=\varepsilon_y=\varepsilon_z=Q$, this becomes
\ba
I_E(Q)&=&Q+(1-Q)\,h\left[\frac{1-3Q/2}{1-Q}\right]\,.
\ea
The corresponding secret fraction (one-way post-processing, no pre-processing and perfect error correction) is $r=1-h(Q)-I_E(Q)$, which goes to 0 for $Q\approx12.61\%$.

The calculation is slightly more complicated for \textit{BB84}, because there only $\varepsilon_x$ is measured; therefore, there is still a free parameter, which must be chosen as to maximize Eve's information. The simplest way of performing this calculation consists in writing $\lambda_1=(1-\varepsilon_z)(1-u)$, $\lambda_2=(1-\varepsilon_z)u$, $\lambda_3=\varepsilon_z(1-v)$, $\lambda_4=\varepsilon_zv$, where $u,v\in[0,1]$ are submitted to the additional constraint \ba(1-\varepsilon_z)u+\varepsilon_zv&=&\varepsilon_x\,.\label{constrapp}\ea Under this parametrization, $H(\underline{\lambda})=h(\varepsilon_z)+(1-\varepsilon_z)h(u)+\varepsilon_z h(v)$ and consequently
\ba
I_E(\underline{\lambda})&=&(1-\varepsilon_z)h(u)+\varepsilon_z h(v)
\ea 
to be maximized under the constraint (\ref{constrapp}). This can be done easily by inserting $v=v(u)$ and taking the derivative with respect to $u$. The result is that the optimal choice is $u=v=\varepsilon_x$ so that
\ba
I_E(\underline{\varepsilon})&=&h\left(\varepsilon_x\right).
\ea 
The usual case is $\varepsilon_x=\varepsilon_z=Q$, which however here does not correspond to the depolarizing channel: the relations above imply $\varepsilon_y=2Q(1-Q)$, which corresponds to the application of the so-called ``phase-covariant cloning machine'' \cite{gri97,bru00}. The corresponding secret fraction (again for one-way post-processing, no pre-processing and perfect error correction) is $r=1-h(Q)-I_E(Q)$, which goes to 0 for $Q\approx11\%$.

\section{Elementary estimates for quantum repeaters}
\label{appqmem}

\subsection{Quantum memories}

A quantum memory is a device that can store an incoming quantum state (typically, of light) and re-emit it on demand without loss of coherence. A full review of the research in quantum memories is clearly beyond our scope. Experiments are being pursued using several techniques, like atomic ensembles \cite{jul04,cho07}, NV centers \cite{chi06}, doped crystals \cite{ale06,sta07}.

Two characteristics of quantum memories are especially relevant for quantum repeaters. A memory is called \textit{multimode} if it can store several light modes and one can select which mode to re-emit; multimode memories are being realized \cite{sim07}. A memory is called \textit{heralded} if its status (loaded or not loaded) can be learned without perturbation; there is no proposal to date on how to realize such a memory, and repeater schemes have been found that work without heralded memories \cite{dua01}.

\subsection{Model of quantum repeater}

\begin{figure}
\includegraphics[scale=0.2]{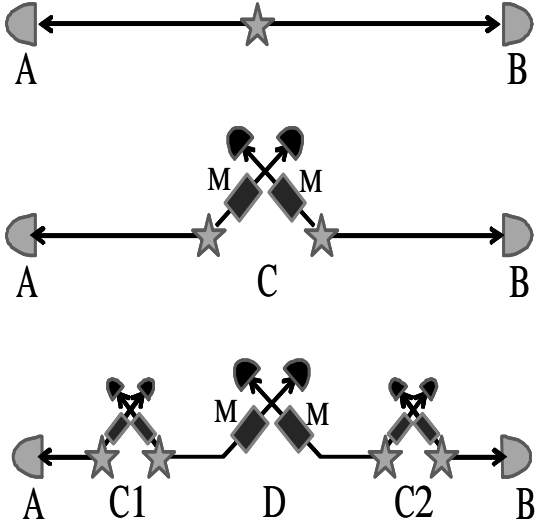}
\caption{Three configurations for quantum repeaters: direct link, two-link repeater and four-link repeater.}\label{fig:repet}
\end{figure}

Here we present a rapid comparison of the \textit{direct link} with the \textit{two-link repeater} and discuss the advantages and problems that arise in more complex repeaters. We consider the architecture sketched in Fig.~\ref{fig:repet}, corresponding to the original idea \cite{bri98}.

\subsubsection{Definition of the model}

Our elementary model is described as follows:
\begin{itemize}
\item Source: perfect two-photon source with repetition rate $\nu_S$;
\item Quantum channel: the total distance between Alice and Bob is $\ell$. The channel is noiseless; its losses characterized by $\alpha$, we denote $t=10^{-\alpha \ell/10}$ the total transmittivity.
\item Detectors of Alice and Bob: efficiency $\eta$; neglected dark counts, dead-time and other nuisances.
\item Quantum memories: multimode memories that can store $N$ modes. We write $p_M$ the probability that a photon is absorbed, then re-emitted on demand (contains all the losses due to coupling with other elements). The memory has a typical time $T_M$, that we shall consider as a life-time\footnote{That is, photons may be lost but do not decohere in the memory. Note that this can be the case even if the atoms, which form the memory, do undergo some decoherence \cite{sta07}.}.
\item Bell measurement: linear optics, i.e. probability of success $\frac{1}{2}$. Fidelity $F$, depolarized noise (i.e. a detection comes from the desired Bell state with probability $F$, from any of the others with equal probability $(1-F)/3$). The detectors have efficiency $\eta_M$ and no dark counts.
\end{itemize}

\subsubsection{Detection rates}

For the \textit{direct link}, the key rate is just the detection rate in our simplified model:
\ba
K_1=R_1&=&\nu_S t \eta^2\,.
\ea

In the \textit{two-link repeater}, the central station (Christoph) holds the two sources and the memories. Consider one of the links, say with Alice. The source produces groups of $N$ pairs, each pair in a different mode; one photon per pair is kept in the memory, the other is sent to Alice. Alice announces whether she has detected at least one photon: if she has, Christoph notes which one; if she has not, Christoph releases the memory and starts the protocol again. The same is happening on the other link, the one with Bob, independently. As soon as both partners have announced a detection, Christoph releases the corresponding photons, performs the Bell measurement and communicates the result to Alice and Bob, who post-select their results accordingly\footnote{Recall that there is no time-ordering in quantum correlations: so, this procedure gives exactly the same statistics as the ``usual'' entanglement swapping, in which the Bell measurement is made beforehand.}. Note that the memories need not be heralded in this scheme.

Here is the quantitative analysis of the two-link repeater. Any elementary run takes the time for the photon to go from the source to the detector, then for the communication to reach back Christoph, i.e. $\ell/c$. In each run, the probability of a detection is $1-(1-\sqrt{t}\eta)^N\approx N\sqrt{t}\eta$. Then, in average, the Bell measurement will be performed after a time\footnote{In fact, let $x=1-(1-\sqrt{t}\eta)^N$: the probability that Alice's (Bob's) detector is activated by the $m$-th group of $N$ pairs is $p_1(m)=x(1-x)^{m-1}$. Therefore, the probability that both links are activated exactly by the $n$-th repetition is $p(n)=2p_1(n)p_1(<n)+p_1(n)^2=x(1-x)^{n-1}[2-(2-x)(1-x)^{n-1}]$ with $p_1(<n)=\sum_{m=1}^{n-1}p_1(m)$. Finally, the number of repetitions needed to establish the link is $\moy{n}=\sum_n np(n)=\frac{1}{x}\left(\frac{3-2x}{2-x}\right)$.} $\tau\approx \frac{3}{2}\frac{\ell/c}{\,N \sqrt{t}\eta}$. Consequently,
\ba
R_2&=&\left\{\begin{array}{cl} \tau^{-1}\,\demi p_M^2\eta_M^2 &\mbox{if }\tau<T_M\\
0&\mbox{otherwise}\end{array}\right.
\ea
where we have supposed that the memory time $T_M$ defines a sharp cut, which is another simplification. This is the expected result: $R_2$ scales with $\sqrt{t}\eta$ and not with $t\eta^2$, because each link can be activated independently.
Finally, in our model, the error rate is uncorrelated with the other parameters and only due to the fidelity of the Bell measurement; so \ba
K_2&=&R_2\,\left[1-2h(\varepsilon)\right]\ea with $\varepsilon=\frac{2}{3}(1-F)$ because one of the ``wrong'' Bell states gives nevertheless the correct bit correlations. In particular, the fidelity of a Bell measurement must exceed 83.5\% to have $K_2>0$.

\begin{figure}
\includegraphics[scale=0.6]{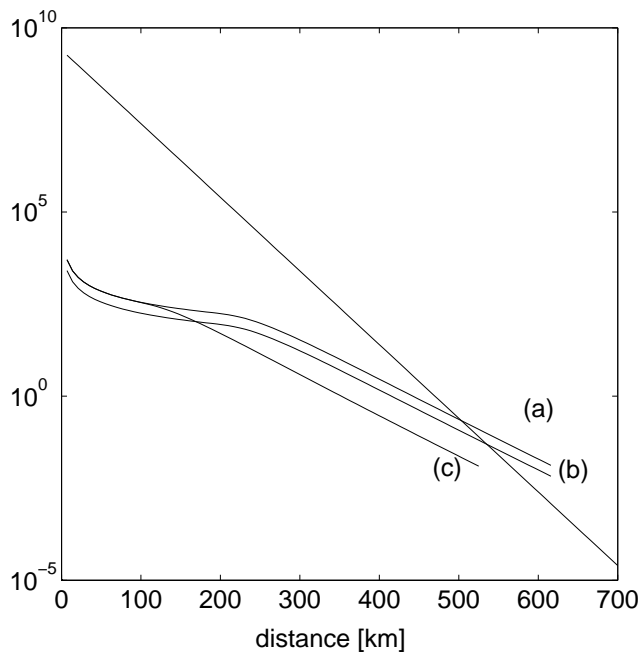}
\caption{Comparison of $K_1$ (straight line) and $K_2$. For all curves: $\nu_S=10$GHz, $\eta=0.5$, $\eta_M=0.9$, $p_M=0.9$, $\alpha=0.2$dB/km (fibers), $T_M=10$s. Line (a): best case, $N=1000$, $F=0.95$; line (b): $N=1000$, fidelity reduced to $F=0.9$; line (c): supported modes reduced to $N=100$, $F=0.95$.}\label{fig:qmem}
\end{figure}

Some plots of $K_1$ and $K_2$ as a function of the distance are shown in Fig.~\ref{fig:qmem}. The chosen values are already optimistic extrapolations of what could be achieved in a not too distant future. We notice that quantum repeaters overcome the direct link for $\ell\gtrsim 500$km in fibers; with $\eta=0.5$ and $N=1000$, this requires $T_M\approx 10$s. Also, the number of modes supported by the memory is a more critical parameter than the fidelity of the Bell measurement. This analysis provides a rough idea of the performances to be reached in order for quantum repeaters to be useful.

For the next step, the \textit{four-link repeater}, we content ourselves with a few remarks. The four-link repeater allows in principle to reach the scaling $R_4\propto t^{1/4}$. The requirements for a practical implementation, however, become more stringent: the four memories must be released before $T_M$; there are three Bell measurements, so $\varepsilon<11\%$ requires $F\gtrsim 95\%$; also, $p_{M'}\approx p_M t^{1/4}$. Moreover, it is easy to realize that the basic scheme (Fig.~\ref{fig:repet}) requires heralded memories, although other schemes do not \cite{dua01}.

\end{appendix}

\bibliographystyle{apsrmp}


\end{document}